\documentclass[epj,nopacs,twocolumn,final,numbook]{svjour}
\usepackage{latexsym}
\usepackage{graphics}
\usepackage{dblfloatfix}

\makeatletter
\newinsert \@kludgeins
\global\dimen\@kludgeins \maxdimen
\global\count\@kludgeins 1000
\gdef \enlargethispage {%
   \@ifstar
     {%
      \@enlargepage{\hbox{\kern\p@}}}%
     {%
      \@enlargepage\@empty}%
}
\gdef\@enlargepage#1#2{%
   \@tempskipa#2\relax
   \ifdim \@tempskipa>.5\maxdimen
     \@latexerr{Suggested\space extra\space height\space
                (\the\@tempskipa)\space dangerously\space
                large}\@eha
   \else
     \ifdim \vsize<.5\maxdimen
       \@bsphack
         \insert\@kludgeins{#1\vskip-\@tempskipa}%
       \@esphack
     \else
       \@latexerr{Page\space height\space already\space
                  too\space large}\@eha
     \fi
   \fi
}
\makeatother

\newcommand{\Vo}{$\omega$}
\newcommand{\Vr}{$\rho$}
\begin{document}
\title{e$^+$e$^-$-pair production in Pb-Au collisions at 
158 GeV per nucleon}
\authorrunning{G.~Agakichiev et al.}
\titlerunning{e$^+$e$^-$-pair production in Pb-Au collisions at 158~GeV/nucleon }
\author{G.~Agakichiev\inst{1}\thanks{\emph{Present address:
2.~Physikalisches Institut der Universit\"at Giessen, 35392 Giessen,
Germany}}, H.~Appelsh\"auser\inst{2}\thanks{\emph{Present address:
Institut f\"ur Kernphysik der Universit\"at Frankfurt, 60438
Frankfurt, Germany}}, J.~Bielcikova\inst{2,3}\thanks{\emph{Present
address: Nuclear Structure Laboratory, Yale University, New Haven,
Ct.~06511, USA }}, R.~Baur\inst{2}, P.~Braun-Munzinger\inst{1},
A.~Cherlin\inst{4}, S.~Damjanovic\inst{2}\thanks{\emph{Present
address: Department PH, CERN, Geneva 023, Switzerland}},
A.~Drees\inst{5}, S.~Esumi\inst{2}\thanks{\emph{Present address:
National Laboratory for High Energy Physics, Tsukuba 305, Japan }},
U.~Faschingbauer\inst{3,2}, Z.~Fraenkel\inst{4}, Ch.~Fuchs\inst{3},
E.~Gatti\inst{6}, P.~Gl\"assel\inst{2}, G.~Hering\inst{1}, C.P.~de los
Heros\inst{4}\thanks{\emph{Present address: Division of High Energy
Physics, Uppsala University, 75121 Uppsala, Sweden}},
P.~Holl\inst{7}\thanks{\emph{Present address: PN Sensor GmbH c/o 
MPI Halbleiterlabor, 81739 M\"unchen, Germany}}, Ch.~Jung\inst{2}, 
B.~Lenkeit\inst{2}, A.~Mar\'{i}n\inst{1}, F.~Messer\inst{2,3}, 
M.~Messer\inst{2},
D.~Mi\'{s}kowiec\inst{1}, O.~Nix\inst{3}\thanks{\emph{Present
address: Deutsches Krebsforschungszentrum, 69120 Heidelberg, Germany}},
Yu.~Panebrattsev\inst{8},
A.~Pfeiffer\inst{2}\thanks{\emph{Present address: CERN,
Geneva 023, Switzerland}}, J.~Rak\inst{3}\thanks{\emph{Present
address: Department of Physics and Astronomy, Iowa State University,
Ames, Ia 50011, USA}}, I.~Ravinovich\inst{4}, S.~Razin\inst{8},
P.~Rehak\inst{7}, M.~Richter\inst{2}, M.~Sampietro\inst{6},
H.~Sako\inst{1}\thanks{\emph{Present address: Japan Atomic Research
Institute~(JAERI), Tokai-mura, Ibaraki-ken 319-1195, Japan }},
N.~Saveljic\inst{2}, W.~Schmitz\inst{2}, J.~Schukraft\inst{9},
W.~Seipp\inst{2}, S.~Shimanskiy\inst{8},
E.~Socol\inst{4}\thanks{\emph{Present address: School of Electrical
Engeneering, Tel Aviv University, Ramat Aviv 69978, Israel}},
H.J.~Specht\inst{2}, J.~Stachel\inst{2},
G.~Tel-Zur\inst{4}\thanks{\emph{Present address: Physics Department,
NRCN Beer Sheva 84190, Israel}}, I.~Tserruya\inst{4},
T.~Ullrich\inst{2}\thanks{\emph{Present address: Physics Division,
Brookhaven National Laboratory, Upton, New York 11973-5000, USA }},
C.~Voigt\inst{2}, S.~Voloshin\inst{2}\thanks{\emph{Present address:
Department of Physics and Astronomy, Wayne State University, Detroit,
Mi 48202, USA }}, C.~Weber\inst{2}, J.P.~Wessels\inst{2,10},
T.~Wienold\inst{2}, J.P.~Wurm\inst{3}, V.~Yurevich\inst{8} (CERES
Collaboration) }
\institute{Gesellschaft f\"ur Schwerionenforschung (GSI), 64291 Darmstadt,
Germany
 \and 
Physikalisches Institut der Universit\"at Heidelberg, 69120
Heidelberg, Germany
 \and 
Max-Planck-Institut f\"ur Kernphysik,
69229 Heidelberg, Germany 
 \and 
Department for Particle Physics,
Weizmann Institute of Science, Rehovot 76100, Israel
 \and 
\hbox{Department of Physics and Astronomy, State University of New York at
Stony Brook, Stony Brook, New York 11974, USA}
 \and 
Politecnico di Milano, Istituto di Fisica, 20133 Milano, Italy
 \and 
Instrumentation Division, Brookhaven National Laboratory, Upton, New York 11793-5000, USA
 \and 
Laboratory for High Energy (JINR), 141980 Dubna, Russia
 \and 
CERN, Geneva 023, Switzerland 
 \and 
Institut f\"ur Kernphysik der Universit\"at M\"unster, 48149 M\"unster, 
Germany 
}
\date{Received:  2 March 05}
\abstract{
We present the combined results on electron-pair production in
158\,GeV/n \mbox{Pb-Au} ($\sqrt{s}$=\,17.2\,GeV) collisions taken at
the CERN SPS in 1995 and 1996, and give a detailed account of the data
analysis.  The enhancement over the reference of neutral meson decays
amounts to a factor of
2.31$\pm0.19\,(stat.)\pm0.55\,(syst.)\pm0.69\,(decays)$ for
semi-central collisions (28$\%$ $\sigma/\sigma_{geo}$) when yields are
integrated over $m>$\,200\,MeV/$c^2$ in invariant mass. The measured
yield, its stronger-than-linear scaling with $N_{ch}$, and the
dominance of low pair $p_t$ strongly suggest an interpretation as {\it
thermal radiation} from pion annihilation in the hadronic
fireball. The shape of the excess centring at
$m\approx$\,500\,MeV/$c^2$, however, cannot be described without
strong medium modifications of the $\rho$ meson. The results are put
into perspective by comparison to predictions from Brown-Rho scaling
governed by chiral symmetry restoration, and from the
spectral-function many-body treatment in which the approach to the
phase boundary is less explicit.
\PACS{
      {PACS-key}{describing text of that key}   \and
      {PACS-key}{describing text of that key}
     } 
} 
\maketitle
\section{Introduction}
\label{intro}
Strongly interacting matter under extreme conditions of temperature
and density is being created by colliding heavy nuclei in fixed target
experiments at the Super-Proton Synchrotron (SPS) at CERN and at the
Relativistic Heavy-Ion Collider (RHIC) at BNL. The motivation derives
from the quest to discover `quark matter', the quark gluon plas\-ma
(QGP), in which quarks and gluons are deconfined and chiral symmetry
is restored~\cite{quark-matter}, two complementary facets of the phase
transition predicted by Quantum Chromodynamics (QCD). This transition
is expected to occur at a critical energy density of
$\epsilon\approx$\,0.7~GeV/fm$^3$ and a temperature of 170-180~MeV as
finite-temperature lattice calculations have shown~\cite{karsch02}.
The initial energy density at full SPS energy is appreciably larger
and reaches about 3~GeV/fm$^3$ in central Pb-Pb collisions, adopting
Bjorken's longitudinal expansion scenario for a simple
estimate~\cite{bjorken83}.  While the existence of the QGP has not yet
been proven, there is circumstantial evidence that its transient
formation is imprinted, in one way or another and to varying degrees,
on measured distributions of final-state hadrons. Since the fireball
terminates in an exploding multi-hadron final state, QGP signatures,
if not collective in character, are prone to be masked by hadronic
interactions.

\begin{figure*}[ttt!]
\resizebox{0.9\textwidth}{!}{%
  \includegraphics{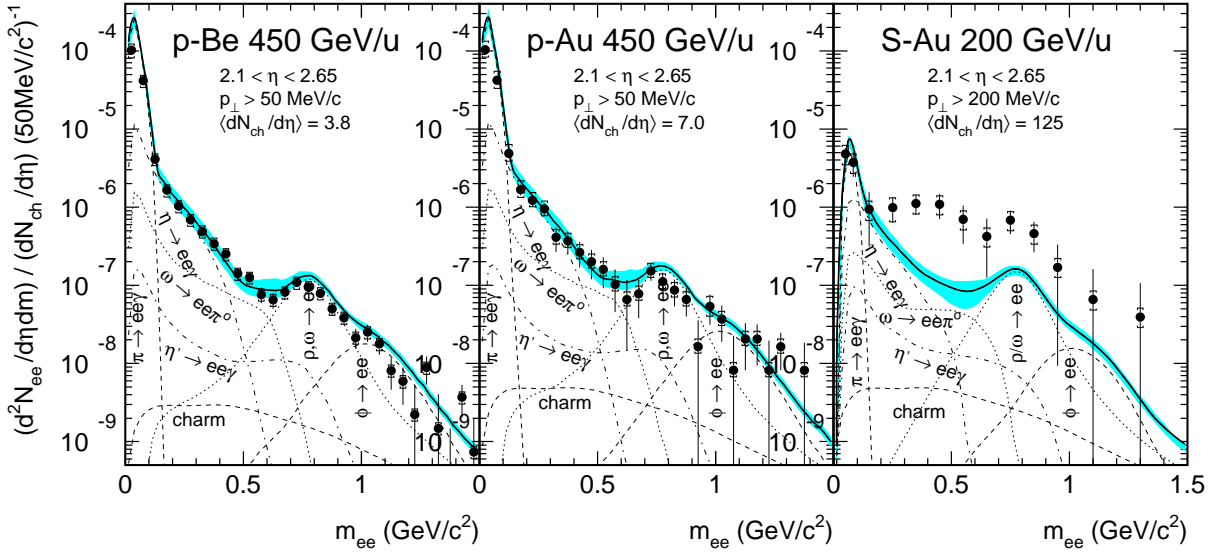} } 
\caption{CERES inclusive $e^+e^-$ mass
  spectra of 450~GeV p-Be, p-Au, and 200~GeV/n
  S-Au collisions ~\cite{neutral-meson-pBe,prl1995}.  Plotted is the number of
  electron pairs per charged particle, both in the acceptance and per
  event. Contributions from various hadron decays as expected from p-p
  collisions are shown together with their sum (thick line), and the
  systematic error on the latter is indicated by the shaded area.  In
  S-Au a higher single-electron $p_t$ cut reduces the $\pi^\circ$
  Dalitz component. See text and footnote.}
\label{fig:1.1}
\end{figure*}

Photons and dileptons are potentially more direct pro\-bes of the early
collision stages since they escape from the impact zone nearly
undisturbed by final-state interactions and have their largest
emission rates in hot and dense matter.  Moreover, according to the
vector dominance model~\cite{sakurai}, dilepton production is mediated
in the hadronic phase by the light neutral vector mesons $\rho,
\omega$, and $\phi$ which mark the low-mass region by
distinctive resonance peaks. Among these, especially the short-lived\\
$\rho(770)$ meson ($\tau=$~1.3~fm/$c$) has acquired a key role as test
particle for `in-medium modifications' of hadron properties close to
the QCD phase
boundary~\cite{brownrho1991,rapp-wambach2000,b-r03}. Changes in
position and width of the $\rho$ have been advocated already 20 years
ago as precursor signals of the chiral
transition~\cite{pisarski1982}.  Restoration of chiral symmetry in hot
and dense matter has become one of the heavily discussed and
exciting issues in non-perturbative QCD
thermodynamics~\cite{rapp-wambach2000,b-r03,hats96,b-r96,wilcz00} as
the melting of the chiral condensate should cause rather drastic
changes of the properties of the light vector mesons and thereby on
the structure of dilepton spectra. Spectral function calculations on
the lattice seem still far from providing model-independent guidelines
for the study of thermal modifications of hadron
properties~\cite{karsch02}.

An enhancement of low-mass lepton pair production was first reported
by CERES~\cite{prl1995,wurm-qm95} and
HELIOS-3~\cite{masera1995,angelis1998} on the basis of 200~GeV/n S-Au
and S-W data, respectively. The CERES $e^+e^-$ mass spectrum is shown
in Fig.~\ref{fig:1.1} together with the reference spectra for
450~GeV p-Be and p-Au
collisions~\cite{neutral-meson-pBe}\footnote{The notation of
the ordinate in Fig.~\ref{fig:1.1} is no longer in use; it should
read as in all other mass spectra shown in this paper. The
`cocktail' has received minor adjustments in the meantime which do not
affect any of the conclusions drawn.}. While the p-A data are
reproduced within errors by final state Dalitz and direct decays of
neutral mesons as known from p-p collisions, electron pairs from S-Au
collisions reveal a substantial enhancement in the mass range
above 250~MeV/$c^2$.

At top SPS energy and close to the critical temperature, the prime
candidate for `thermal radiation' from the hadronic phase of the
fireball~\cite{pipi-anni,mclerran85} is pion annihilation,
\begin{equation}
\pi^+\pi^-\rightleftharpoons\rho\rightarrow\,e^+e^-.
\end{equation}
This thermal process with a threshold at $2\,m_\pi$ is dynamically
enhanced via the electro-magnetic form factor of the pion by the
$\rho$ resonance~\cite{cleymans1993}, with a dilepton branching of 1
in $10^4$.  Yet, the $\rho$ serves not only as test particle, but its
strong coupling to the $\pi\pi$ channel makes it also a major
constituent of hot hadronic matter. Numerous theoretical approaches
incorporating pion annihilation using {\it vacuum properties} of the
$\rho$ meson, failed without exception to describe the
data~\cite{drees1996}.

This suggested in-medium changes of the $\rho$ spectral function that
shift dilepton strength down to lower masses.  The first
calculations that were successful in describing both the CERES and the
\mbox{HELIOS-3} data made use of the scaling conjecture of Brown and 
Rho~\cite{brownrho1991}, which postulates that the mass of non-strange
vector mesons decreases in dense matter together with the scalar
quark condensate, the order parameter of the chiral transition.  This
`dropping mass' scenario received independent support by work on QCD
sum rules~\cite{hatsuda-lee1992}, and it turned out like a tailor-made
concept: linked to a fireball model~\cite{likob95,likobs96} or
embedded into transport calculations~\cite{casseheko95,cassehekr96},
it gave excellent fits to the data.

But it was also pointed
out~\cite{dey90,rapp-chanfray-w1996,harada97,mishra02} that chiral
symmetry considerations alone would only require that masses of chiral
partners, here the vector meson $\rho$ and the axial vector meson
$a_1$, become degenerate, but by no means necessarily
massless. Moreover, significant mixing must accompany any mass shift
when approaching the phase transition along
$T$~\cite{kim-rapp-brown-rho1999}. The `dropping mass' scaling idea
has been very recently revisited by the original authors~\cite{b-r03}
welcoming an alternative scheme of how chiral symmetry might be
restored, the Georgi vector limit~\cite{georgi90} in which the chiral
partner of the (longitudinal) $\rho$ is the Goldstone pion, both
becoming massless in approach of the chiral
transition~\cite{halasz97,harada01}.  Thermal modifications of hadron
properties in general, and of in-medium spectral functions in
particular, have not yet come within reach of QCD lattice calculations
with finite baryon density~\cite{karsch02}. Very preliminary results
suggesting dropping vector meson masses have been
reported~\cite{muroya2003}.

An alternative approach to explain the low-mass dilepton enhancement
in the CERN SPS data focused on the calculation of spectral functions
in a hot and strongly interacting hadron resonance gas at finite
baryon density by conventional many-body
techniques~\cite{rapp-wambach2000,cassing-brat1999}.  These confirmed
earlier calculations of the two-pion self-energy in nuclear matter
which indicated that the $\rho$ spectral function suffers significant
broadening but only negligible shift in
mass~\cite{herm-frim-noe1993,chan-schu1993}. The many-body
calculations combined with suitable reaction models  describe
the observations actually quite
well~\cite{klingl-weise1996,rapp-chanfray-w1997,cassing1997,rw98}: the
low-mass wing of the broadened in-medium $\rho$ spectral function
receives strong thermal Bose enhancement which results in an amplified
dilepton strength considerably {\it below} the vacuum $\rho$ position,
while simultaneously the yield at the vacuum position was depleted,
consistent with approximate unitarity~\cite{rapp-gale1999}.

Our discussion so far was limited to aspects of dilepton emission
rates with possible modifications by the medium.  However, total pair
yields derive from space-time integration over {\it a priori} unknown
density and temperature profiles which are usually modelled by
hydro-dynamical~\cite{sollfrank97,hung98} or microscopic
transport~\cite{cassing-brat1999,likobs96,casseheko95} calculations,
or fireball models~\cite{likob95,renk02}. Certainly, the external
inputs, e.g.\@ to the hydro-dynamical and fireball calculations, have to
conform with whatever knowledge there is on initial conditions,
depending on collision geometry, and on the ($\mu$, $T$)
coordinates of the trajectory in the phase diagram. The medium
modifications of the $\rho$ suggested by the SPS results seem to
require a strongly interacting, hot and dense hadronic fireball with
sufficient time spent between hadronisation and thermal freeze-out. If
this time would be insufficient, the enhanced dilepton production may
have stronger links to the hadronisation stage or the plasma phase
than hitherto assumed.

The CERES Collaboration measured dilepton production in 158\,GeV/n
Pb-Au collisions in 1995 and 1996 with a greatly improved
setup~\cite{baur1996,ullrich-qm96} compared to the sulfur-beam
experiment.  The main objective of the Pb runs has been achieved: to
corroborate with a large statistics sample the enhanced dielectron
production at low masses for the heavy Pb-Au collision
system~\cite{ullrich-qm96,ravinovich-qm97,plb422,lenkeit-qm99,lenkeit-paris}.
Improved background rejection was achieved which was compulsory
in an environment of very large rapidity density of hadrons and
secondary photons. Among the physics goals considered most important
for further insight into the nature of the processes at work was the
centrality dependence of the enhancement. 

The significance of the baryon chemical potential for in-medium
modifications at SPS energies prevailing over that of pion number,
or temperature, as found in most
calculations~\cite{cassing-brat1999,rapp-wambach2000}, but not in
all~\cite{bleicher00}, was given experimental support by the recent
finding of an even somewhat larger enhancement measured in the CERES
Pb-Au low energy run at 40~GeV/n~\cite{adamova-ee03,damjanovic-phd},
compared to that at 158\,GeV/n.

This paper presents the combined results of all data on electron pair
production in 158\,GeV/n Pb-Au collisions taken by the CERES
Collaboration in the years 1995 and 1996. Most of the analyses were
performed in the course of Doctoral Dissertations in Heidelberg, Darmstadt
and Rehovot dealing with the 1995~\cite{voigt-phd,socol-phd} and
1996~\cite{socol-phd,lenkeit-phd,hering-phd} data.  Publications of
analysis results of the 1995
data~\cite{ullrich-qm96,ravinovich-qm97,plb422} and the 1996
data~\cite{lenkeit-qm99,lenkeit-paris} are superseded by the combined
results presented here. There are no major deviations of the unified
results reported in this paper to those published previously for the
separate data sets.

The paper starts with a description of the experimental setup and
summarises the instrumental means CERES has at its disposal to cope
with background. A detailed description of the data analysis is given
in sect.~\ref{sec:3} which concludes with the centrality
determination. The Monte-Carlo simulation method applied for measuring
reconstruction efficiency and optimising the rejection of
combinatorial background is addressed in sect.~\ref{sec:4}. The
`cocktail' of hadron decays which serves as an important reference for
electron-pair production is discussed in sect.~\ref{sec:5}. Results of
both data sets are presented in sect.~\ref{sec:6} which also includes
a discussion of statistical and systematic errors. The final mass and
transverse momentum spectra are presented and compared to the hadronic
cocktail in sect.~\ref{sec:7}.  Section~\ref{sec:8} contains a physics
discussion, from an experimentalist point of view, on the comparison
of data to current theoretical models.  The paper concludes by
summarising what has been achieved and which issues are still open but
might be clarified in the not too distant future.

\begin{figure*}[t!]
\begin{center}
\resizebox{0.65\textwidth}{!}{%
  \includegraphics{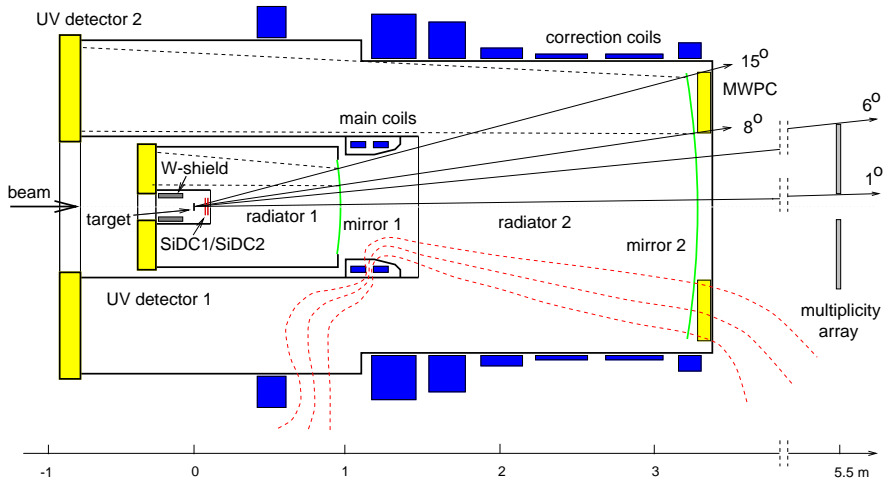} } 
\end{center}
\caption { CERES setup during the 1995/1996 Pb-beam runs}
    \label{fig:2.1}
\end{figure*}

\section{The CERES experiment in 1995/96}
\label{sec:2}

CERES is dedicated~\cite{proposal88} to the measurement of electron
pairs in the low-mass range from 50~MeV/$c^2$ up to about
$\sim$1.5~GeV/$c^2$; an upper mass limit is imposed by counting
statistics due to the rapid decline in cross section. The spectrometer
covers the pseudo-rapidity region close to mid-rapidity, 2.1 $<$ $\eta$
$<$ 2.65. It is axially symmetric around the beam and has 2$\pi$
azimuthal coverage. Transverse pair momenta are accepted down to
$\sim$20~MeV/$c$ for masses above $\sim$400~MeV/$c^2$. These are great
assets when investigating soft processes.  CERES maintains
transpa\-rency for hadrons and photons as strictly as possible.

A schematic view of the spectrometer in the run periods 
1995 and 1996 is shown in Fig.~\ref{fig:2.1}.  At the heart of the
spectrometer are the two coaxial ring-imaging Cherenkov detectors
\cite{rich_set}, one (RICH-1) within the other (RICH-2) along the
beam and separated by a compact super-cond\-ucting solenoid for
momentum analysis. A doublet of sili\-con-drift chambers
(SiDC)~\cite{chen93} replaces the previously used single SiDC to
enable precise charged-particle tracking into RICH-1 for rejection of
close tracks, and for off-line measurement of charged multiplicity. A
multi-wire proportional counter (Pad Chamber) with pad readout was
added behind the mirror of RICH-2 to provide external tracking
downstream of RICH-2. The new tracking detectors outside the field
were added to cope with the high multiplicities of Pb-Au
collisions~\cite{pb-proposal94,cogne95}. These upgrades are described
in Ref.~\cite{baur1996} and Ref.~\cite{holl96} for RICH and SiDC,
respectively.  A multiplicity detector (MD) of plastic scintillators
behind the Pad Chamber serves as first-level trigger device. Below we
introduce the individual detector components in order of their
arrangement along the beam.

\subsection{The target area}
\label{subsec:1}

\begin{figure}[b!]
\begin{center}  
\resizebox{0.47\textwidth}{!}{%
  \includegraphics{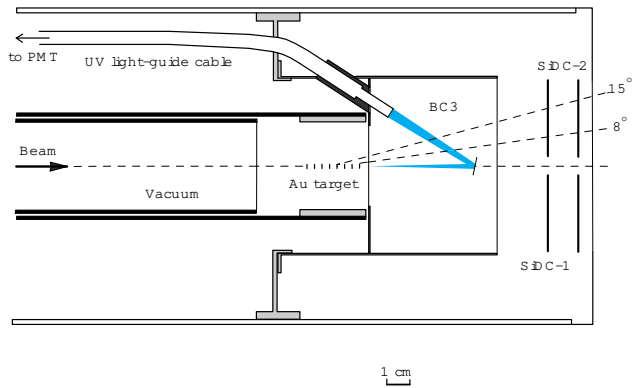} } 
\end{center}  
\caption
  {Schematic cross section of the target area along the beam direction
  showing the segmented target, the doublet of SiDC's, and in between
  the small mirror of BC3 which directs Cherenkov light from every
  passing Pb ion into a light guide for vetoing the interaction
  trigger. The acceptance of the RICH detectors is indicated (dashed
  lines).} \label{fig:2.2}
\end{figure}

The target area, hardly visible in Fig.~\ref{fig:2.1} and enlarged in
Fig.~\ref{fig:2.2}, comprises the segmented target, the doublet of
SiDC's, the light-collecting parts of the interaction-vetoing beam
counter BC3 (see sect.~\ref{subsec:6}) and interfaces to the adjacent
parts of the spectrometer. It is housed within a hollow cylindrical
recess into
RICH-1 which is lined by a cylindrical tungsten mantle of 20~mm
thickness which shields the UV-detectors from heavily ionising
particles emerging in backward direction from the target. The recess
is hermetically closed towards the radiator of RICH-1 by a
double-walled aluminised mylar window (2 x 50~$\mu$m), the
intermediate volume being ventilated with nitrogen.  The SiDC's and a
thin mirror for BC3 are mounted within a double-walled tube of 160~mm
diameter, made of aluminium and carbon fibre which is water cooled and
provides a laminar flow of dry nitrogen for cooling the SiDC's and the
front-end chips mounted on the same mother board. A light guide transfers
the Cherenkov light received from a thin mirror. 

The beam enters the
target area via an evacuated Al tube that reaches until a few
millimetres short of the segmented target and is sealed by a thin
mylar window .

The segmented Au target consists of 8 discs of 25$\mu$m thickness and
600$\mu$m diameter spaced uniformly by 3.2~mm along the beam.  The
total target thickness of 200$\mu$m Au corresponds to an interaction
length of $ \lambda/\lambda_I= 0.83\%$ for Pb-Au collisions. The
segmentation assures that only a fraction of the total radiation
length is effectively seen by photons and electrons propagating into
the spectrometer acceptance. Under worst conditions of full
illumination, there is only a chance of about 20$\%$ that the next
downstream disc is within the acceptance. The effective
radiation length is $X/X_\circ\approx$ 0.55$\%$, compared to $3\%$ for
an unsegmented target of identical interaction length.

The Au discs are supported by 2.5~$\mu$m thick mylar foils, and the
entire assembly is contained in a thin-walled (0.5~mm) carbon fibre
tube of 60~mm diameter; the Au discs are accurately aligned to the
axis of the tube.  The Au discs contain 90$\%$ of the beam over the
full length of the target once collimators and magnet settings of the
beam line have been optimised. The beam diameter was measured to have
a Gaussian envelope of $\sigma$= 220~$\mu$m.

\subsection{The silicon-drift telescope}
\label{subsec:2}

The SiDC's fulfil several purposes in the overall
concept: (i) to locate the interaction vertex within the segmented
target, (ii) to maintain a precise two-point charged-particle tracking
free of ambiguities, improving thereby the momentum resolution of the
spectrometer, (iii) to assist in the ring pattern recognition of
RICH-1, and (iv) to reject conversions and Dalitz pairs which are not
resolved in the RICHes, by detecting a double pulse-height signal or
two close tracks.

The SiDC telescope made up of two closely spaced cylindrical
SiDC's was the major element in upgrading CERES for
the Pb-beam experiments~\cite{cogne95}. CERES is the first experiment
which successfully implemented~\cite{chen92} this detector
concept~\cite{gatti847}, and it did so from the start, however, with
only a single detector~\cite{chen93,fasch93}. This allowed to check
that the electron pair originates from a common vertex, but was of
course insufficient for full tracking. 

Since cylindrical SiDC's seem not very well known, we insert here a
short description. The active area is practically the full area of a
3-inch-diameter wafer 280~$\mu$m thick which has a central hole of
about 6~mm diameter for the passage of the beam. The signal electrons
generated by ionisation of charged particles traversing the wafer
drift radially outward to an array of 360 anodes located at the
periphery of the detector. The radial coordinate $r$ of the point
where a charged particle crossed the detector plane is measured by the
drift time $t$ of the electron cloud. The charge sharing between
neighbouring anodes measures the azimuthal coordinate $\phi$. The pair
of coordinates $(r,\phi)$ is provided for each crossing charged
particle for events with a total charged multiplicity of several
hundred.  The longest drift distance is about 3 cm. The nominal value
of the drift field of 500~V/cm results in a maximum drift time of
about 4 $\mu$s.  The drift field is provided by means of 240
concentric $p^+$ electrodes with 130~$\mu$m pitch on both sides of the
detector, suitably biased by an implanted voltage divider. A major
innovation in the design was a `sink anode' providing a path for the
leakage current generated at the Si-SiO$_2$ interface away from the
signal anode, i.e. without contributing to the anode leakage current.

The ideal circular electrode shape could not be designed in early 1990
due to limitations in the software controlling mask
production. Electrodes were shaped instead as regular polygons of 120
sides. We were surprised to see that the distribution of hits
displayed peaks at every third anode~\cite{fasch96}.  The `efficient'
anodes were those located in the central part of each 3$^\circ$
triangle forming the polygon.

The telescope implemented in 1995 consisted of a 3-inch detector
followed by a 4-inch detector, the latter of the novel AZTEC
design~\cite{holl96} which eliminated the focusing problem.  The
spacing of the two detectors was 14.3 (15.0)~mm and the distance from
the target centre to SiDC-1 amounted to 98.5 (110) ~mm in 1995 (1996).

In production of the 3-inch detector\footnote{produced by SINTEF, 0134
Oslo 3, Norway}, the front and back lithographies were rotated with
respect to each other by 1.5$^\circ$ to reduce the non-radial
components in the drift field. This trick reduced to negligible levels
the focusing effect of the central anodes that had been a severe
obstacle for reaching the design azimuthal hit
resolution~\cite{fasch96}.

In 1996, both detectors were of the 4-inch type\footnote{produced by
EURISYS M\'{e}sures, F-67380 Lingolsheim, France}. The larger anode
radius of 42~mm allowed to increase the distance to the target for
lower hit occupancy, especially at small radii. The sensitive area of
the detectors is increased from 32~cm$^2$ for the 3-inch devices to
55~cm$^2$ for the 4-inch detectors. The `field cage' polygons consist
of 277 concentric $p^+$ electrodes each having 360 instead of 120
sides before, and the respective $p^+$-implantations on the two wafer
sides are rotated by 0.5$^\circ$ with respect to each other for near
perfect radial field geometry. Another novel property of the AZTEC
detectors is the interlaced anode structure: each one of the 360
anodes is subdivided into 5 segments, two of which (extending over
16$\%$ of the anode pitch of 1~degree) are interlaced to the closest
of the neighbouring anodes to enforce charge sharing.  This provides a
more accurate azimuthal position measurement when calculating the
centre of gravity of the distribution.  Compared to the '95 runtime,
the new design improves the azimuthal resolution from 2.5 to
1~mrad. The resolution in radial direction is 30 $\mu$m both for the
95 and 96 set-ups.

Charge signals from 3-inch detectors are amplified by 32-channel
front-end OLA chips placed on the detector motherboards which were
developed to test ALICE silicon-drift prototype
detectors~\cite{dabrowski93} and were produced in a custom bipolar
process\footnote{Owned by Tektronix at the time of production; since
then Maxim Integrated Products}. They consist of a charge-sensitive
preamplifier, a quasi-Gaussian shaper\footnote{The time constant
$\tau$= 38~ns, about twice the design value, deteriorated the
potential double-pulse resolution, but avoided a large ballistic
deficit over the full drift.}  and a symmetrical line driver. The
rather high gain (30~mV/fC) gave rise to wild collective oscillations.
Stable operation was achieved only with the introduction of
20~$\Omega$ damping resistors connected in series into the
50~$\Omega$-terminated output lines.

For the readout of the 4-inch detectors, new 16-channel front-end
chips had been designed along a CMOS concept~\cite{gramegna1997} which
incorporated bipolar drivers and were well adjusted in shaping time
(37~ns), gain (9~mV/fC), dynamic range (5~$mips$\footnote{minimum
ionising particles}), and low equivalent noise charge (140~e$^-$) to
meet our requirements.\footnote{produced by AMS in 0.8$\mu$m biCMOS
technology.}

An outside buffer stage transmits the bipolar signals over 40~m flat
cables to the FADC's ({\it flash analog to digital
converters})\footnote{Series DL300 of Fa.~B.~Struck, Tangstedt near
Hamburg} which sample the data with 50~MHz and store it 256 bytes
deep.  This corresponds to a drift time range of 5.12~$\mu$s.
Digitisation is 6~bit with non-linear characteristics.\footnote{${\rm
Channel.No.(0...63)= 256\cdot U_{in}/(0.2+ 3\cdot U_{in})}$, input
voltage ${\rm U_{in}}$ in Volt.}  Data are continuously sampled until
interrupted by an external `stop' signal so that the channel memory
contains always the last 5.12~$\mu$s of data.

Since the trigger signal and clock are asynchronous, there is a random
phase difference producing a time jitter of 20~ns/$\sqrt{12}$ rms.  It
is measured with a TDC ({\it time-to-digital converter}), and 
correction is done off-line.

Each of the four FADC crates per detector houses a SIM ({\it scanner
interface module}) which scans the data after the trigger was received
for contents above a predefined threshold ({\it readout
threshold}). Readout is activated whenever the threshold is surpassed
in two successive time bins, and stopped, if contents in two
successive time bins fall below it. The contents of five preceding
channels are also readout ({\it pre-samples}) for off-line
reconstruction of the baseline. 

\subsection{The RICH detectors}
\label{subsec:3}

The radiators are filled with methane at atmospheric pressure. The
high Cherenkov threshold of
$\gamma_{thr}\simeq$~32~\footnote{different in RICH-1 and RICH-2 by
about one unit.} ensures that more than 95$\%$ of all charged
particles pass without creating Cherenkov light (`hadron blind
tracking'). The Cherenkov light is reflected backward onto
2-dimensionally position-sensitive gas detectors which are separated
from the radiator volume by UV-transparent windows.  By their upstream
position with respect to the target, the UV-detectors are not exposed
to the huge forward flux of charged particles. The price to be payed
for this geometry is the limited acceptance in polar angle $\Theta$,
indicated by the lines in the upper part of Fig~\ref{fig:2.1}.

High-energy electrons produce Cherenkov rings with asymptotic radius,
R$_\infty$= 1/$\gamma_{thr}\simeq$~ 30~mrad. The difference in
radiator lengths (86~cm and 175~cm for RICH-1 and RICH-2,
respectively) is partially compensated by better UV transmission in
\mbox{RICH-1} (CaF$_2$ window) compared to RICH-2 (quartz window), so
that the asymptotic number of photons per ring, 10.8 and 11.5, for
RICH-1 and RICH-2, respectively, come out rather
similar\footnote{The numbers of Ref.~\cite{rich_set} measured with
ethane are increased by one photo-electron due to the larger bandwidth
in methane.}. By the same reason, photon detection in RICH-1 reaches
farther into the UV.

In both RICHes spherical mirrors focus the Cherenkov photons radiated
from a straight trajectory back onto a ring image in the focal plane
of the UV detectors. As the mirror in RICH-1 is traversed by all
electrons before the second ring image for momentum measurement is
taken in RICH-2, there are stringent physics reasons to keep the
radiation length as low as possible: besides reducing the number of
external conversions, it is the multiple scattering of low-momentum
electrons which reduces the detection efficiency for soft pairs and
deteriorates the momentum resolution. The mirror of RICH-1
therefore is made very thin (1.1~mm) so that it adds only 0.4$\%$ of a
radiation length. It is based on a laminated carbon fibre structure
which defines the spherical geometry\footnote{ manufactured by MAN
Technologie AG}. An evaporated coating of aluminium protected by
magnesium fluoride achieved persistent UV reflectivity of 80$\%$ at
300~nm.

The UV-detectors consist of a conversion space followed by two
parallel-plate avalanche stages and a multi-wire proportional
chamber. The originally planned mode of running only with parallel-plate
amplification was abandoned in favour of an added multi-wire stage,
following a painful learning process on spark break down problems in 
pure parallel-plate schemes~\cite{spark94}. 

The operating gas is He + 6$\%$ CH$_4$ at atmospheric pressure +
TMAE-saturated\footnote{Tetrakis-di-Methyl-Amino-Ethylen} vapour at
40$^\circ$C as photon converter. The use of TMAE demands that the UV
detectors be kept hot to avoid condensation. To avoid temperature
gradients across the delicate UV-transparent windows separating the
detectors from the radiators, the entire spectrometer is kept hot at
about 50$^\circ$~C. The UV detectors operate at a total gain of about
2$\cdot$10$^5$ for high photon detection efficiency (${\rm
\sim 85\%}$).  The ion clouds produced in the last wire amplification
stage induce signals on the pads. The latter form a grid of pitch 2.74~mm
and 7.62~mm, and the resulting total number of pads is 53,800 and
48,400 in RICH-1 and RICH-2, respectively~\cite{pad_read}.

The UV detectors have been operated without opening since 1991.
During the 1995 run, UV-1 degraded in performance. The detector could
not be operated at the desired gain of 2$\times$10$^5$ without an
excessive spark rate. Early in 1996, the UV-1 detector was opened and
all mesh electrodes, in particular the cathodes and the multi-wire
plane showed some kind of deposit. Most mesh electrodes were exchanged
and the wire anode subjected to ultrasonic cleaning. The refurbished
detector performed very well during the 1996 run~\cite{socol-phd}.

\subsection{Deflection in the magnetic field}
\label{subsec:4}

The magnetic field for momentum analysis is generated by two
super-conducting solenoids carrying currents in opposite
sense. Charged particles experience an azimuthal deflection between
the two RICHes which is inversely proportional to the momentum,
\begin{equation}
\Delta\phi= \frac{\phi_0}{~p}~\left (\frac {\rm mrad}{{\rm GeV}/c}\right ),
\end{equation}
\noindent
and the sense of which, for fixed polarity of the field, defines the
charge sign. The constant is $\phi_\circ$=~146~mrad\,GeV/$c$.  Between
SiDC and the Pad Chamber the deflection is only 66$\%$ of this value,
$\phi_\circ$=~96~mrad\,GeV/$c$.\footnote{The RICHes measure the change
in {\it local} angle, while deflection in the Pad Chamber is derived
from the displacement relative to the distance from the vertex.} To first
order, the polar angle is not affected. Particles deflected in azimuth
by $\Delta\phi$ encounter a small second-order deflection towards the
beam axis which amounts to
\begin{equation}
 \Delta\theta= -78~(\Delta\phi)^2~\left (\frac{\rm mrad}{\rm rad^2}\right ). 
\end{equation}
\noindent
Two sets of warm correction coils are tuned to achieve a field-free
radiator in \mbox{RICH-1} and to align the field lines in the radiator
of RICH-2 parallel to the particle trajectories from the target. This
way straight trajectories inside both radiators are
achieved. Moreover, the absence of deflection in the first RICH
detector allows to identify conversion and Dalitz pairs by their small
opening angles.

\subsection{The Pad Chamber}
\label{subsec:5}

The last tracking detector is a multi-wire proportional chamber
located closely behind the mirror of RICH-2, at a distance of about
3.3~m from the target. It has an inner and outer radius of 42~cm and
85~cm, respectively, and is free of radial spokes. The Pad Chamber
covers the fiducial pseudo-rapidity interval ${\rm 2.05<\eta<2.65}$ of
the CERES spectrometer. It was added before the run period in 1995 as
an external tracking device behind RICH-2 to assist the ring pattern
recognition and reduce the fake-ring background in the
high-multiplicity environment of Pb-Au collisions. By providing an
absolute reference for the silicon and the RICH detectors, the Pad
Chamber, with an angular resolution of about 0.6~mrad (in $\theta$),
proved a powerful tool in the geometrical inter-calibration of the
detectors which helped to improve momentum resolution.

The Pad Chamber is operated with a 90/10$\%$ Ar/CO$_2$ mixture. The
multi-wire anode is at equal distance (5~mm) to the upstream mesh
cathode and the downstream pad cathode. Only about half the ionisation
charges are collected during the $2\mu$s charge integration time of
the pad readout electronics. The electronic avalanche produced by a
charged particle traversing the Pad Chamber induces a signal in some
of the $\approx$~29000 pads of the pad cathode. The pad size is the
same as in RICH-2, and the pad readout electronics was adopted from
RICH-2.

\subsection{The trigger}
\label{subsec:6}

A system of beam counters (BC) has been specifically developed to meet
the requirements of minimal mass exposure in the beam and target
region and sufficient radiation hardness~\cite{volodya97}.  The
trigger system is based on three small Cherenkov beam counters
operating in air, one (BC1) about 60 m upstream,
monitoring incident beam particles, and one (BC2) about 6~m downstream
of the target detecting ions passing through the spectrometer.  The
\begin{figure*}[h]
    \begin{flushleft} 
\resizebox{0.4\textwidth}{!}{%
    \includegraphics{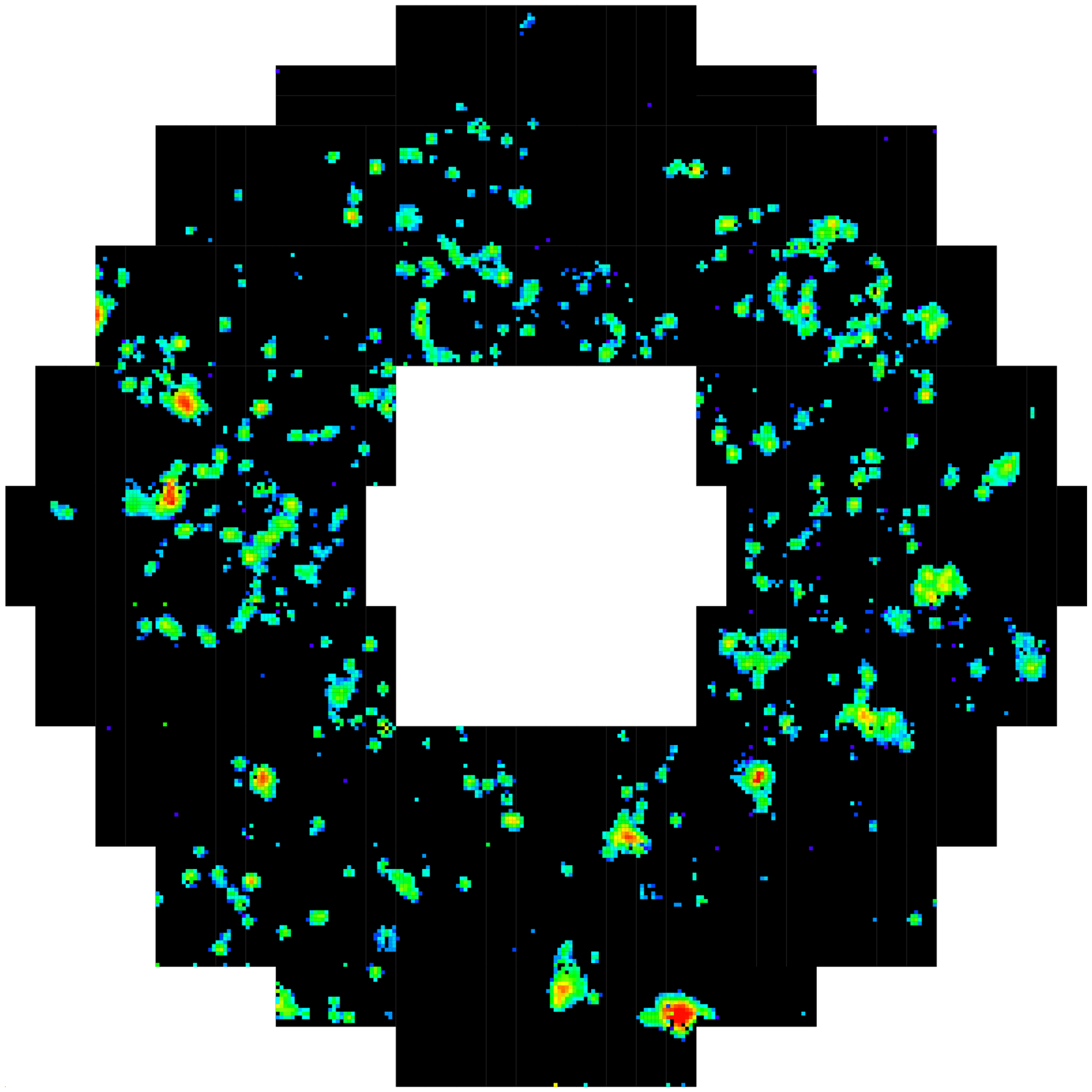}}
    \end{flushleft}

    \vspace{-7.2cm}
    \begin{flushright} 
\resizebox{0.4\textwidth}{!}{%
    \includegraphics{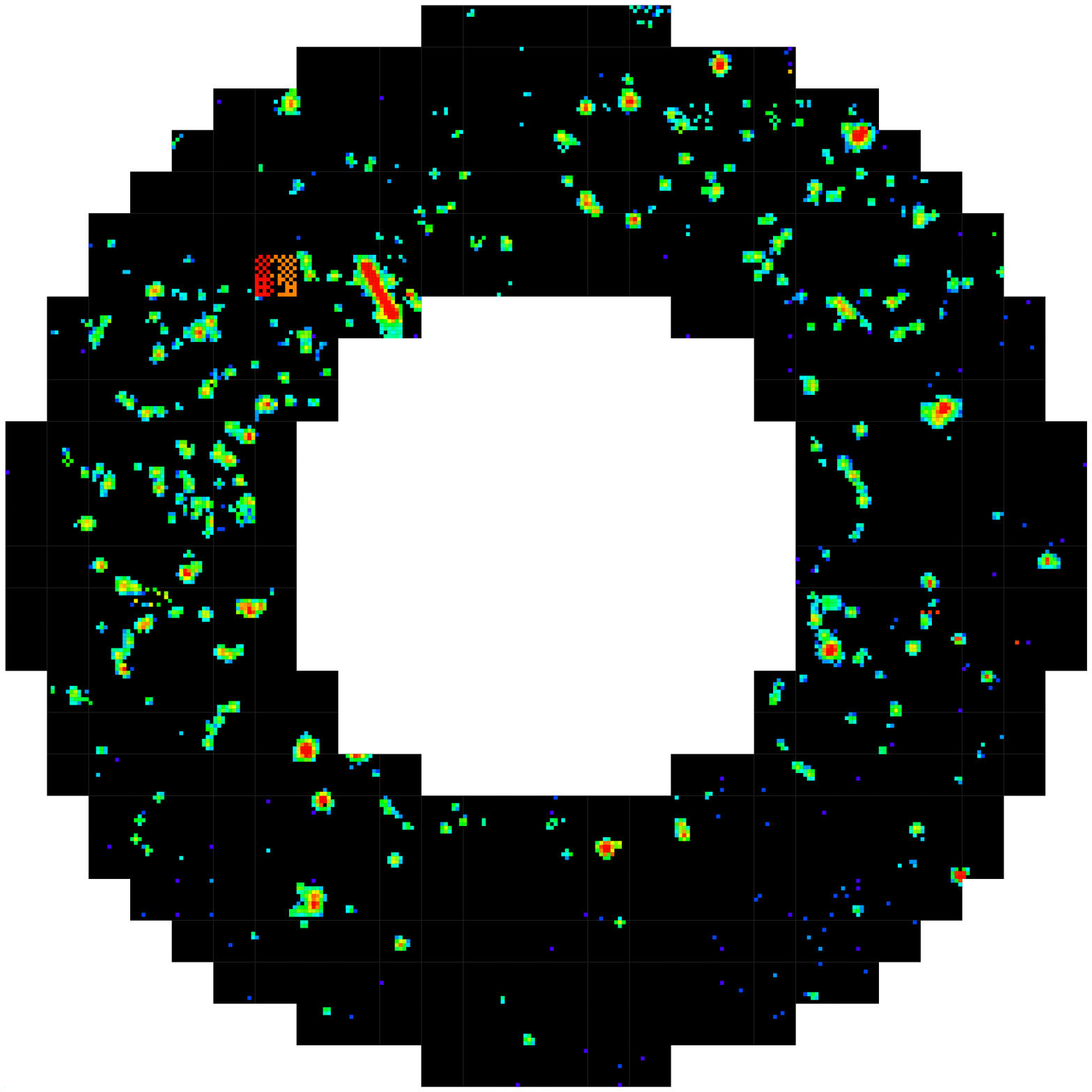}}
    \end{flushright}

    \begin{flushleft} 
\resizebox{0.4\textwidth}{!}{%
    \includegraphics{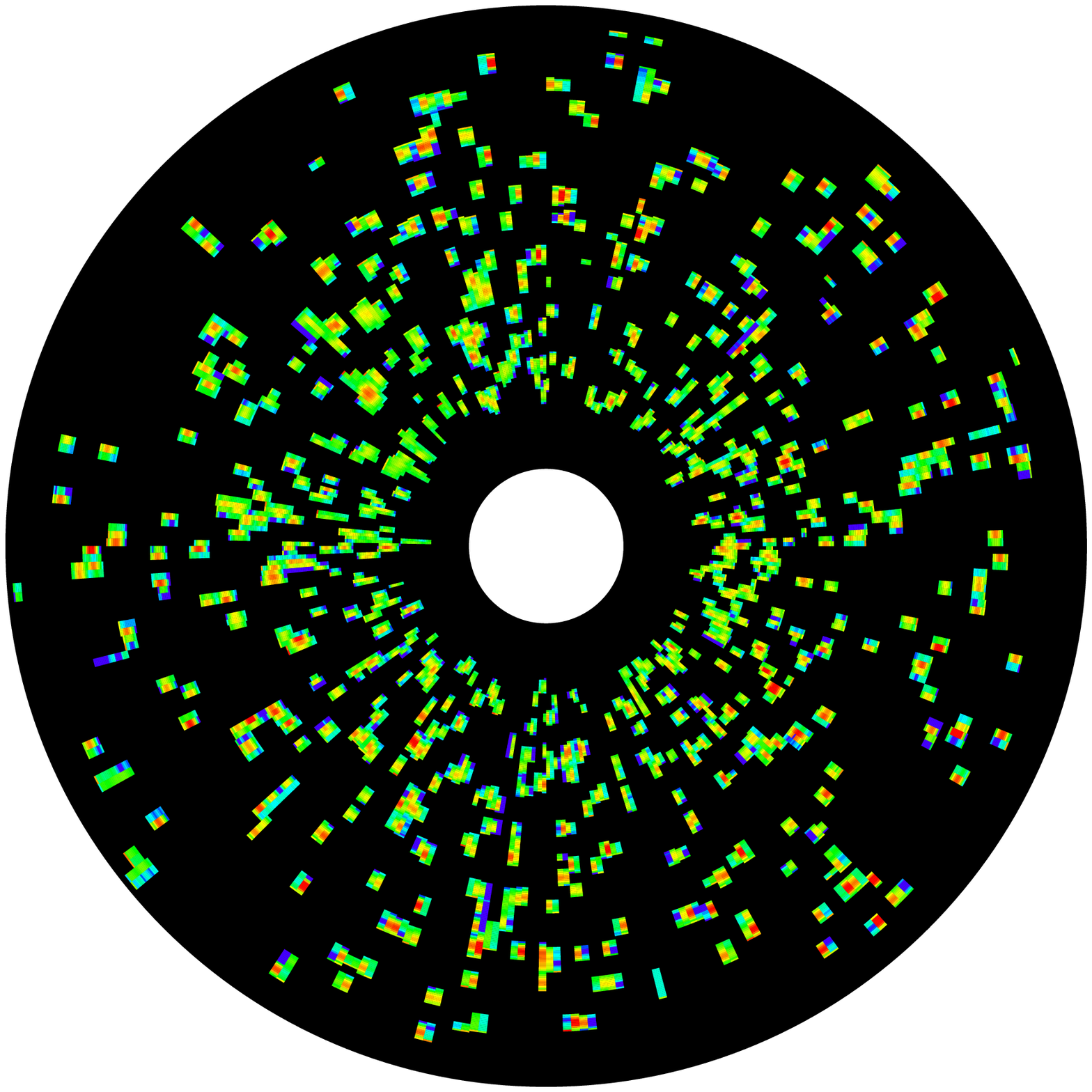}}
    \end{flushleft}

    \vspace{-7.2cm}
    \begin{flushright} 
\resizebox{0.4\textwidth}{!}{%
    \includegraphics{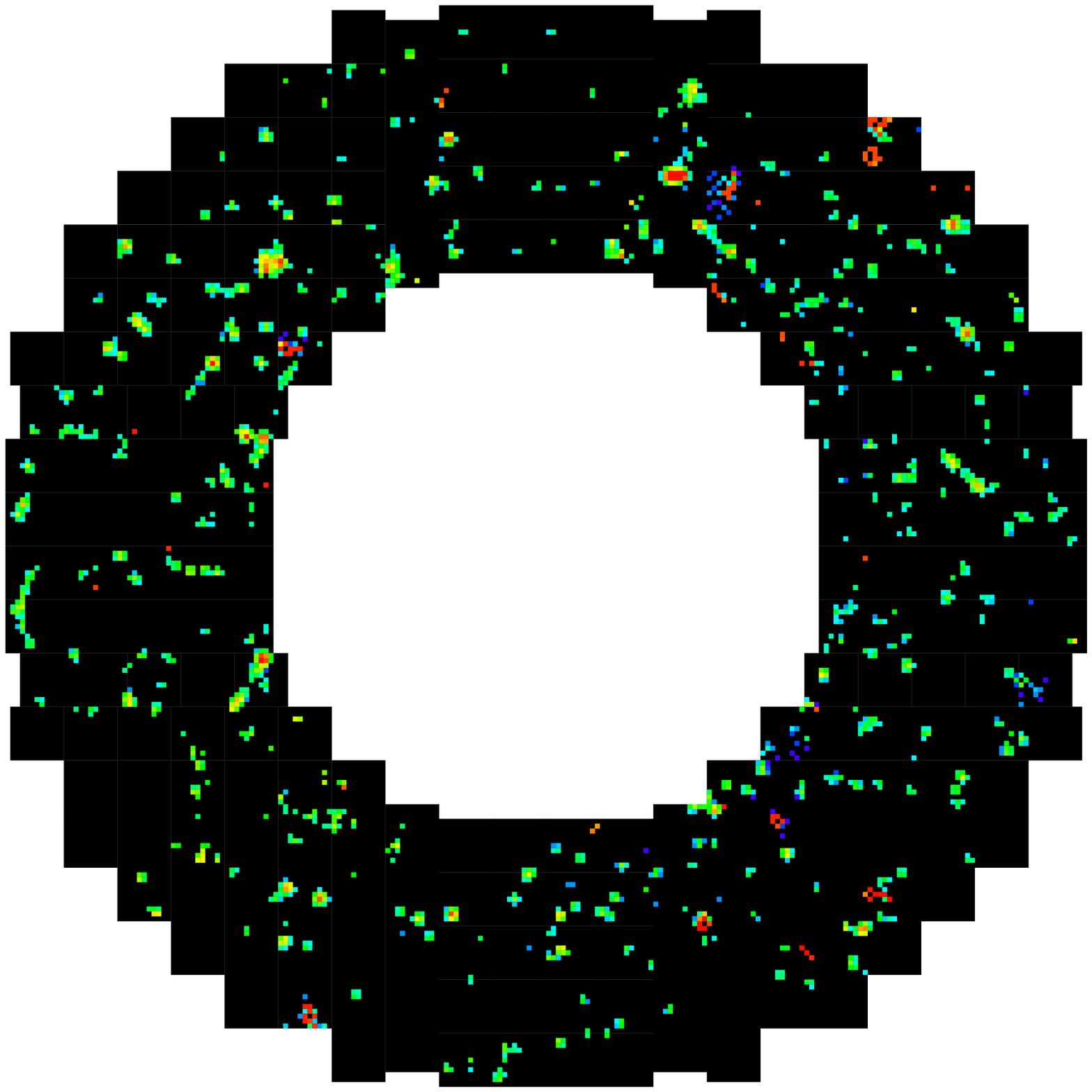}}  
    \end{flushright}
    \begin{minipage}{0.95\linewidth}     
    \caption{\label{fig:2.3} Event displays for RICH-1 (top left) and
RICH-2 (below), SiDC-1 (top right) and Pad Chamber (below).
  }
    \end{minipage}
\end{figure*}
third Cherenkov counter (BC3) registers each intact Pb ion downstream
of the last target in order to veto the interaction trigger.  To this
purpose, Cherenkov photons (2700 per cm air) emitted in a narrow
forward cone are reflected away from the axis onto a UV light guide by
means of a tiny (6~mm diameter) aluminised mylar mirror on axis about
6~cm downstream of the target. The Cherenkov light is fed into a
photo-multiplier just outside the spectrometer (Fig~\ref{fig:2.2}).

A plastic scintillator in front of the spectrometer (VC) is used to
veto upstream interactions.  The interaction trigger is defined as the
logical AND of BC1 and the veto of BC3 and VC, INT=
BC1$\cdot\overline{\rm VC}\cdot\overline{\rm BC3}$.  Centrality is
selected with the multiplicity detector (MD), an array of 24 plastic
scintillator paddles downstream of the RICH detectors at
$\eta=$2.9-4.7, the light output of which serves as a measure of the
number of ionising particles that have passed. The centrality trigger
is defined as INT$\cdot$MD, and a hardware threshold is set at 100
{\it mips}. The accuracy of the trigger threshold and its stability
over time is limited, mostly due to gain variations in the
photo-multiplier tubes. In the off-line analysis, a precise
multiplicity measurement is provided by the two SiDC's. The trigger
selection corresponds roughly to the top 30$\%$ of the geometrical
cross section~\cite{claudia}. A more precise calibration will be
presented in sect.~\ref{subsec:19}.

\subsection{Data acquisition}
\label{subsec:7}

The on-line response of the detectors is shown in Fig.~\ref{fig:2.3}
for a semi-central 158\,GeV/n Pb-Au event recorded during data taking
in 1996.

The large amount of information per event contained in five highly
granular detectors is collected by a fast data acquisition
system. Data reduction is performed already at the hardware level when
individual detectors are readout with zero suppression and pedestal
subtraction. The data from the detectors are split into several
separate readout chains which are processed in parallel for higher
readout speed and transferred into special Memory modules where the
data is compressed using Huffmann coding.

A set of two (in 1995, three in 1996) CPU modules in VME technology is
used to collect the data from the Memory modules for each event
during the burst period (4.8~s) in a round-robin mode and writing it
to tape during the intervals between bursts (14~s). As the write
speed of the Digital Audio Tape (DAT) drives used is rather low,
each CPU has three drives connected to it, allowing for an
effective aggregated write speed of about 1.5 MByte/s per CPU (in 1995,
about 8 MB/s in 1996). This allowed to record on average
 550 events/burst of 40-45 kByte each in 1995; the upgraded
DAQ together with optimised detector settings and therefore smaller
event sizes (30-35 kB) recorded about 1000 events/burst in 1996.

\subsection{Data taking in 1995 and 1996}
\label{subsec:8}

CERES/NA45 had data taking runs of 9 days in fall 1995 and of 27 days
in fall 1996 at the CERN SPS with a 158~GeV/n Pb beam on Au
targets.  The average beam intensity in both years was
about 1$\times$ 10$^6$ ions per burst of duration 4.8~s.

Because of the small target and beam dimensions, a readjustment of the
beam position was one of the regular shift duties.  To keep the
spectrometer efficiency at the designed level, the gains in the UV
detectors were monitored continuously and held within limits of about
30$\%$ at 2$\cdot10^5$ by adjusting high voltage upon changes of
atmospheric pressure. 

Collision events were selected with the interaction trigger
threshold set on 100 mips equivalent in the multiplicity array. The
trigger contained an admixture of downstream interactions on the level
of 15\% of the target interactions. These were discarded in the
off-line analysis on account of the silicon-drift track multiplicity.
The latter served for more accurate centrality definition and also
revealed some difference in the effective calibration of the
multiplicity array between the two runs.  

In 1995 8.5 million events were collected with average multiplicity
$\langle dN_{ch}/d\eta\rangle$= 220 corresponding to the top
33$\%$ of the geometrical cross section. In 1996, the total sample was
42 million events of average charged particle multiplicity 250, or
26$\%$ of the top geometrical cross section. The multiplicity refers to
the number of tracks in the SiDC's and is averaged over the pseudo-rapidity
range $\eta$= 2-3.  In the middle of the runs the polarity of the
magnetic field was switched.  

\subsection{Instrumental means of coping with background}
\label{subsec:9}

We shortly review here the instrumental means by which CERES recovers
a weak signal of low-mass electron pairs from high levels of
combinatorial background.

Approximate `hadron blindness' is achieved by using two Ring Imaging
Cherenkov (RICH) detectors with a high threshold
$\gamma_{th}\approx$~32. While electrons with momenta above 16~MeV/$c$
produce Cherenkov light, pions overcome the threshold only at
4.5~GeV/$c$. More than 95\% of all charged hadrons pass without
producing Cherenkov light.

The radiation length within the spectrometer acceptance has been kept
at $X/X_0$ $\sim$ 1 \%. This is the level where the number of
conversions is about equal to the number of Dalitz pairs.  It is the
result of persistent efforts in the design of all spectrometer
components to reduce the detector materials, among which
the thin mirror of \mbox{RICH-1} and the segmented target are the
most important.

Our physics sample are electron pairs with mass above
200~MeV/$c^2$. Below 200~MeV/$c^2$, photon conversions and $\pi^0$
Dalitz decays shoot up in yield which diminishes the sensitivity to
interesting physics. In pursuit of the goal to recognise soft pairs of
conversions and $\pi_0$-Dalitz pairs with highest possible efficiency,
the CERES spectrometer provides two powerful handles: the SiDC doublet
detects close tracks by double-dE/dx response in pulse height, or by
resolved close hits. The fact that the radiator of RICH-1 is free of
magnetic field allows to see electron pairs of small opening angle
undeflected, i.e.\@ as close, resolved Cherenkov rings, or rings with
a larger number of photons when the two tracks are separated by less
than $\sim$8~mrad. Soft electron tracks from conversions and
$\pi^\circ$-Dalitz pairs are strongly deflected by the magnetic field
between the two RICHes, electrons and positrons in opposite sense. By
setting an upper limit to the azimuthal deflection between RICH-1 and
RICH-2, or between SiDC and Pad Chamber, a cut on track $p_t$ is
implemented which is one of the most effective measures to reject
background.

To achieve a high efficiency for reconstruction of soft pairs, all
detectors have full azimuthal coverage and the vetoing detectors
have a slightly larger (minimum 1.9 $<$ $\eta$ $<$ 2.8) fiducial rapidity
acceptance than the detectors after the magnetic field (2.15 $<$
$\eta$ $<$ 2.6). The more subtle details of the rejection strategy are
discussed in sect.~\ref{sec:3}.

\section{Data analysis}
\label{sec:3}
\subsection{Overview}
\label{subsec:10}

In this overview we sketch the strategy of the data analysis up to the
pairing level when identified electron tracks in a given event are
selected for combination to pairs. The presentation is
mainly based  on data and only rarely refers to Monte-Carlo
simulations. The latter are, however, implicit in the choice of
various quality and rejection cuts and will be treated in sect.~\ref{sec:4}.

\subsubsection{The analysis strategy}
\label{subsec:11}

The invariant mass squared of the pair is given by the squared sum of
the electron 4-momenta,
\begin{equation}
 m_{ee}^2\,c^2= ({\bf p_{e^+}} + {\bf p}_{e^-})^2\,=
2\,p_{e^+}\,p_{e^-}\,(1 - cos\,\Theta_{ee}).
\end{equation}
For the standard single-electron cut $p_t\geq$~200~MeV/$c$ used in the
data analysis, the dynamic range of the electron momenta is somewhat
restricted so that the dynamic range of the invariant mass is largely
determined by that of the laboratory opening angle $\Theta_{ee}$
between the electron tracks.

\begin{figure}[t!]
\begin{center}  
\resizebox{0.35\textwidth}{!}{%
    \includegraphics{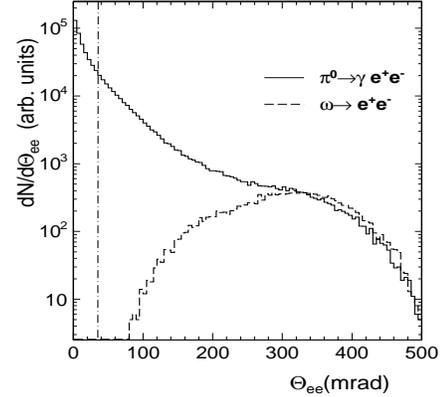}}
\end{center}  
  \caption{\label{fig:3.1} Opening-angle and single-track
  \mbox{p$_t$} distributions  of \mbox{$\pi^\circ$}-Dalitz
  pairs (full histograms) and \mbox{$\omega$} mesons (dashed
  histograms) in the CERES acceptance; Monte-Carlo
  simulations. Standard cuts (vertical lines)
  are not applied.} 
\end{figure}

Signal electron pairs with $m\geq$~200~MeV/$c^2$ have opening angles
considerably larger than the asymptotic Cheren\-kov ring radius of
30~mrad; the massive pairs from $\rho$, $\omega$, and $\phi$ decays
are opened about ten times wider (Fig.~\ref{fig:3.1}, upper panel).
Our operational definition of {\it signal} or {\it open pairs}
includes an opening angle cut $\Theta_{ee}\geq$~35~mrad.\footnote{It
reduces the number of Dalitz pairs. As an important side effect, the
cut enforces more uniform track distributions in polar angle of pairs
from different sources.} For $\eta-$Dalitz decays, the mean opening
angle is about~60~mrad.  Conversion and $\pi^\circ$-Dalitz pairs have
small masses\footnote{Photon conversion pairs are nearly massless but
may acquire a small apparent mass by multiple scattering.} and average
opening angles below 2~mrad and 20~mrad (rms), respectively. The
sample of pairs with masses below\\ 200~MeV/$c^2$ and
$\Theta_{ee}\ge$~35~mrad will be referred to as the {\it Dalitz
sample} and is also used for checks on reconstruction efficiency and
absolute yields.

The $p_t$ distributions of electron tracks from conversions and
$\pi^\circ$-Dalitz  pairs are steeper than those of open 
signal pairs. This feature provides the 
only rejection handle at the {\it track level}, albeit a very 
powerful one: the 200~MeV/$c$ cut on track $p_t$  
reduces close-pair tracks much stronger than signal 
tracks as seen in Fig.~\ref{fig:3.1}, lower panel.

The most severe problem of the experiment is the enormous
combinatorial background. We do not know which electrons belong to a
pair and therefore we accept combinations of all tracks that
qualify. However, when we find a pair with $m<$~200~MeV/$c$,
its tracks are excluded from further pairing. Because the S/B ratio
for these pairs is usually very good, we can declare a fully
reconstructed Dalitz or conversion pair with good confidence.

\begin{figure}[tt!]
\resizebox{0.52\textwidth}{!}{%
\includegraphics{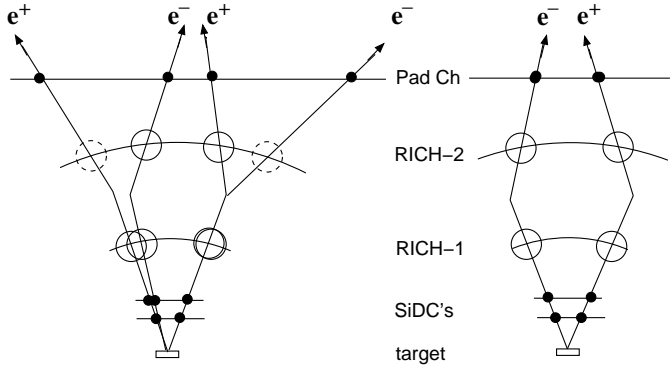}}
   \caption { \label{fig:3.2}  Left: A $\pi^\circ$-Dalitz decay, $\pi^\circ \rightarrow e^+e^-\gamma$, 
and a {\it V track} indicating a conversion pair,
 $\gamma\rightarrow e^+e^-$, show up in RICH-1 as close doublets or
 single rings, followed in RICH-2 by two separate rings where one is
 missing (dashed), due to detector inefficiency or deflection out of
 the acceptance. Right: The remaining single tracks are combined to a
 fake open pair which looks like a signal pair, opened already in
 RICH-1.}
\end{figure}

The background arises whenever low-mass pairs are only partially
reconstructed and the remaining tracks are combined, as visualised in
Fig.~\ref{fig:3.2}. Clearly, combinatorial pairs cannot be
distinguished from genuine signal pairs and contribute
to the entire mass range of interest, exceeding the signal by three
orders of magnitude if all low-mass pairs would contribute. The final
pair sample at masses above 200~MeV/$c^2$ is still only about 10$\%$
of the residual level of combinatorial background pairs. This is why
already small inefficiencies in reconstruction of soft pairs,
acceptance losses, etc., give rise to large relative levels of
combinatorial background. Multiple scattering of conversions and
$\pi^\circ$-Dalitz tracks is a further important source of losses.

A mild $p_t\geq$~50~MeV/$c$ cut is applied during the production stage
(see below) to limit the search area in azimuth.  Many of the
conversion and Dalitz pairs with only one leg above 200~MeV/$c$ can be
reconstructed and later rejected this way. Such pairs are ten times
more numerous than pairs with both electrons above 200~MeV/$c$, so
that many stiff electron tracks are taken out before entering the
pairing stage. Applying the strong $ p_t$ cut {\it before} the filter
would have kept those tracks in the combinatorics.  Tracks attributed
to pairs with opening angles $\Theta_{ee}\leq$ 35~mrad - the logical
complement of the opening-angle cut for signal pairs - are marked to
be excluded from further pairing.

The deflection of electrons and positrons by the magnetic field
between the RICHes provides the unique search pattern of {\it V-tracks},
i.e. one ring in RICH-1, possibly somewhat blurred as it contains UV
photons from two close tracks, and two separated rings in RICH-2 at
about the same polar angle (see also Fig.~\ref{fig:3.2}).

Rejection of combinatorial background is optimised by tuning various
cuts. Since high rejection power and high signal efficiency are
competing requirements, an appropriate measure of signal quality is
required.  The optimisation has to be kept rigorously free of bias.
A critical discussion of this important issue is given in 
sects.~\ref{subsec:15} and \ref{subsec:16}.

\subsubsection{The analysis stages}
\label{subsec:12}

Without a higher-level trigger in the CERES Pb-beam experiments, the
actual data volume is much larger than that of interesting events, and
a primary data reduction, the `production' stage, is called for.
Implementation of an effective production filter requires an accurate
geometrical inter-calibration of all detectors.  The availability of
the Pad Chamber since 1995 allowed precise local tuning of the
entire spectrometer using samples of high-$p_t$
pions~\cite{ceretto-phd}.

During production, the full analysis chain for electron track
reconstruction is at work, albeit under loose quality criteria. Events
which contain at least two electron tracks with $p_t\geq$ 50~MeV/$c$
are stored in a database for further processing.  As millions of
events have to be processed with sophisticated pattern recognition
algorithms, the production is time consuming. The 1995 production on
the CS2 parallel computer at the CERN CN division with 32 SUN-SPARC2
processors took about 10 weeks; the 1996 data were preprocessed on a
PC farm in several turns with readjusted production filters during
4~months.  The final data analysis mainly deals with the optimisation
of the pair sample.

\subsection{Reconstruction of electron tracks }
\label{subsec:13}
The reconstruction of electron tracks in the present analysis takes
full advantage of the external tracking detectors, the doublet of
SiDC's before the RICH spectrometer and the Pad Chamber after it.

\subsubsection{Coordinate systems}
{\it Raw-data detector coordinates}

In RICH detectors and in the Pad Chamber, hits are encoded as pad
amplitudes in a two-dimensional mesh of $(x,y)$ coordinates of the
`pad plane'. The natural unit is the pad size of 2.74~mm in
RICH-1, and 7.62~mm in \mbox{RICH-2} and the Pad Chamber, which is
used up to the reconstruction of rings and ring centres in the
RICHes, and of hits in the Pad Chamber.

For the SiDC's, the symmetry of the radial drift field and the
circular ring of anodes is maintained at the raw data level: the
intrinsic coordinates are anode numbers (0-359) for the azimuth
location and time bins (0-255) of 20\,ns for the drift time.

\vspace{0.2cm}
\noindent
{\it Local detector coordinates}

RICH detectors measure {\it angles} of particle trajectories that
connect centres of Che\-renkov rings with the vertex point, the units
are $\Delta\theta= \Delta s/f$, expressed by pad size $\Delta s$ and
focal length $f$; these are $\Delta\theta$=~2.18~mrad and 1.82~mrad in
RICH-1 and RICH-2, respectively. The local coordinates in paraxial
approximation are expressed by the tangents of the polar angle $\theta$.
Because of spherical aberration by the mirrors, the expression using
$\theta$ instead turns out to be a much better approximation,
\begin{equation}
 x= f\,\,\theta~{\rm cos}\,\phi
+ x_\circ,\hspace{0.2cm} y= f\,\theta~{\rm sin}\,\phi + y_\circ, 
\end{equation}
\noindent
where $x_\circ, y_\circ$ are the coordinates
of the origin of the pad plane. In the off-line analysis, the angles
$\theta$ and $\phi$ are evaluated by ray tracing using look-up tables
to correct for local modulations in focal length.

Hits in the silicon-drift detectors are given by their radial position
$r$ (referring to the centre of the wafer) and azimuth angle $\phi$
which is defined as in the global laboratory system. The
transformation from drift time to radius requires knowledge of the
electron drift velocity with possible spatial as well as temporal
changes.

\vspace{0.2cm}
\noindent
{\it Laboratory coordinates}

Once track segments of several detectors are to be joined, each
detector is put into a three-dimensional global laboratory system with
its own geometric calibration parameters, i.e. small $(x,y)$-shifts of
detector axes away from the optical axis, rotations around $z$, tilts,
etc. This global {\it laboratory system} uses left-handed Cartesian
coordinates with the $z$-axis along the beam and the $x$- and $y$-axes
pointing to the right and upward, respectively, when looking with the
beam. The origin is in the centre of SiDC-1.  As we do not
measure space points within the short magnetic field, it is convenient
to work with straight trajectories all along from vertex to the Pad
Chamber, and store the particle's momentum and charge sign derived
from magnitude and sign of the azimuthal deflection.

\vspace{0.2cm}
\noindent
{\it Event coordinates}

Eventually, in the global {\it event coordinate system}, the event
vertex $ (x_V, y_V, z_V)$ is taken as the origin, and 
tracks are described by polar coordinates $\theta$, $\phi$ at the vertex.

\subsubsection{Hit finding in the SiDC's}
 
The 360 readout channels of {\it anodes} and the 256 time slices, or
{\it time bins}, sampled by the FADC, span a matrix each cell of which
is assigned a 6-bit non-linear raw-data amplitude. The amplitudes
are linearised and pedestals are subtracted. 

\begin{figure}[b!]
\begin{center}
\resizebox{0.38\textwidth}{!}{%
    \includegraphics{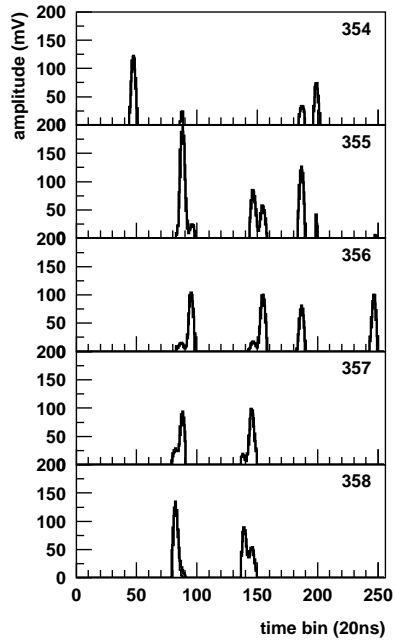}} 
\end{center}   
\caption{\label{fig:3.3}
    A collection of pulses on adjacent anodes. 1996 data.}
\end{figure}

The data field is then searched for contiguous regions of cells with
nonzero amplitude. Each cell is attributed to one such {\it
cluster}. All cells of fixed anode number within a cluster form a time
sequence of amplitudes called a {\it pulse}. The signal of a
minimum-ionising particle produces pulses of typically 5 time bins and
is spread over 2.2 anodes on average.  Pulses of less than 3 cells
above a hardware threshold are discarded.  Examples of such pulses on
a few neighbouring anodes are displayed in Fig.~\ref{fig:3.3}. Due to
complementary signal transmission, the pulses are
free of pickup which plagued previous runs.

The centres of gravity of the pulses are calculated by Gaussian
regression, or by Gauss fits taking the known, drift time dependent
widths from a table. After time $t$, the drifting electron cloud
arriving at the anodes has developed a time-spread due to diffusion of
$\sigma_t^{in}= \sqrt{2\,D\,t/v_{drift}}$. Here
$D\simeq$~35$\times$\,10$^{-4}$\,mm$^2/\mu$s is the electron diffusion
constant in silicon and $v_{drift}$ denotes the drift velocity which
varied in the range 6.0~-~8.5\,mm/$\mu$s depending on the voltage
setting. The charge pulse is folded with the quasi-Gaussian response
of the shaper, $\sigma^{shaper}$, so that the time spreads add in
quadrature. The shaping introduces a {\it ballistic deficit} in
amplitude $\delta=\sigma^{shaper}/\sigma_t^{in}$ which is corrected
for.  Pulse heights saturated in the peak cells are approximately
reconstructed.

The stop-pulse correction mentioned in sect.~\ref{subsec:2} removes the
random phase jitter. The conversion from drift
time to drift distance is presented in sect.~\ref{subsec:13}.

Pulse trains on neighbouring anodes of the same cluster are
merged into a {\it hit} if centres of gravity differ by less than one
time bin. The hit coordinate in azimuthal direction is calculated as
the centre of gravity of the contributing pulses, weighted by their
peak amplitudes. The hit  amplitude is the sum of the pulse amplitudes.

\subsubsection{Hit finding in RICHes and Pad Chamber}

\noindent
Signals of the RICH detectors and the Pad Chamber are read out from
the checkerboard-like arrangement of pads which receive the charge
amplified in multi-wire proportional chambers.  The same
hit-reconstruction algorithm is used for these detectors.

Adjacent pads with amplitudes above a readout threshold are
connected to \emph{clusters} some of which are caused by
background and electronic defects. Typical background clusters, like
long and thin stripes from ionising particles on oblique trajectories,
or clusters with many pads in saturation, are removed with the help of
various cleaning algorithms. Remaining clusters are split into
regions containing one local maximum and are identified as UV-photon
{\it hits} in the RICHes or charged-particle hits in the Pad Chamber.
Hit centres are calculated as the centres of gravity of the
contributing pads.

\subsubsection{Ring candidates in the RICH detectors}

\begin{figure}[ttt!] 
  \resizebox{0.35\textwidth}{!}{%
    \includegraphics{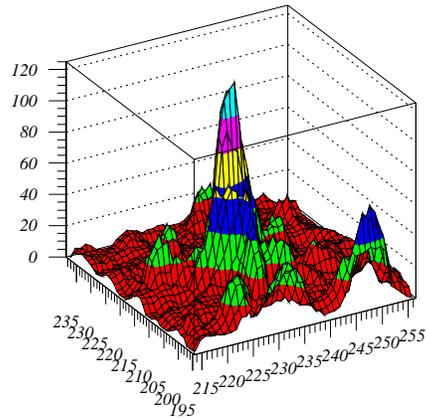}}  
\caption{\label{fig:3.4}(a) A Cherenkov
  ring in the RICH-1 pad plane, indicating pad amplitudes. (b) First
  Hough transformation. (c) Second, non-linear Hough transformation.}
\end{figure}

RICH detectors require an additional step of pattern recognition to
search for rings with asymptotic radius produced by
electrons~\cite{robust96}.  A typical RICH-1 ring is shown in
Fig.~\ref{fig:3.4}. The ring search is done using the Hough
transformation~\cite{hough62} from the data field of the pad plane
into the `Hough array' (or `parameter space') which has the same
dimension.  A given cell is the `image' of all data points (hits) on a
circle around that cell, and its `Hough amplitude' is the sum of the
data amplitudes on that circle. A ring candidate shows up as a
peak in the `Hough array' which is the higher the more photon hits, or
illuminated pads, lie on the ring. In practice, a `digital' Hough
transformation is used: all pads with signal amplitudes above a
defined threshold enter with unit weight into the sum amplitudes,
irrespective of their amplitudes.

The event display in Fig.~\ref{fig:3.4} demonstrates that rings can be
formed also by random arrangement of single photon hits, some of them
produced by pions near the Che\-renkov threshold.  The first step in
ring pattern recognition is a linear Hough transformation of the pad
plane onto the parameter plane. Besides the real maximum in the Hough
array, there are other local maxima connected to fake ring centres. As
a counter measure, a second, non-linear Hough transformation is
performed which assigns a relative weight to each cell in the Hough
array such that each hit counts most for its most favourable
ring-centre.  This way fake rings are suppressed, as can be seen from
Fig.~\ref{fig:3.4}, by a larger gap that has developed between the
amplitudes of the real and the fake rings.  The parameter to select
real rings at this stage of the analysis is the amplitude after the
second Hough transformation.

All surviving ring candidates are assigned their final centre
coordinates by a robust estimation which is based on iterative
re-weighted least-squares fits of circles with asymptotic radius to the
hits. A second fit with variable ring radius should eliminate charged
pions which have a non-asymptotic radius.  The function minimised is a
modified $\chi^2$ where the fit-potential varies in a Gaussian way
(instead of quadratically) with the distance to the minimum.  For an
extensive description of the fitting algorithm see
Refs.~\cite{ullrich-phd,ring_fit}. Other parameters besides the
amplitude in the Hough array, like the number of hits and the spread
of the hits around a perfect circle, can be used for fake-ring
suppression.

Ring reconstruction efficiency is determined by Monte-Carlo
simulations with a cut on the number of photon hits, assumed to be
Poisson-distributed. At the stage described up to here, ring
reconstruction efficiencies of about 85$\%$ are still confronted with
many fake rings, about 10 to 20 for one reconstructed ring that is
real. It was in anticipation of such alarming majority of fake rings
that the collaboration decided in 1994 to implement full external
tracking~\cite{pb-proposal94}.

\subsubsection{Calibration of the SiDC telescope}

To first approximation, the relation between
the drift distance and drift time is linear. From the known radial
extension $\Delta R$ of the active area and the total drift time
$\Delta t$, the drift velocity can be calculated as
\begin{equation}
 v_{drift}= \Delta R/\Delta t, ~~~~~\Delta R= R_{max}- R_{min},
\end{equation}
where $R_{min}$ is the inner edge of the active area and $R_{max}$
corresponds to the anode radius.\footnote{The anode radius is 32~mm
and 42~mm for the 3-inch and 4-inch detectors, respectively, $R_{min}$
is typically 10~mm and the total drift time 3 to 5~$\mu$s.}.  The
corresponding drift time is
\begin{equation}
\Delta t= t_{max}-t_{min},
\end{equation}
 where
\begin{figure}[t!]
\begin{center}
     \resizebox{0.32\textwidth}{!}{%
    \includegraphics{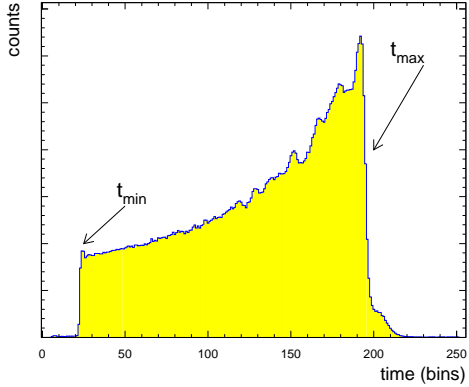}} 
\end{center}    
    \caption{\label{fig:3.5} Drift time spectrum,
    reflecting the pseudo-rapidity density of charged particle 
    production. Small structures are from local field variations.  
    One time bin =~20~ns, $v_{drift}\approx$~9.4~$\mu$m/ns. SiDC-2.}
\end{figure}
$t_{min}$ is the time corresponding to the shortest drift path
(ionisation directly under the anode), and $t_{max}$ the drift time of
particles starting at $R_{min}$.  A typical hit distribution as a
function of drift time is displayed in Fig.~\ref{fig:3.5}. By fitting
the edges of the drift-time spectrum, the values of $ t_{min}$ and $
t_{max}$ are determined.

Following such preliminary calibration, the geometrical alignment of
both detectors and the determination of the interaction vertex is
performed. Other corrections, such as the stop-pulse correction,
corrections on drift-velocity variations due to temperature changes,
etc., are done while maintaining the reconstructed vertex position.

\subsubsection{Vertex reconstruction}

Once hits are reconstructed, the information of hit positions in the
laboratory coordinate system is used to combine hits to {\it tracks}
and find the {\it vertex position} to which almost all tracks of a
given event point to. The procedure to minimise the quadratic sum of
hit mismatches between the two detectors and its iteration is
extremely time consuming. We used therefore a robust vertex fitting
approach~\cite{robust97}: all hits in SiDC-1 and SiDC-2 (typically
more than 100) are combined to straight track segments and a weighted
sum of their projected distances to the assumed vertex position is
calculated. In the next iteration, this centre of gravity becomes a
new starting value for the vertex position and each track segment gets
a new weight according to its deviation from the mean value in the
step before. After the position of the vertex is determined, its
$z$-position is redefined as the exact position of the closest vertex
disc.
 
\begin{figure*}[t!]
     \resizebox{0.99\textwidth}{!}{%
    \includegraphics{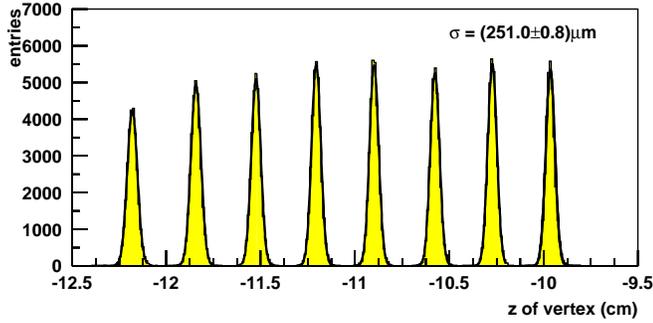}}    
     \caption{\label{fig:3.6}
        Reconstruction of interaction vertices. (a) Density of
$z$-positions with peaks corresponding to the eight target discs. (b,c) Vertex 
resolution in the {\it x-y} plane determined from 
        high-$p_t$ pions. 1996 Pb-Au data.}
\end{figure*}

Figure~\ref{fig:3.6} displays the density of reconstructed vertex
positions along the beam.  The peaks reveal the positions of the eight
discs of the segmented target assembly with a spacing of 3.2~mm. On
average, the reconstruction was done with 160 charged particles per
event. The data shown in Fig.~\ref{fig:3.6} was accumulated over $
2.6\times$10$^5$ Pb-Au events, or six hours, and demonstrates a
certain long-term stability.  The resolution\footnote{i.e. the
standard deviation in the mean  $z$ position} of
$\sigma_z= 250~\mu$m is sufficient to identify the correct target disc
without ambiguity\footnote{due to the unknown location of
the vertex inside the 25~$\mu$m thick foil, there is an rms error of
25/$\sqrt{12}$, about 7~$\mu$m.} .

More critical for tracking accuracy is the precision with which the
vertex can be localised in the plane transverse to the beam, as there
are no fixed points. To measure it we used stiff pion tracks with
$p_t\geq$~1.2~GeV/$c$ to minimise multiple scattering.  The scatter
plot in Fig.~\ref{fig:3.6} accumulates over many events the distance
in $x$ and $y$ between the actual event vertex (determined from all
tracks) and the point where a given pion track intersects the
respective target disc. Projecting on the axes, we obtain the 1-dim
distribution shown to the far right; a lateral resolution of 28~$\mu$m
($x$ and $y$) is derived. The transverse vertex resolution is
therefore $\sigma_r=
\sqrt2\sigma_x\simeq 40~\mu m$.  By choosing the vertex position in
the $x,y$-plane as the origin of the event coordinate system, we
account for event-by-event displacements within the diameter of the
target.

\subsubsection{Charged particle tracks}

Silicon-drift track segments are constructed by
connecting the vertex point to hits in SiDC-2 which lie within the
fiducial acceptance.  A track segment is accepted if there is at least
one hit in SiDC-1 within a predefined window around the point of
intersection; for more than one hit, the centre of gravity is
taken. Once the interaction vertex and the SiDC track segments are
reconstructed, the trajectories of charged particles are extra\-polated
downstream to the Pad Chamber. If a pad hit is found within a certain
fiducial window, a track candidate is created. 

The sizes of the fiducial windows are expressed as multiples of
the rms widths of the corresponding matching distributions; usually
these are 5$\,\sigma_{match}$ during production and 3$\,\sigma_{match}$
during the final analysis. The fiducial windows were
taken momentum-dependent to reduce efficiency losses at low momentum
due to multiple scattering. 

The matching of tracks in $\phi$ direction between detectors separated
by the magnetic field requires special attention since such tracking
window corresponds to a momentum cut. To avoid loss of tracks by
multiple scattering for large $\phi$ deflection, the fiducial window
in $\theta$ direction is opened to assume the shape of a butterfly.
During the production stage, we use a tracking window in $\phi$ of
$\pm$~0.6~rad. This corresponds to the p$_t$-cut of 50~MeV/$c$
mentioned already.  During off-line analysis, the size of the
butterfly is approximated by a sector of 100~mrad in $\phi$,
corresponding to a lower momentum cut-off of 1~GeV/$c$, the standard
$p_t$ cut of 200~MeV/$c$, and 3~mrad in $\theta$ direction. Matching
distributions will be discussed in sect.~\ref{subsec:13}.

The contribution of background hits was estimated by applying detector
rotations. Since the hits in the SiDC's are highly correlated, the
main background contribution comes from random combinations of SiDC
track segments with hits in the Pad Chamber. By rotating these
hits in a given event by a random angle with respect to
the silicon detectors, the true physics signal is destroyed and only
background tracks remain.

By requiring that track elements from all five detectors match within
three standard deviations of the detector resolutions combined with
the rms spread due to multiple scattering, fake tracks are reduced to
a negligible level.

\subsubsection{Straight pion tracks for calibration}

 Straight tracks of high-momentum pions are an important tool for
 fine-tuning the spectrometer calibration~\cite{ceretto-phd}.  Pion
 identification uses external tracking by the SiDC's and the Pad
 Chamber to predict ring centre positions in the RICHes. Pions have
 been selected under tight quality criteria regarding matching, number
 of photons on Cherenkov rings, and clean environment around rings.

To extend the SiDC track segments into the Pad Chamber, only a narrow
window of $\pm30$~mrad in azimuth is searched for a matching hit,
variations in deflection being very small.  In polar direction, the Pad
Chamber is searched merely over the matching window between the two
detectors since multiple scattering is negligible. All hits found are
candidates.  From the coordinates before and after the field, the ring
centres in the RICH detectors are predicted. The pointing to RICH-1,
without deflection, is unproblematic. With the momentum
information derived from the azimuthal deflection between SiDC and Pad
Chamber, the tracks are extrapolated into RICH-2, behind the field.

Photon hits that fall into the vicinity of the predictors are
collected and used as input for a robust fitting
algorithm~\cite{robust96}.  To find centre and radius of the ring, the
rms deviation of hit positions from a circle with radius $R$ around the
ring centre is minimised in a Gaussian fit potential with three free
parameters. Once the radius $R$ is found, the pion momentum is
re-evaluated from the relation 
\begin{equation}
\sqrt{p^2+ m^2}= 
\gamma_{thr}\,\frac{m}{\sqrt{1-(R/R_\infty)^2)}},
\end{equation}
where 
$R_\infty$ stands for the asymptotic ring radius.

\subsubsection{Spectrometer calibration}

For the analysis of the 1995 and 1996 data every spectro\-meter
component was calibrated. Several successive steps were
\begin{figure}[t!]
      \resizebox{0.48\textwidth}{!}{%
    \includegraphics{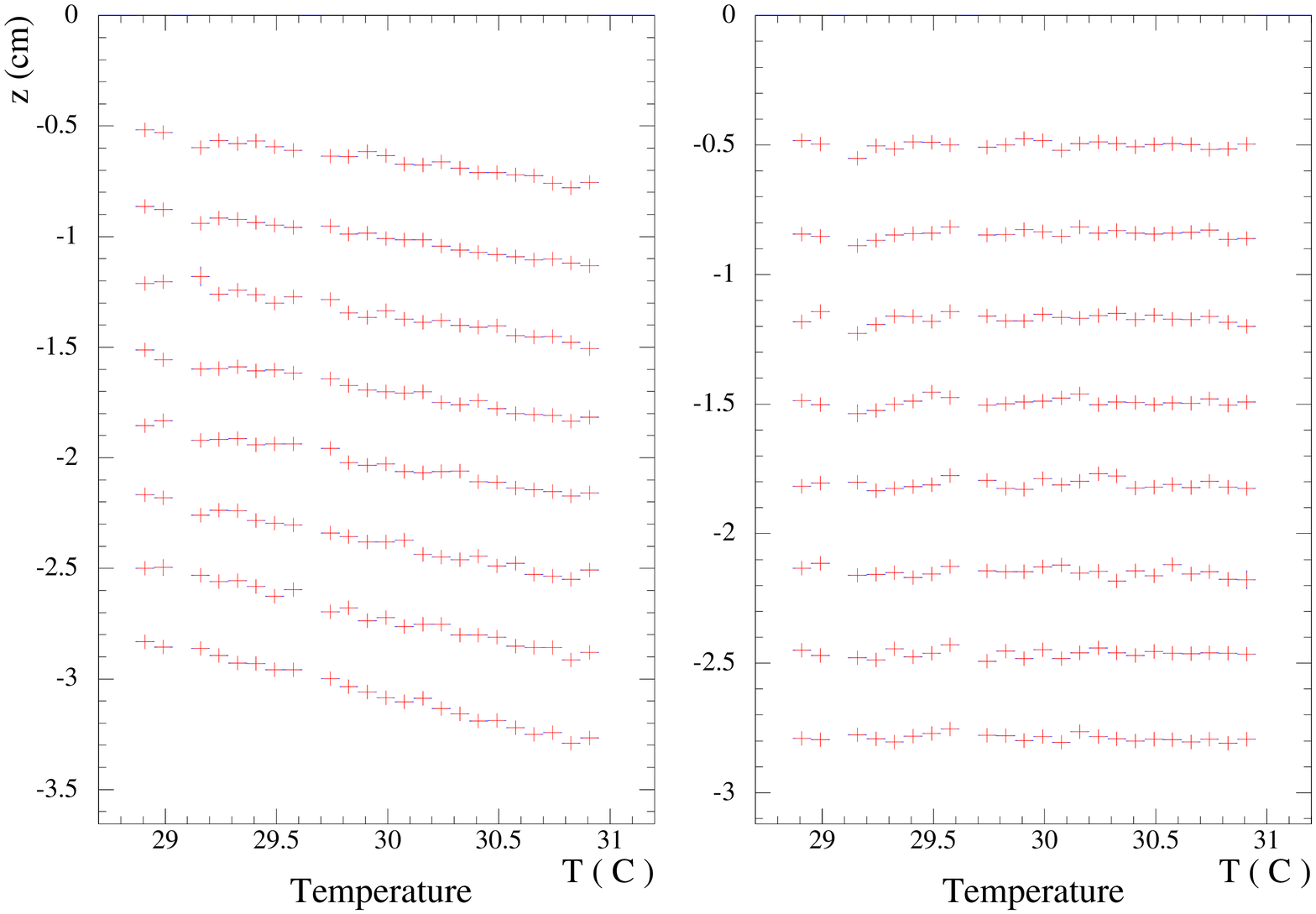}}    
    \caption{\label{fig:3.7} Vertex positions of the 8 target discs
    taken when cooling of the SiDC's was switched
    off (left), and  after stabilising on the drift time
    spectrum (right). SiDC-2, 1995 data.}
\end{figure}
necessary to align all detectors and to derive appropriate correction
functions for the analysis chain. A first rough calibration was
performed to adjust detector offsets and to correct small rotations
and tilts.

Since electron mobility in silicon strongly depends on
temperature\footnote{The electron mobility $\mu= v_{drift}/E$ depends
on temperature as $ \mu\propto T^{-2.4}$.}, the calibration is
strongly affected by temperature variations.  To keep the calibration
stable over periods of hours, we employed the simple and fast method
by which the position of the upper edge of the drift time spectrum is
monitored to provide an `online' drift-velocity stabilisation. The
feedback procedure successfully stabilises the calibration as can be
seen from the resulting stability of vertex positions in a set of test
measurements taken while the cooling was switched off,
Fig.~\ref{fig:3.7}. Under normal running conditions, the temperature
variations were typically below 0.5$^\circ$C over 12 hours.

\begin{figure}[t]
       \resizebox{0.48\textwidth}{!}{%
    \includegraphics{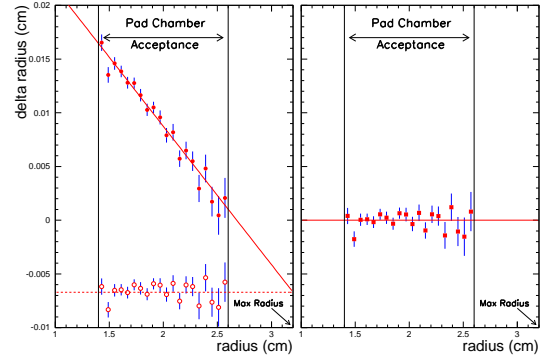}}    
    \caption{\label{fig:3.8} 
Deviations in radial hit position with respect to reference 
of stiff $\pi$ tracks; before calibration
(left, full circles), after tuning the $v_{drift}$-parameter  
 (left, open circles), and the $t_{min}$ parameter
    (right). SiDC-1, 1995.}
\end{figure}

On even longer time scales, drifts in the calibration of the SiDC's
were prevented employing the fixed reference provided by the Pad
Chamber using high $p_t$ high-statistics pion samples.  The method
registers the deviations of hits in either one of the SiDC's from the
straight line that connects the event vertex with a selected hit in
the Pad Chamber; it works fine in a pre-calibrated system. The effects
of fine tuning the parameters $v_{drift}$ and $t_{min}$ of SiDC-1 can
be seen in Fig.~\ref{fig:3.8}.

Response from RICH-1 when given close scrutiny by the absolute
reference grid of the Pad Chamber revealed slight deformations of the
spherical mirrors which caused non-linearities in polar
angle. Following the observation that the radius of curvature of
mirror-1 decreases towards the rim, application of a small
$\theta$-dependent variation in the focal length removed this problem.

\subsubsection{Matching quality}

\begin{figure}[b!]
        \resizebox{0.50\textwidth}{!}{%
    \includegraphics{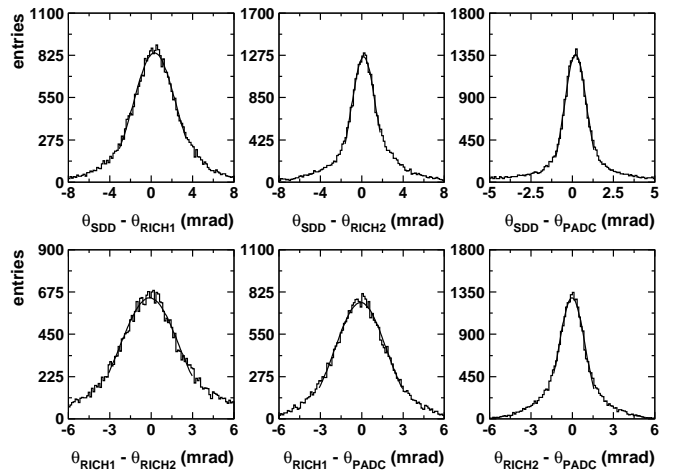}}    
\caption{\label{fig:3.9}
The $\theta$-matching distributions for various detector combinations 
determined with high-p$_t$ charged pions.}
\end{figure}

Starting with an internal calibration of the SiDC vertex telescope,
the remaining detectors are aligned to the centre of SiDC-1. Internal
consistency and quality of the readjusted calibration can be evaluated
from the residual offset in centres of
\begin{figure}[t!]
\begin{center}
          \resizebox{0.42\textwidth}{!}{%
    \includegraphics{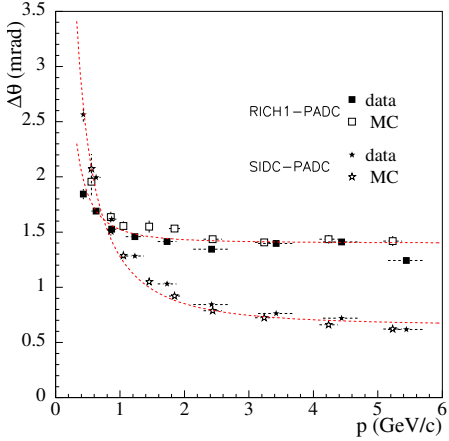}}  
\end{center}  
    \caption{\label{fig:3.10} Quality of track matches $\Delta\theta$ 
    of Pad Chamber to RICH-1 and to SiDC {\it vs}  momentum. Shown are
    1996 data and Monte-Carlo simulations. See text.}
\end{figure}
gravity, the widths of the matching distribution close to the peak,
and the amount of background in the tails, for all detector
combinations. Examples of matching distributions are shown in
Fig.~\ref{fig:3.9}.

At low momenta, the matching quality deteriorates because of multiple
scattering in the material, and it is no longer determined by the
detector resolutions alone.  The widths of
the matching distributions in $\theta$, plotted for
SiDC~-~Pad\,Chamber and RICH-1~-~Pad\,Chamber shown in
Fig.~\ref{fig:3.10} for measured and simulated spectra, exhibit very
well the dominance of the multiple-scattering contribution increasing
with $1/p$ over the momentum-independent detector resolution towards
small momenta. The simulations with the actual spectrometer parameters
are shown by dashed lines.  Final track selection is achieved
by appropriate momentum-dependent cuts in the matching distributions.

\subsection{Momentum analysis and mass resolution}
\label{subsec:14}

Since particles passing the SiDC's and RICH-1 have
 experienced no magnetic field, these detectors are used as
 zero-deflection reference.  
\begin{figure}[b!]
         \resizebox{0.48\textwidth}{!}{%
         \includegraphics{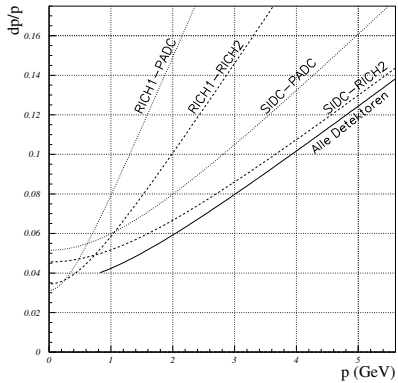}}
         \caption{\label{fig:3.11} Momentum resolution obtained with
         different detector combinations {\it vs} momentum; MC
         simulations. The solid line combines the information of all
         five detectors.}
\end{figure}
The deflections $\Delta\phi$ of charged particles of momentum $p$
between the RICHes and between the SiDC's and the Pad Chamber were
given in sect.~\ref{subsec:4}. The best resolution is obtained from all
possible detector combinations, weighted by the respective accuracy.
The result of the simulation with the measured detector resolutions as
input is shown in Fig.~\ref{fig:3.11}.  By detailed Monte Carlo
simulation which realistically describes the individual detector
resolutions and the quality of the matching distributions, the overall
resolution is described by the function
\begin{equation}
\Delta p/p= \sqrt{(2.3\% \cdot p)^2+ (3.5\%)^2},
\end{equation}
 with $p$ in GeV/$c$. The momentum dependent part is due to 
 detector resolutions, the constant part due to multiple
 scattering. The mass resolution at the $\rho/\omega$ is
 about 6$\%$, at the $\phi$ about 7$\%$.

\subsection{Rejection of combinatorial background}
\label{subsec:15}

With a $p_t$ cut of only 50~MeV/$c$ at the
production level, an important part of the rejection is already
achieved: very soft tracks are removed, the tracks of all
reconstructed conversion and Dalitz pairs are marked and taken out of
the further analysis, including pairs with only one
electron of $p_t>$~200~MeV/$c$. Only then is the single-track
$p_t$-cut tightened from 50 to 200~MeV/$c$ which reduces the
combinatorial pair background by about a factor of 10 while keeping
97\,$\%$ of signal pairs.  Tracking cuts are narrowed down from
5$\sigma$ to 3$\sigma$. Together with increased requirements in track
and ring quality, this results in a drastic suppression of fake tracks
as mentioned before.

\begin{figure}[t!]
         \resizebox{0.48\textwidth}{!}{%
    \includegraphics{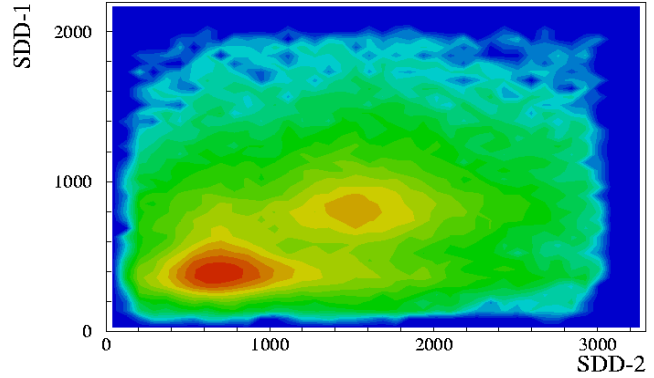}}  
\caption{\label{fig:3.12} Correlated  SiDC
    amplitudes in 5~mrad summing window for sample enriched in photon
    conversions. Majority of particles have single tracks in both
    SiDC's (pions), or double amplitudes which 
    signify electron pairs from photon-conversion in the target.}
\end{figure}

The most powerful rejection tools on the pair level derive from the
fact that target conversions and a large fraction of $\pi^\circ$
Dalitz decays produce close tracks. These are either registered as
single hits of `double-dE/dx' in the SiDC's when the separation is
less than about 3~mrad, or as a `double ring' in \mbox{RICH-1} when the
two rings coalesce into a single structure for separations
below 8~mrad; or as more or less well resolved tracks, still
considerably closer than the mean spacing of charged particle tracks
in the SiDC's or of electron tracks in RICH-1.

To maintain high rejection power also in the region between perfect
overlap and resolved double hits, we use a `summation' window in the
SiDC's of 5~mrad in which hit amplitudes are summed up\footnote{This
allowed to relax demands on double-hit resolution and thereby to avoid
excessive (and mysterious) `hit splitting'.}. Figure \ref{fig:3.12}
shows a two-dimensional plot of the hit response in SiDC-1 {\it vs}
SiDC-2 obtained this way.  It demonstrates that with an appropriate
two-dimensional rejection-cut, photon conversions in the target are
efficiently rejected without cutting much into the Landau tail of
single-hit distributions.  RICH-1 identifies close tracks of
conversions and Dalitz pairs by the number of photon hits, which
suffers severely from pile-up losses, however. Alternatively, we use
directly the analog sum of the gain-corrected pad amplitudes to
identify `double rings'. Photon conversions occurring in SiDC-1 differ
in response from target conversions only by producing on average a
single-hit signal in SiDC and can be recognised by such signature.

To be specific, we use the following four rejection steps
which evolve in complexity with the information gained 
along the trajectories:

\begin{enumerate}
\item Tracks  are rejected if the amplitudes of the 5~mrad summation
window in {\it both} SiDC's are larger than the typical single hit
response encoded by parameters \\$S1_{high}$ and $S2_{high}$.

\item Conversions in SiDC-1 are rejected by the requirement that
the amplitude in SiDC-1 is below $S1_{low}$ and the amplitude in SiDC-2
above $S2_{high}$,  summed over a 7.5-mrad window; and by observation
of a double ring in RICH-1.

\item A track with a double ring in RICH-1 is rejected already
if $S_{high}$ is surpassed in only one of the SiDC's. Besides improving
rejection efficiency in general, this cut rejects conversions in SiDC-2.
 
\item Tracks are rejected if another ring in RICH-1
is found within a wider window of 35~mrad.  To avoid excessive
vetoing by accidental structures, the second ring is required to be of
high quality and to connect to a track segment in both SiDC's.
\end{enumerate}

A few comments are in order. The search in the SiDCs for hits near
\begin{figure}[t!]
\begin{center}
          \resizebox{0.48\textwidth}{!}{%
    \includegraphics{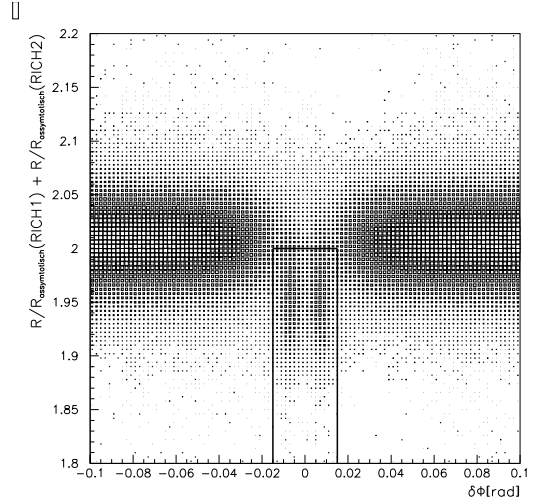}}  
\end{center}  
    \caption{\label{fig:3.13}
    Scatter plot of the sum of the reduced ring radii
    \mbox{$R/R_\infty$} in RICH-1 and RICH-2 over
    deflection in the field
 for reconstructed electron tracks with asymptotic
    radius \mbox{$ R_\infty$}.
 Vertical stripes are from charged pion tracks of both charge signs. 
The pion rejection
    cut used is indicated. }

\end{figure}
tracks cannot be extended to larger distances because of the
increasing chance to find a pion, vetoing the signal. For the purpose
of `Dalitz rejection', the veto is better based on RICH-1 where the
inspection area can be opened up to 35~mrad without much signal loss
because the density of {\it electron} tracks is so much lower.
The strategy described above is quite successful in rejecting
conversions but even with the close-ring cut it is of limited power
in rejecting $\pi^0$-Dalitz pairs because of the larger opening
angles involved.

In the last step we also reject surviving charged pion
tracks. Figure~\ref{fig:3.13} is a scatter plot of track deflection
$\Delta\phi$ as abscissa, and the sum of the reduced ring radii in
both RICHes on the ordinate; this quantity only depends on the Lorentz
$\gamma$ factor.  Because of different mass compared to electrons,
pions of same momentum (deflection) have a different $\gamma$
(radius).  It is seen that pion tracks even with asymptotic radius are
clearly distinguished from electron tracks by virtue of the much
smaller deflection in the field, and they can be rejected by a
two-dimensional cut in both RICH detectors on ring radius {\it vs}
deflection.  This cut was used only in the '96 analysis; it improves
the signal-to-background ratio at large masses and has little impact
at lower masses.

\subsection{Optimising signal quality}
\label{subsec:16}

In tuning the parameters of quality and rejection cuts, we strictly
avoided to optimise on the data itself.  Such practice is known to
increase the risk of statistical fluctuations producing spurious,
misleading results. Rather, the signal efficiency was determined by
overlaying Monte-Carlo-generated tracks chosen from the hadronic
cocktail on real events and measuring the rate of successful
reconstructions using the quality and rejection cuts as in the current
step of the data analysis. The signal efficiency may also be monitored
and optimised using a sample of fully reconstructed Dalitz
pairs~\cite{socol-phd,damjanovic-phd}. The rejection steps were also
tuned by Monte-Carlo techniques with generated
conversions and $\pi^0$-Dalitz decays. Such a Monte-Carlo
simulation of the background is quite demanding and was only achieved
in the 1996 data analysis~\cite{lenkeit-phd}. In the 1995 analysis,
the background was taken from the data sample itself.

By these procedures, each analysis cut is arbitrated objectively
according to how much in background rejection is gained for how little
loss in signal efficiency. A clear understanding of what is meant by
an `optimal' balance is provided by the `equivalent background-free
signal'\footnote{It is equal to the number of signal pairs of the same
statistical significance if there would be {\it no background}.}
\begin{equation}
 S_{eq}= \frac{S^2}{S+2B}.
\end{equation} 
For a given rejection cut which reduces the signal from $S$ to
$S^\prime= \varepsilon\,S$ and rejects a fraction $(B-B^\prime)/B=\,r$ of the
background, the ratio 
\begin{equation}
\frac{S_{eq}^\prime}{S_{eq}}\approx\frac{\varepsilon^2}{1-r}
\end{equation} 
should be maximised.

In summary, the rejection steps together reduce the number of
background pairs by factors of 15 and 12 for mass above and below
200~MeV/$c^2$, respectively, while signal pairs are reduced by only
20$\%$.

\subsection{Pairing and subtraction of combinatorial background}
\label{subsec:17}

The combinatorial background $B$ of unlike-sign pairs can be accounted
for by the like-sign pair sample because by lack of any physics source
the latter is purely combinatoric,
\begin{equation}
B= 2\,\sqrt{N_{++}N_{--}}.
\end{equation}
Here, $N_{++}, N_{--}$ denote the numbers
of $e^+e^+$ and $e^-e^-$ pairs, respectively. The relation
is exact when charge symmetry is fulfilled.  The number of signal
pairs~$S$ is obtained by subtracting the number of background pairs
from the number of unlike-sign pairs,
\begin{equation}
 S= N_{+-} - B.
\end{equation}
The variance in $S$ is approximately\footnote{Combinatorial pairs are
non-Poissonian, their variance exceeds the mean.  Because
the density is only of order $10^{-4}$ per event, the deviation
from eqn.~(3.14), however, is below 5$\%$.} 
\begin{equation}
 \sigma_S^2= N_{+-} + B.
\end{equation}
Typically, the open pair signal $S$ accounts for only a
small fraction of our total unlike-sign pair sample for masses above
200~MeV/$c^2$.

The asymmetry induced by Compton electrons and ${\textstyle K_{e3}}$
decays is below 1$\%$ and would alter the factor of 2 in eqn.~(3.12) by
less than $1\times 10^{-3}$.  It is worthwhile to note that eqn.~(3.12)
\begin{figure}[t!]
\begin{center}
        \resizebox{0.48\textwidth}{!}{%
    \includegraphics{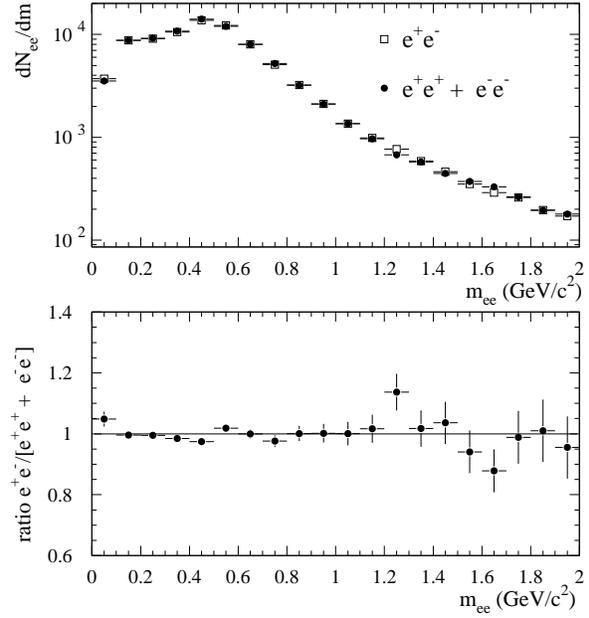}}   
\end{center} 
    \caption{\label{fig:3.14} Comparison of mass spectra of
    strictly combinatorial pairs generated from clean data samples of
    conversions and Dalitzes (see text).  Neither in form nor
    in absolute yield does a significant charge asymmetry become visible.
    }
\end{figure}
remains a very good assumption also in case of a small imbalance in
the number of tracks of positive and negative charge, be it due to
different acceptance or reconstruction efficiencies, or to a generic
imbalance~\cite{socol-phd}. For an asymmetry of 10$\%$ in the number
of positive and negative tracks, the factor 2 in the above expression
would decrease by 2$\%$ only.

The asserted symmetry has been tested by generating purely combinatorial
pairs of either type, like-sign and unlike-sign, from data by
`reversing' the rejection cuts. The recipe is the following: prepare a
clean sample of conversions and Dalitz decays and recombine all tracks
such that none meets its original partner. This sample is void of
signal pairs and in every respect resembles the experimental pair
background. The mass spectra of both components of this artificial
background sample are compared with each other in
Fig.~\ref{fig:3.14}. No significant charge asymmetry is visible
between the mass spectra of the like-sign and the unlike-sign pair
background, neither in shape nor in absolute yield. Over the
full mass range, the ratio of the integral yields is
\begin{equation}
Y_{e^+e^-}/(Y_{e^+e^+}+Y_{e^-e^-})= 0.996\pm 0.005,
\end{equation}
which excludes an asymmetry larger than 1$\%$ at 90$\%$ confidence.
We shall return to this topic in sect.~\ref{subsec:28}.

Let us take an alternative route of estimating combinatorial
background. A philosophically correct way is to generate unlike-sign
pairs by combining strictly uncorrelated tracks chosen from different
events under the same kinematic constraints as applied to the
data. This is commonly referred to as the {\it mixed event}
method. Practically, efficiency losses that occur in the true data for
close tracks have not been accounted for.  A comparison of combinatorial
mass spectra obtained by the two methods from the full 1996 data set
\begin{figure}[t!]
           \resizebox{0.48\textwidth}{!}{%
    \includegraphics{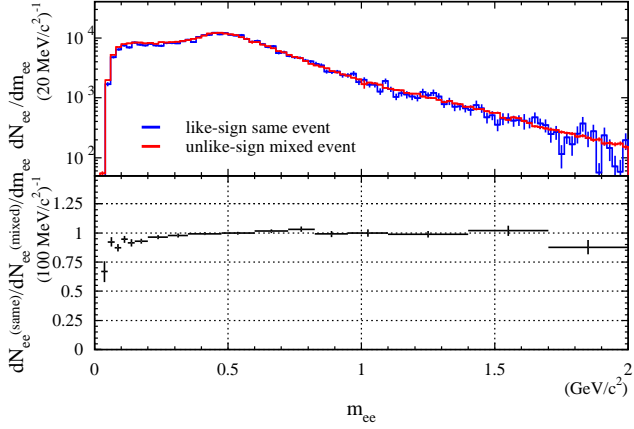}}    
\vspace{-0.cm}
    \caption{\label{fig:3.15} Comparison of combinatorial background
    spectra from same-event like-sign pairs and, alternatively, from
    unlike-sign pairs of different events. Top: Comparison on log
    scale.  Bottom: ratio `same'/`mixed' on linear scale. 1996 data.}
\end{figure}
is displayed in Fig.~\ref{fig:3.15}.  The coarse comparison on a
log scale (top panel) does not show deviations. On a finer linear
scale (bottom panel), the ratio indicates that the like-sign pairs
loose up to 30$\%$ compared to the mixed-event generated unlike-sign
pairs - which is, however, 
\begin{figure}[b!]
           \resizebox{0.48\textwidth}{!}{%
    \includegraphics{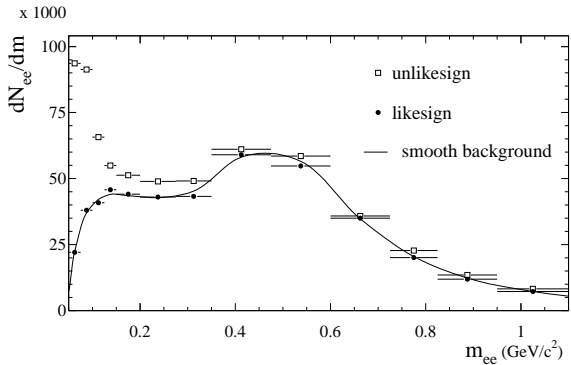}}    
    \caption{\label{fig:3.16}  Unlike-sign (open squares) and like-sign
     mass spectra  (full circles) after all rejection cuts.
    Like-sign combinatorial background shown as smooth curve is
    obtained by random pairing of tracks with the same
    kinematical composition as in data.
    The drop in yield below 400~MeV/$c^2$
    originates from the single track $p_t$-cut of 200~MeV/$c$.
1996 data.}
\end{figure}
due to the mentioned loss in detection efficiency at close track
separation; an effect not present in the current realization of the
mixed-event method. Yet, the ana\-lysis allows to conclude that above
$m\approx$~300\,MeV/$c^2$ the ratio stays at unity within a band of
less than $\pm$1.5$\%$.

Measured mass spectra of like-sign and unlike-sign pairs are shown in
Fig.~\ref{fig:3.16}. For comparison, a like-sign pair background has
also been generated by random pairing of electron tracks with momenta,
polar angles and transverse momenta chosen at random from measured
distributions. The result of such simulation is shown in the figure
and compares very well with the measured like-sign mass spectrum. The
simulated opening-angle distribution (not shown) also agrees well with
the measured distribution.

Figs.~\ref{fig:3.14}, \ref{fig:3.15} and \ref{fig:3.16} display the
typical shape of the combinatorial mass spectrum for the applied cut
on track ${ p_t}$.  This kinematical condition produces a falloff
towards low masses and a broad peak at about twice the minimum track
$p_t$. The peculiar shape of the combinatorial mass spectrum peaking
around mass of 500~MeV/$c$$^2$ $-$ where the enhancement dominates $-$
raises doubts whether the combinatorial background has been correctly
assessed and subtracted. We return to this issue in
sect.~\ref{subsec:28} to assert an unbiased background handling.

\subsection{Spectra and statistical errors}
\label{subsec:18}

The raw signal counts and their relative statistical
errors are obtained by subtracting the combinatorial background 
 channel by channel from the spectrum of unlike-sign
pairs,
\begin{equation}
{\textstyle s_{\textstyle i}= (n_{\,i,+-} - b_{\,i}),}
~~~~{\textstyle \sigma_i/\textstyle s_i= 
\sqrt{n_{\,i,+-} + b_{\,i}}/(n_{\,i,+-} - b_{\,i}).}
\end{equation}
The counts per channel add up to the total counts known already from
eqns.~(3.12) and (3.13),
\begin{equation}
 \sum_i  n_{\,i,+-}= N_{+-},~~~~~ \sum_i b_{\,i}= B.
\end{equation}
Pair yields per event, corrected for
reconstruction efficiency and normalised to $\langle N_{ch}\rangle$,
are written symbolically
\begin{equation}
{\textstyle y_{\textstyle i}= 
g\,(1/\epsilon)_{\textstyle i}\,s_{\textstyle i},}
\end{equation}
with  $g$ a scaling factor and  $1/\epsilon$ for efficiency correction,
possibly channel dependent.  
Relative statistical errors are determined by the raw
data counts, 
\begin{equation}
\sigma_{\textstyle y,i}\,/y_{\textstyle i} = 
 \sigma_{\textstyle i}/s_{\textstyle i}.
\end{equation}

To reduce staggering in signal counts due to channel fluctuations in
the background spectrum, `smoothed' combinatorial pair spectra free of
bin-to-bin fluctuations have been subtracted. The method
allows, moreover, to limit the fluctuations in data points of
background-subtracted spectra to truly statistically independent
errors; it is discussed in the following.

The Monte-Carlo generation of background spectra is described in the
previous section.  The subtraction from the unlike-sign pair spectrum
is done bin by bin,
\begin{equation}
s_{\textstyle i}^{smooth}= 
n_{ i\,,+-}- b_{\textstyle i}^{MC}= 
n_{i\,,+-}- f_{\textstyle i}\,B.
\end{equation}
\noindent
The fractions $f_i$ add up to $1$.  The bin-to-bin errors
\begin{equation}
\left(\sigma_{\textstyle y,i}\,/y_{\textstyle i}\right)_{smooth} = 
\sqrt{n_{i\,,+-}}/(n_{i\,,+-} - b_{\textstyle i})
\end{equation}
in our case of $B/S\gg 1$ amount to a fraction quite accurately of
$1/\sqrt{2}$ of the full error of (3.19). The finite sample error
in $B$, to which the smooth background spectrum is normalised, was
neglected so far. It contributes to each channel a share of the same
magnitude as the bin-to-bin error itself\footnote{because these
errors add in quadrature.}. 

Since the `normalisation' errors common to all channels differ in size,
according to the magnitude of $f_i$ in eqn.(3.20), they cannot be
represented as an error in scale.  However, there is no need to show
them separately; we only have to keep in mind that the data points
have normalisation errors which are of the same size as the
statistical bin-to-bin errors which will be displayed, but are
tightly correlated within the entire spectrum.
 
\subsection{Centrality determination}
\label{subsec:19}

The measurement of absolute cross sections is hampered by the fact
that the number of particles that actually pass the segmented target
is not precisely defined; this is because the diameter of the Au disks
is not much larger than the beam diameter.  We decided therefore to
calibrate the centrality of collisions by the shape of the
pseudo-rapidity density of charged particles as measured by the
SiDC's. Once the measured $N_{ch}$ distribution is corrected for
various instrumental distortions, it is used to calibrate the charge
density obtained from minimum-bias UrQMD
calculations~\cite{urqmd}. These still differ from the measured
distribution at the low-$N_{ch}$ tail due to the non-ideal response of
the interaction trigger. The final step consists in describing the
triggered $N_{ch}$ distribution by a linear combination of UrQMD
calculations belonging to different impact parameters. These steps are
described below.
\begin{figure}[t!]
\begin{center}
             \resizebox{0.50\textwidth}{!}{%
    \includegraphics{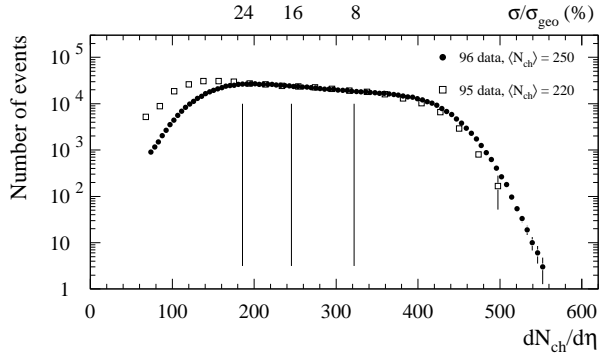}}
\end{center}    
    \caption{\label{fig:3.17}
      Distribution of centrality triggers for the two data samples of
      1995 and 1996. Centrality is expressed by the number of
      reconstructed charged particle tracks in the two SiDC's in the
      range $2\leq y\leq 3$, corrected for pile-up and efficiency.
      The selection corresponds to the most central 33$\%$ and 26$\%$ of
      the total inelastic cross section for 1995 and 1996, respectively.}
\end{figure}

\subsubsection{Charged-particle density}

Charged-particle density in the range 2$\,\leq \eta\leq
\,$3 is derived off-line from the number of tracks that emerge from
the event vertex and intersect the two SiDC's sufficiently close to
reconstructed hits.  The determination of the track reconstruction
efficiency is a rather complex problem.  To the percent accuracy
required, it is influenced not only by hardware imperfections
(i.e. dead anodes, electronic noise, pulse shape), but also by the
quality of hit and track reconstruction. In particular, `pile-up'
effects due to finite double-hit resolution and artificial hit
splitting are the most important effects to be corrected for. To that
goal, we simulate charged particle tracks in the SiDC's generated from
UrQMD events. The analysis of the Monte-Carlo events uses the standard
data analysis software, besides larger matching windows of
5~$\sigma_{match}$. Details are described in sect.~\ref{sec:4}.

The functional dependence of the inverse of the reconstruction
efficiency $\varepsilon$ on the measured number of charged particles
is of the type
\begin{eqnarray} 
\varepsilon^{-1}(N_{ch,measured}) & = & N_{ch,true}/N_{ch,measured}= 
\nonumber\\
a\,+ b\,N_{ch,measured}. 
\end{eqnarray} 
The intercept at $N_{ch,measured}= 0$ corresponds to the inverse
`static' efficiency $a^{-1}$ and includes not only losses due to dead
anodes, but also gains from artificial hit splitting. It amounted to
97$\%$ and 93$\%$ in 1995 and 1996, respectively. The parameter $b$ is
about (6-7)$\times$10$^{-4}$ which corresponds to about 10$\%$
relative losses, considerably more than expected from pile-up alone.

The corrected multiplicity distribution of the trigger-selection is
shown in Fig.~\ref{fig:3.17}. In the middle part and at the
high-$N_{ch}$ side, the shape of the distribution is supposed to be an
undistorted image of partial cross sections, but not so at low
$N_{ch}$ where the {\it trigger profile} of the MD with threshold set
at 100 ${\it mips}$ becomes visible (see sect.~\ref{subsec:6}). The
thresholds of the centrality trigger corresponded roughly to 30\% for
'96 and 35$\%$ of the total inelastic cross section for '95.

\subsubsection{Calibrating the centrality using
UrQMD calculations}

A more precise calibration of the centralities is described in the
following. Minimum-bias distributions obtained by UrQMD (version 1.3)
have been scaled slightly to make their upper corners coincide with
the measured trigger distributions.  From the UrQMD minimum-bias
collisions several centrality classes are sorted out according to
certain impact parameter ranges.  To reproduce the shape of the
trigger profile at the lower edge of the $N_{ch}$
\begin{figure}[t!]
           \resizebox{0.48\textwidth}{!}{%
    \includegraphics{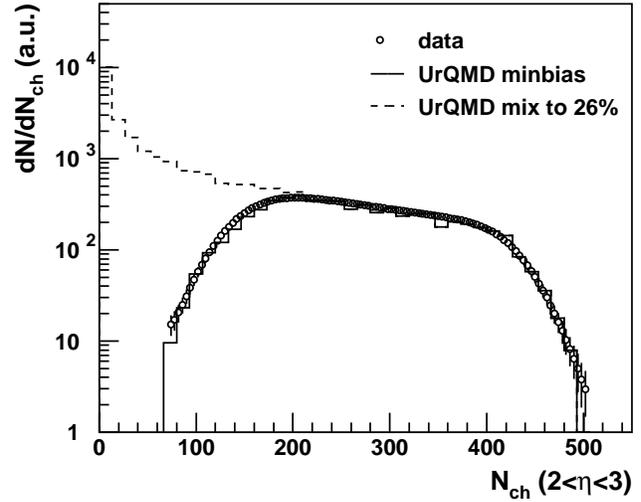}}    
\caption{\label{fig:3.18} 
UrQMD: minimum-bias $N_{ch}$ distribution scaled to fit the upper
corner (dashed histogram), and distribution of three centralities
composed to describe the trigger profile at the lower corner (solid
histogram); compared to the triggered $N_{ch}$ distribution (string of
small circles), see text. 1996 data.}
\end{figure}
distribution, a suitable linear combination of the corresponding
$N_{ch}$ distributions is used. The resulting distributions are
displayed in Fig.~\ref{fig:3.18}.  The total inelastic cross section
of 6.94~barn, and the trigger fractions were calculated with a
geometrical overlap model. The resulting $x\equiv \sigma/\sigma_{inel}$
 are:
\begin{equation}
x('95)=~0.6\times35\% + 0.2\times33\% + 0.2\times28\%=\,33\%
\end{equation} \begin{equation}
x('96)=~0.16\times35\% +0.15\times30\%+0.69\times23\%=\,26\%.
\end{equation} 
The systematic error from variation in the linear combinations alone
is estimated as $\pm$1.5$\%$ absolute. For the unified data, an
average centrality of 28$\%$ of the top geometrical cross section has
been adopted with an estimated uncertainty of about $\pm$2$\%$.

\section{Monte Carlo simulation}
\label{sec:4}
\subsection{Detector simulation}
\label{subsec:20}

The spectrometer with all material in proper geometry has been
implemented in the GEANT~\cite{geant} detector simulation package
version 3.15.  The present version of GEANT was not able to
describe the number of photons per electron ring correctly, which
forced us to write our own function.

The detector response is simulated by taking into account the
particular signal generation including all known effects that
influence position, width and amplitude.  In the SiDCs these include
the charge division among adjacent anodes, local variations in
drift-velocity over radius, diffusion along the drift path in radial
and orthogonal direction, anode-wise gain variations including dead
anodes, electronic noise from the readout, and digitisation errors.
In the RICH detectors we account for chromatic aberration and mirror
quality. In the UV detectors and the Pad Chamber, simulations include
the transverse diffusion in the conversion zone and the first
amplifying gap, the digitisation effects in the last multi-wire
amplification, and the noise of the readout electronics.

Most crucial but difficult to achieve is a realistic simulation of the
background in which the simulated signal is to be embedded.  The
background is caused by charged particles, gamma rays
and electro-magnetic radiation produced directly or indirectly by the
collision, $\delta$-electrons, and particles passing the UV-detectors.
\begin{figure}[tt!]
\begin{center}
        \resizebox{0.42\textwidth}{!}{%
    \includegraphics{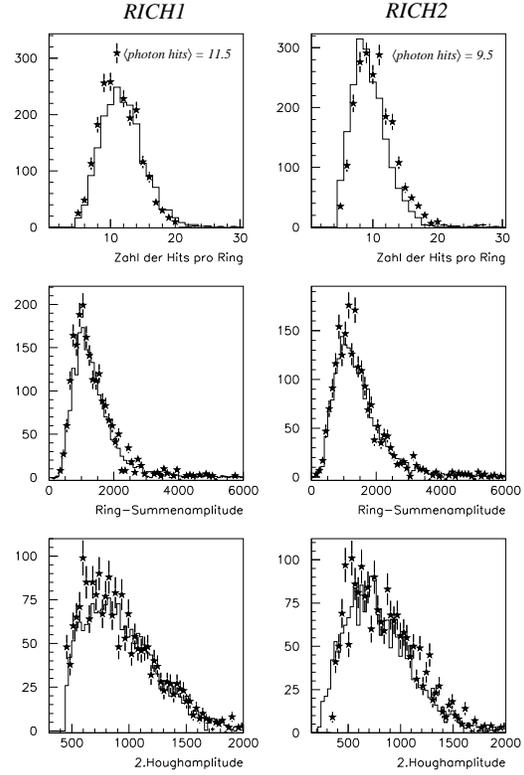}}  
\end{center}  
     \caption{\label{fig:4.1} Comparison of
    reconstructed Cherenkov rings in Monte-Carlo simulations
    (stars) and in data analysis (histogram) regarding the number
    of hits per ring (top row), the sum amplitude (middle row), and
    the amplitude of the second Hough transformation (bottom row).}
\end{figure}
In addition, the entire spectrometer or parts thereof might deviate
from response linear with produced signal charges due to transient
saturation effects of various kinds.  No sufficiently reliable
simulation of the background was possible up to now. We use
real data events into which the simulated Monte Carlo (MC) signals are
embedded. This is done by adding the MC amplitude for each
single cell, or pad, on top of the data amplitude. The systematic
error introduced by increasing the detector occupancy has been
estimated to be negligible.

The MC signal was reconstructed using algorithms identical with those
of the data analysis.  By comparing the signals of reconstructed MC
tracks and reconstructed data tracks we fine-tuned the parameters of
the detector simulations to achieve the best description of the
data. The characteristics employed for the comparison were the amplitude
distributions of hits, the widths, the numbers of
pads/time-bins/anodes contributing to a hit, the number of hits
belonging to a ring and the local variations of all these parameters
over the detectors.  An example of the agreement achieved between MC
simulation and data analysis in reconstruction of
Cherenkov rings is shown in Fig.~\ref{fig:4.1}.

It is crucial to model quantitatively the response to compound signal
patterns, like accuracy in reconstruction of hit positions or ring
centre positions, or the number distributions of reconstructed photons
per ring. Quantities of that kind depend also on characteristics of
the physics input like opening-angle distributions of pairs in the
sample and the polar distribution of electron tracks. They are
therefore best checked on the data itself. For that purpose a special
sample enriched in $\pi^0$-Dalitz decays was extracted from the data
by searching for pairs with opening angles less than
50~mrad~\cite{plb422}. Such pair sample has a typical
signal-to-background ratio of order one.

The properties of this sample compare well with the corresponding
properties of simulated $\pi^0$-Dalitz decays in ring-parameters and
efficiency.  There is but one exception: the measured ring centre
resolution of the first RICH detector is worse by a factor of 1.4
compared to the MC simulation, and also deviates from the expected
resolution for single photon hits.  For lack of understanding, we
introduce an ad hoc Gaussian random smearing of 0.9~mrad into the
simulation to meet the measured value.  The influence on final
momentum resolution is very weak, since the azimuthal track position
measured in the SiDC's is anyhow more accurate and outweighs the RICH
contribution to tracking.

\subsection{Determination of reconstruction efficiency}
\label{subsec:21}

To compare the number of reconstructed electron pairs to any
physics-based expectation, an absolute normalisation is required. To
correct spectra for reconstruction efficiency\footnote{For unit pair
efficiency, all pairs that have both electrons in the acceptance of
the spectrometer are reconstructed.}, measured yields are multiplied
by the inverse pair reconstruction efficiency averaged over the full
sample of pair candidates. We may call this a correction `on a
statistical basis', in contrast to an `event-by-event correction'
where the invariant-mass or $p^{ee}_t$ spectra are incremented with
the inverse efficiency of that event, or of particular track
candidates.

One way to determine the reconstruction efficiency is to measure the
number of pairs from a well-defined physics sample, and adjusting it
to the expected number.  We first explain this `Dalitz method' and
then turn to the alternative Overlay Monte-Carlo technique.

\subsubsection{The Dalitz method}

Our sample of `Dalitz pairs' with $m<$200~MeV/$c^2$, but opening
angles larger than 35~mrad and track $p_t$ above ~200~MeV/$c$ is
considered an excellent choice in place. It consists mainly of
$\pi^0$-Dalitz pairs and a contribution of $\eta$-Dalitz pairs on the
10$\%$ level. The opening-angle cut reduces conversions to (5-10)~$\%$
which can be corrected for.  At present, we cannot exclude the (very
interesting) possibility that a strong source of thermal electron
pairs may compete on the level of a few
percent~\cite{rapp-wambach2000}.

The required average inverse reconstruction efficiency is derived by
dividing the measured number ratio of char\-ged-particles to Dalitz pairs by
the cocktail expectation,\footnote{By lower case $n$ we denote numbers
from physics simulation, by upper case $N$ measured counts.}
\begin{equation}
\left\langle \frac{1}{\varepsilon} \right\rangle=
\left\langle \frac {N_{ch}}{N_{ee}}\right\rangle\left/
\frac{\langle n_{ch}\rangle}{\langle n_{ee}\rangle}\right..
\end{equation}
With the event generator GENESIS~\cite{genesis-old} the number of
electron pairs of the hadronic cocktail is calculated, but it is
normalised to the number of neutral pions; using the
$\pi^\circ$-to-charged-particle ratio, we obtain the cocktail
reference ratio in the required normalisation to charged particles,
\begin{equation}
\frac{\langle
n_{ee}\rangle}{\langle n_{ch}\rangle}=
\left(\frac{n_{\pi^\circ}}{n_{ch} }\right)\,
\frac{\langle n_{ee}\rangle} {\langle n_{\pi^\circ}\rangle}.
\end{equation}
Note that only the cocktail part factorises into averages of electron
pairs and charged particles\footnote{Since all components of the
cocktail scale linearly with $n_{ch}$, ratios are {\it constant}.} 
while the experimental average does not, since $N_{ch}/N_{ee}$ depends,
via reconstruction efficiency on $N_{ch}$.

The recipe that follows is to average the ratio of the number of
charged particles to the number of Dalitz pairs over all events of a
given multiplicity class, and divide by the cocktail ratio to obtain
the factor 1/$\varepsilon$ by which the raw data counts have to be
multiplied.  The discussion emphasises the importance to measure the
reconstruction efficiency as a function of $\langle N_{ch}\rangle$.

\subsubsection{The Overlay-Monte-Carlo method}

A method of wider application is to reconstruct a MC-simulated
sample of electron pairs from the sources under consideration, be it
Dalitz decays with masses below 200~MeV/$c^2$ as discussed below, or
the full {\it hadronic cocktail} of electron pairs, always under
condition of all acceptance and analysis cuts and using standard
analysis software. For realistic background conditions, the simulated
pairs are overlaid one by one on data events of given
\begin{table}[b!]                 
\renewcommand{\arraystretch}{1.35}
  \begin{center}                 
     \begin{tabular}{ | c | l | }  
     \hline                        
    \raisebox{2.5ex}[-1.5ex]{\bf }{\bf } &
     {Reconstruction efficiency}  \\
    \raisebox{.0ex}[-1.5ex]{\bf }{\bf } &
     {\ piecewise\ \ \ \ \ compound}  \\
      \hline
      SiDC1      & \      94 $\%$  \ \ \ \ \ \ \ \ \  \\[0.cm]
      \hline             
      SiDC2      & \     88 \%  \ \ \ \ \ \ \  \\[0.cm]           
      \hline             
      SiDC-tracks & $\Longrightarrow$ 83 $\%$  \ \ \ \ \ \ \ \ \ 84$\%$ \\[0.cm]
      \hline             
      RICH1  & \  81 \%  \ \ \ \ \ \ \  \\[0.cm]
      \hline
      RICH2  & \  86 \%  \ \ \ \ \ \ \  \\[0.cm]
      \hline
      MWPC   & \  96 \% \ \ \ \ \  \ \  \\[0.cm]
      \hline   
      full tracks &$\Longrightarrow$ 55 $\%$  \ \ \ \ \ \ \ \ \ 55$\%$ \\[0.cm]
      \hline   
      pairs  &     $\Longrightarrow$ 31 $\%$  \ \ \ \ \ \ \ \ \ 30$\%$ \\[0.cm]
      \hline             
      \end{tabular}
   \end{center}
\caption{ \label{Deteffi} 
Detector efficiencies separately and combined to full-track and pair
efficiencies (left column). Efficiencies of fully reconstructed tracks
and pairs (right column). Overlay-Monte-Carlo technique, 1996 data sample.}
\end{table}
$\langle N_{ch}\rangle$, and the desired correction factor is given by the
average of the inverse probability of successful reconstruction,
\begin{equation}
\left\langle \frac{1}{\varepsilon} \right\rangle=
\left\langle \frac{1}{p_{rec}} \right\rangle.
\end{equation}

The pair efficiencies can be decomposed into the products of the track
efficiencies of the detectors for a meaningful check on the
reliability of the simulation. Table~\ref{Deteffi} compares full-track
and pair efficiencies for a standard
Dalitz sample with those obtained by piecewise multiplication
of individual detector efficiencies; all efficiencies were measured by
the Overlay-Monte-Carlo technique. The close agreement between the two
indicates that the procedures of track reconstruction and pairing
work properly.

A further reduction in pair efficiency by about a factor of two occurs
in the course of the background rejection discussed in
sect.~\ref{subsec:23}; also these efficiency losses are understood
by the MC simulations.

The Overlay-MC method applied to high-mass electron {\it pairs}
suffers from the low efficiency combined with acceptance losses 
which forbid to collect sufficient statistics samples.  We have
therefore based all corrections on products of {\it track}
\begin{figure}[tt!]
  \resizebox{0.40\textwidth}{!}{%
    \includegraphics{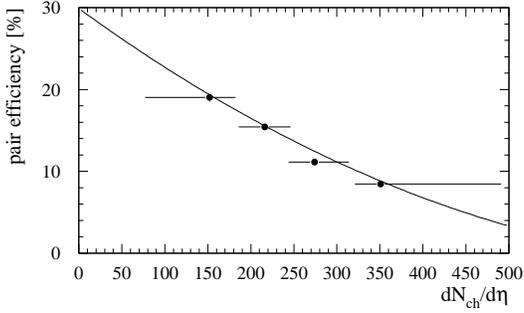}}    
    \caption{\label{fig:4.2} The multiplicity dependence of the
    efficiency to reconstruct low-mass Dalitz pairs under acceptance
    and analysis cuts. The solid line is from Overlay-MC simulation.
    The data points show the decline in efficiency with increasing
    particle density assuming that the Dalitz yields are proportional
    to it. The integral efficiency was normalised to the MC value.
    Statistical errors are comparable to or smaller than data circles,
    horizontal bars show binning of trigger distribution.  1996 data,
    Ref.~\cite{lenkeit-phd}.}
\end{figure}
efficiencies constructed in such a way that on the one hand the
correct $dN_{ch}/d\eta$ dependence of the Dalitz {\it pair} efficiency
as shown in Fig.\ref{fig:4.2} is reproduced, and on the other, that
the specific differences in pair efficiency are taken care of in an
approximate way. Those arise from the $\theta$-dependent hit density
and are treated separately.

\subsubsection{$N_{ch}$ dependent efficiency}

Reconstruction efficiencies for Dalitz pairs with mass below
200~MeV/$c^2$ are displayed in Fig.~\ref{fig:4.2} as a function of
charged-particle rapidity density. The pairs are filtered for opening
angles $\Theta_{ee}\geq$~35~mrad to reduce conversions, and they obey
the condition $p_t\geq$~200~MeV/$c$ on single-electron tracks.  For
the data points, the dependence on $N_{ch}$ is derived by postulating
that sources scale with $ \langle N_{ch}\rangle$ while the absolute
magnitude is fixed by normalising to the MC simulation as explained
above. We observe a loss in MC pair efficiency by a factor of 2.2 when
rapidity density is raised from 150 to 350. This degradation factor
for the three '96-data
analyses~\cite{socol-phd,lenkeit-phd,hering-phd} amounts to
2.30$\pm$0.13; the individual results lie remarkably close considering
the large spread in absolute efficiencies. The causes of the
efficiency loss are twofold.  For one, the recognition of hits in the
SiDC's and of Cherenkov rings in the RICHes is deteriorated by an
increasing number of background hits. In addition, with increasing
track density also the chances increase that rejection of background
tracks accidentally vetoes nearby signal tracks.

The strong decrease of pair efficiency with $\langle N_{ch}\rangle$
agrees very well with the Overlay-MC simulation also displayed in
Fig.~\ref{fig:4.2}. In the '96 data analyses, this curve is used for
efficiency correction on an event-by-event basis: signal pairs are
stored with weights given by the inverse of the efficiency at given
event multiplicity. The same procedure was applied in the '95 data
analysis, but the $\langle N_{ch}\rangle$ dependence is considerably
weaker; the ratio quoted above is only 1.4~\cite{voigt-phd}), possibly
because of a lower event background compared to 1996, but also due to
a different treatment of close hits in the SiDC's.

\subsubsection{$\theta$-dependent efficiency}

\begin{figure}[tt!]
\begin{center}
            \resizebox{0.51\textwidth}{!}{%
    \includegraphics{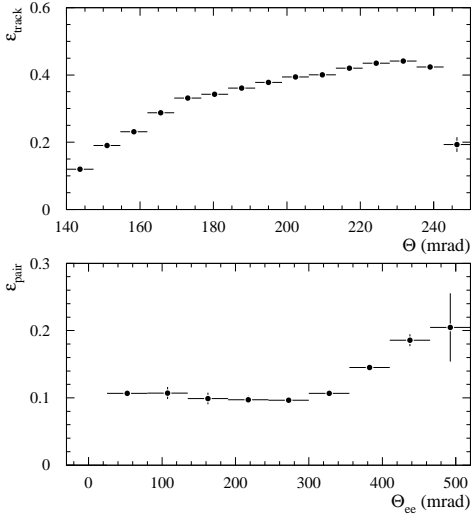}}
\end{center}    
\caption{\label{fig:4.3} Top:
  Track efficiency {\it vs} polar angle $\theta$ drops as track
  density increases towards small $\theta$.  Bottom: Pair
  reconstruction efficiency {\it vs} pair opening angle $\Theta_{ee}$
  rises towards large $\Theta_{ee}$ where tracks of accepted pairs are
  forced to large $\theta$. A shallow minimum occurs where tracks come
  closest to the minimum $\theta$. Overlay-Monte-Carlo technique on
  '96 data.}  
    \end{figure}

With track density increasing towards small polar angles, the track
reconstruction efficiency drops by more than a factor of three over
the acceptance, as shown in the upper panel of
Fig.~\ref{fig:4.3}. Since we will deal only with pair yields
integrated over the acceptance, one is inclined to disregard this
effect.

A $\theta$-dependent efficiency, however, combined with the limited
$\theta$ acceptance, affects pairs of different opening-angle
characteristics in different ways; and we have seen already in
Fig.~\ref{fig:3.1} that $\pi^\circ$-Dalitz pairs and pairs from
$\omega$ decays have indeed very different $\Theta_{ee}$ behaviour.

The acceptance condition affects wide open pairs considerably.  For
the given example of $\omega$ decays with typical
$\Theta_{ee}\approx$~400\,mrad, only pairs with both electrons at
large $\theta$ are accepted and reconstructed with above-average
efficiency. The rise in pair efficiency for very large opening angles
is seen in the lower panel of Fig.~\ref{fig:4.3}. The shallow minimum
at $\Theta_{ee}\approx$~250~mrad is typical for the majority of the
open pairs in our sample. We see also that the efficiency changes only
by very little towards lower $\Theta_{ee}$. The flat region even
includes our sample of Dalitz pairs; owed to the standard cut
$\Theta_{ee}\geq$~35~mrad, their most-probable polar angle is as large
as 190~mrad. So, compared to the 'majority of open pairs' cited above,
the reconstruction efficiency of the sample of Dalitz pairs is almost
undegraded. This is very satisfactory for the 'Dalitz method'
described above.

Yet, in order to protect characteristic features in spectra of pairs
with opening angles comparable to the acceptance of the spectrometer,
the efficiency correction should involve the $\theta$ distributions of
the contributing tracks. We describe this method in the following.

The $\theta$ dependence in track efficiency was taken care of in an
approximate way under the boundary condition to leave the measured
efficiency for Dalitz pairs shown in Fig.~\ref{fig:4.2} unchanged.
The pair efficiency is factorised into track efficiencies
which are modified according to
\begin{equation} 
\varepsilon^{~any\,pair}_{track}~(N_{ch}, \theta)=
\varepsilon^{~Dalitz}_{track}~(N_{ch})\cdot f~(\theta).
\end{equation} 
 The function $f(\theta)$ incorporates the measured
$\theta$ dependence of the track efficiency shown in
Fig.~\ref{fig:4.3}, upper part. 

The procedure is to increment in an event of multiplicity $N_{ch}$ 
for each pair candidate a weight into the like-sign or
unlike-sign spectrum which is
\begin{eqnarray}
 w(N_{ch};\theta_1,\theta_2)= \frac{1}{{\varepsilon^{any\,pair}_{\,track}\,(N_{ch},\theta_1)
~\varepsilon^{any\,pair}_{track}\,(N_{ch},\theta_2)}} &\approx &\nonumber \\
\frac{1}{\varepsilon^{~Dalitz}_{pair}(N_{ch})
\cdot f(\theta_1)\cdot f(\theta_2)}.~~~~~~~~~~
\end{eqnarray} 
The function $f(\theta)$ is suitably normalized to assure that the
Dalitz sample is unchanged. 

The described event-by-event correction was not performed in the '95
data analyses while the '96-analysis of Ref.~\cite{hering-phd} was
corrected for a truly two-dimensional efficiency.

\subsection{Pair acceptance}
\label{subsec:22}

The acceptance of the CERES spectrometer for {\it electron pairs} has
been evaluated by MC simulations assuming uniform input
distributions in mass $m$, pair transverse
\begin{figure}[t!]
          \resizebox{0.425\textwidth}{!}{%
    \includegraphics{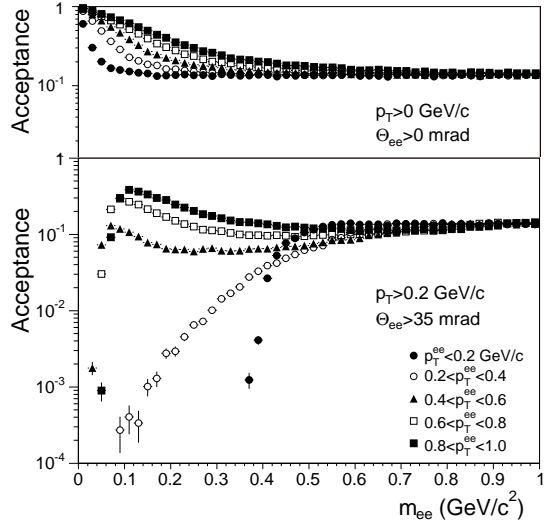}}    
    \caption{\label{fig:4.4} Acceptance of electron pairs under
    standard cut conditions (lower part) and without cuts (upper part)
    {\it vs} invariant mass, for five selected bins in transverse
    pair momentum $p_t^{ee}$.  }
\end{figure}
momentum $p_t^{ee}$, and rapidity $y$. How the CERES acceptance
depends on $m$ and $p_t^{ee}$ is shown in Fig.~\ref{fig:4.4} for the
standard rapidity range 2.1$\leq\eta\leq$2.65; the normalisation is
done cutting also the virtual photon distribution to this range in the
input. We have applied here the standard analysis cuts
$p_t\geq$\,200~MeV/$c$ and $\Theta_{ee}\geq$\,35\,mrad. The results
without cuts are shown for comparison. For masses below 100~MeV/$c^2$,
the acceptance without cuts approaches one, while the angular cut
causes a steep reduction. Higher up in mass, the single-electron $p_t$
cut reduces the acceptance at low pair $p_t^{ee}$, acting like a $m_t$
cut, while at $p_t^{ee}\geq$~400~MeV/$c$, the acceptance becomes more
uniform and even, within a factor of two, independent of mass, due to
some `equalising' effect of the opening-angle cut.  Still higher up,
for $m>$\,500\,MeV/$c^2$, the acceptance is essentially uniform both
in mass and in $p_t^{ee}$.

An acceptance correction for pairs of `new physics' sources is
problematic in principle, since the decay dynamics is not precisely
known and may depend on additional parameters, such as helicity
angle. In comparison of theories to CERES data, proper accounts
of the pair acceptance and other cuts are usually taken by the authors
themselves who have the necessary insight into the dynamics of the
modelled sources to do so.  We therefore {\it correct data only for
 reconstruction efficiency, not for  pair acceptance}.

\subsection{Optimising rejection of combinatorial background}
\label{subsec:23}

To foster confidence in the data analysis and to improve on data
quality it was essential to understand the combinatorial background in
every detail. We have presented in the preceding section how the
measured numbers of background and signal pairs evolve with the
rejection steps applied. Here, we use the tool of full MC simulation
to arrive at a quantitative understanding of the background
sources. It goes without saying that the elaborate simulation of
background rejection supplied a very effective handle to fine-tune
rejection cuts.

\begin{figure}[t!]
\begin{center}
 \resizebox{0.42\textwidth}{!}{%
 \includegraphics{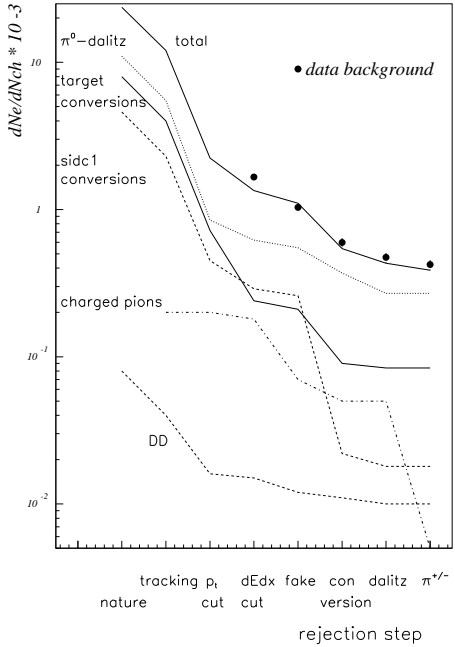}} 
\end{center}
\caption{\label{fig:4.5} MC simulation of number of electron tracks per 
$N_{ch}$ from various physics sources, along sequence of rejection
steps and compared to measured background (full circles, 1996 data).
Individual steps are 1: status after production (mild
rejection, generous quality cuts), 2: tracking, 3: $p_t$ cut, 4: dE/dx
cut, 5: fake rejection, 6: conversion rejection, 7: Dalitz
rejection, 8: $\pi^\pm$ rejection. }
\end{figure}

Figure \ref{fig:4.5} plots the number of electron {\it tracks} from
various physics sources which survive a given rejection step in the MC
simulation.  The decrease is
either due to rejection of single tracks that have unwanted properties
like low $p_t$, bad matching or suspicious environment, or due to
removal of recognised conversion or Dalitz pairs which are still
intact and fulfil all conditions.

The experimental points are numbers of background tracks from the
post-production steps of the 1996 data analysis. They are derived from
the measured total number of background {\it pairs} accumulated in
42.2$\,10^6$ events, assuming Poissonian track statistics. The
agreement between absolute numbers from experiment and simulation is
quite reassuring; also the size in relative suppression is in good
agreement with the simulation. The residual  background remaining 
after all eight  rejection steps is dominated by Dalitz pairs.
 
\section{Hadronic decay sources}
\label{sec:5}

The `conventional' sources contributing to the inclusive mass spectra
of electron pairs in the mass range below 1.5~GeV/$c^2$ are free
decays of light neutral mesons up to and including the $\phi$.  These
contributions have been determined for p-Be and p-Au collisions with
considerable precision in an experiment which combined the
electron-pair spectrometer of CERES with the photon calorimeter of
TAPS~\cite{neutral-meson-pBe}.  In addition to the inclusive
electron-pair yield, the cross sections and $p_t$ distributions of the
$ \pi^\circ, \eta, ~and ~\omega$ mesons have been measured via the
electro-magnetic decay modes $\gamma\gamma$ and $\pi^\circ\gamma$, and
an exclusive reconstruction of the $\pi^\circ$ and $\eta$-Dalitz
decays ${\pi^\circ,\eta\rightarrow e^+e^-\gamma}$. These data,
supplemented by older measurements from NA27~\cite{na27}, allowed us to
simulate the electron-pair mass spectra originating from decays of
neutral mesons. The result was subject to all experimental cuts and
folded with mass resolution.  Within error limits it was concluded
that the inclusive mass spectra of electron-pairs measured in
450~GeV/c p-Be and p-Au collisions are consistently described by the
expected $e^+e^-$ pair `cocktail' of neutral meson
decays~\cite{neutral-meson-pBe}.

The {\it hadronic cocktail} serves as our reference standard of
hadronic sources to expose effects which are specific to
nucleus-nucleus collisions, i.e. spectral shapes and yields which
cannot be described by a mere superposition of nucleon-nucleon
interactions.  Supported by evidence from light collision
systems~\cite{becattini96}, particle ratios were originally assumed
constant from p-p to heavy systems~\cite{drees}, and particle yields
were scaled with charge multiplicity.  This cocktail was instrumental
in gauging the low-mass excess in 200~GeV/$c$ S-Au
collisions~\cite{prl1995} and in 158~GeV/n Pb-Au collisions from the
first (1995) data set~\cite{ravinovich-qm97,plb422}.

The hadronic cocktail is calculated with the Monte Carlo event
generator GENESIS~\cite{genesis-old} described in considerable detail
elsewhere~\cite{neutral-meson-pBe}. It has been improved since then
but the changes with respect to previous CERES analyses of
Refs.~\cite{ravinovich-qm97,plb422,lenkeit-qm99,lenkeit-paris} are
subtle. These publications are superseded by the present paper.

We sketch here only the main points. Mesons are produced with cross
sections scaled up from p-p or p-A, taking into account the modified
$p_t$ and rapidity distributions. Open charm contributions have
been neglected on the basis of reliable
estimates~\cite{open-charm98}. Mesons are then allowed to decay with
known branching ratios. Decay kinematics are simulated for the Dalitz
decays $PS\rightarrow e^+e^-\gamma$ \ of the pseudo-scalar mesons
\begin{table*}[t]
\begin{center}  
     \begin{tabular}{ | l | c | c | c | c |}  
         \hline
         {\bf decay} &  {\boldmath $\sigma_{tot} /
         \sigma_{\pi^0, tot}$ } & {\bf branching-} & {\bf $T_\circ$} [MeV]   &
         {\bf Form factor} \\ 
         {} & {} &  {\bf ratio} &{} & \\
         \hline
         $\pi^0 \rightarrow e^+e^- \gamma$ & 1.
         & ($1.198\pm0.032) \times 10^{-2}$ & 100 /         230 & pole approximation \\ 
         {} & {} & {} & {} & {$b$=5.5~GeV$^{-2}$} \\         
         \hline
         $\eta \rightarrow e^+e^- \gamma$ & 0.085 & ($6.0\pm0.8) \times 10^{-3}$
         & 238 &         pole approximation \\ 
         {} & {} & {} & {} & {$b$=(1.9$\pm$ 0.4)~GeV$^{-2}$} \\    
         \hline
         $\rho \rightarrow e^+e^-$ &  0.094 &($4.67\pm 0.09) \times 10^{-5}$
         & 263 & \\
         \hline
         $\omega \rightarrow e^+e^-$ &  0.069 &($7.14 \pm 0.13) \times 10^{-5}$
         & 265 & \\ 
         $\omega \rightarrow \pi^0 e^+e^-$ & &($5.9\pm 1.9) \times 10^{-4}$
         & & Breit-Wigner \\
         {} & {} & {} & {} & {$m_{\rho}$=0.65, $\Gamma_{\rho}$=\,0.05~GeV} \\   
         \hline
         $\phi \rightarrow e^+e^-$ &  0.018
         &($2.98\pm 0.04) \times 10^{-4}$ & 292 & \\
         \hline
         $\eta^\prime \rightarrow e^+e^- \gamma$ &  0.0078 &$\approx 5.6\times 10^{-4} $
         &         285 & Breit-Wigner \\
         {} & {} & {} & {} & {$m_{\rho}$=0.76, $\Gamma_{\rho}$=\,0.01\,GeV} \\ 
         \hline
         \end{tabular}
\caption{ \label{tab:decays} Summary of the parameters used in the generator
 to simulate
         the contributions from hadron decays.}
\end{center}
\end{table*}
$\pi^\circ,\eta~and~\eta^\prime$, for the direct decays of the light
vector mesons, $V\rightarrow e^+e^-$, and for the $\omega$ Dalitz
decay ${\omega\rightarrow e^+e^-\pi^\circ}$.  Electron momenta are
Lorentz-transformed into the laboratory system and convoluted with the
experimental resolution profile.  The simulated events are subject to
the same filters concerning acceptance, $p_t$ and opening angle.

Production cross sections of the neutral mesons, their rapidity and
transverse-momentum distributions are essential ingredients for
simulating the decay contributions to the dilepton spectrum.  When
data on hadron production from Pb-beam experiments at the SPS (NA44,
NA49, NA50, WA98) became available in time for the 1996 data analysis,
we used this input from Pb-beam data whenever
possible~\cite{lenkeit-qm99,lenkeit-paris,lenkeit-phd}. Information
not directly available was derived from the statistical model which
describes ratios of integrated hadron yields at chemical freeze-out
very well with only two fit parameters, the temperature and the baryon
chemical potential; the particular values used are $T$= 170~MeV and
$\mu_b=$~68~MeV~\cite{chem-eq99}. The collision system is modelled in
a state of collective transverse expansion which is based on the
observation~\cite{wessels96} that inverse-slope parameters $T$ of
transverse momentum spectra, except for pions, systematically increase
with mass~\cite{stachel-paris98}.

Data on $p_t$ distributions of pions have been measured by
CERES~\cite{bielcikova,ceretto-tsukuba}, NA49~\cite{appels-np98},
NA44~\cite{kaneta-99}, and WA98~\cite{aggarwal-98-01}.  They are
exceptional in that they can not be described by a single exponential.
The $\pi^0$ spectrum is generated with two slopes: at $m_t
\leq$~200~MeV where it is dominated by secondaries, with $T_\circ$=~100~MeV,
and for the higher part with an inverse slope of $T_\circ$=~230~MeV.  The
inclusive $m_t$ distribution for neutral pions measured by WA98 is
extrapolated to small $m_t$ using the charged pion distributions from
NA44.  At small $m_t$ the spectra are dominated by secondary decays
from heavier mesons. The decay $\eta\rightarrow 3\pi^\circ$ is added
separately by hand to the $\pi^0$ distribution.

The rapidity distribution~\cite{sikler99} of negatively charged
hadrons, described as a Gaussian centred at $y_{max}$=~2.9 with
$\sigma_y$=~1.5, has been adopted for all mesons.  While the widths of
the hadron rapidity distributions decrease with particle mass in
proton induced collisions, this is not observed in lead-induced
collisions~\cite{hoehne99}.\footnote{Particle ratios taken at
mid-rapidity are therefore the same as those from partially or fully
integrated yields.}

The parameters from Pb-beam data and the statistical model are given
in Table~\ref{tab:decays}. Meson production ratios are implemented
relative to the number of $\pi^\circ$'s and include feeding from
heavier resonances. Compared to the earlier reference to proton
induced collisions~\cite{neutral-meson-pBe}, the production ratios of
heavier hadrons are enhanced as, e.g. the $\eta$/$\pi^0$
\cite{peitzmann98} and the  ~$2\phi/(\pi^+ +\pi^-)$ ratio
\cite{jouan-falco98,puehlhofer98}.
The meson yields are normalised to the charged-particle density
by the ratio
\begin{equation}
\langle{\,N_{\pi^\circ}\,}\rangle/\langle{\,N_{ch}}\,\rangle=~0.44.
\end{equation}
Brackets denote averaging over the CERES acceptance.

The decay branching ratios given in Table~\ref{tab:decays} are from
Ref.~\cite{pdg2004}.  To simulate the Dalitz decays, the
Kroll-Wada expression~\cite{kroll-wada} is multiplied by the
electro-magnetic transition form factors fitted to the LEPTON-G
data~\cite{lepton-g}.  The pole approximation $F(M^2)= (1 -
b\,M^2)^{-1}$ is used for the determination of the form factors of the
$\pi^\circ$ and the $\eta$ Dalitz decays~\cite{lepton-g}. For the
$\omega$ and the $\eta^\prime$, the form factors are determined by
fitting a Breit-Wigner function
\begin{equation} 
|F(M^2)|^2= \frac{m_{\rho}^4}{(M^2 - m_{\rho}^2)^2 +
m_\rho^2\Gamma_\rho^2} 
\end{equation} 
 to describe the resonant behaviour according to the vector dominance
 model.  The parameters used are listed in Table~\ref{tab:decays}. The
 direct decays of the vector-meson were generated following Gunaris and
 Sakurai~\cite{gounaris-sakurai}.

The 2-body decay of the $\rho$ meson has been
re-evalua\-ted~\cite{frimann-knoll00} for the new GENESIS
code~\cite{genesis-new}.  We defer details of the revised formula as
it is implemented in the 2003 version of the New GENESIS to the
appendix.  The resulting mass distribution, due to a Boltzmann-type
phase space factor $e^{-M/T}$ and a momentum dependent phase space,
both omitted in the previous code, receives a shoulder on the
low-mass side and a steeper falloff to higher masses.

All decays where assumed isotropic in the rest frame of the decaying
meson except for the Dalitz decays to e$^+$e$^-$$\gamma$ which follow a
1 + cos$^2$($\theta$) distribution, where $\theta$ is measured with
respect to the virtual photon direction.

\begin{figure}[t!]
\begin{center}
  \resizebox{0.5\textwidth}{!}{%
    \includegraphics{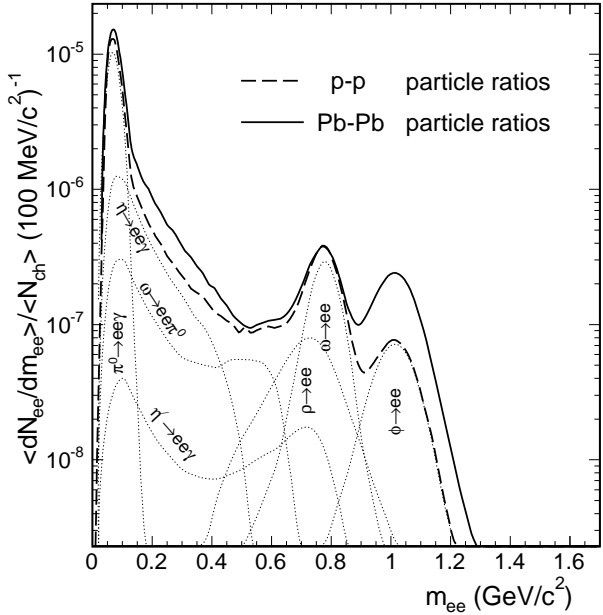}}    
\end{center}
 \caption{\label{fig:5.1}
    New GENESIS hadronic cocktail based on Pb-beam data and the
    thermal model compared to GENESIS based on p-p
    collisions.  Shown are mass spectra from simulated neutral mesons
    decays with acceptance, $p_t$ and opening-angle cuts and after
    folding with 1996 mass resolution.}
\end{figure}

Figure \ref{fig:5.1} depicts the cocktail based on the Pb-beam data
with all corrections. It is the mass spectrum of pairs in the CERES
acceptance $2.1\leq \eta\leq 2.65$ with standard cuts
$\Theta_{ee}\geq$~35~mrad and $p_t\geq$~200~MeV/$c$ . The generator
output has been folded with the mass resolution function of the 1996
data set, which includes both the momentum resolution as the dominant
source of smearing as in eqn.~(3.9), as well as the resolution in pair
opening angle; the latter is approximately
\begin{equation} 
\sigma^2_{\Theta_{ee}}\approx (\sqrt{2}\sigma_{\theta})^2 +
\overline{sin^2\theta}(\sqrt2\cdot\sigma_{\phi})^2,
\end{equation} 
where $\sigma_{\theta}\approx$~0.6~mrad and
$\sigma_{\phi}\approx$~3.0~mrad are the angular track resolutions
taking information from all detectors together. The effects of
bremsstrahlung emission by electrons traversing the detector material
have been included in the simulations, but due to the low material
budget in the acceptance of $X/X_{\circ}\approx$~1\,$\%$, they are
hardly noticeable. The overall mass resolution is about 6$\% $ in the
$\rho/\omega$ region and 7$\% $ in the region of the $\phi$ (see 
sect.\,3.3).

The cocktail for Pb-Au collisions is compared in Fig.~\ref{fig:5.1} to
the cocktail of hadronic sources up-scaled from p-p interactions; the
latter is also folded with the CERES resolution obtained for the '96
data analysis.

The systematic errors of the normalised decay cocktail, relevant for
the numerical comparison to the experimental pair data in sect.\,7,
are discussed separately for the mass regions below and above
200\,MeV/$c^2$.

The low-mass region is dominated by the $\pi^\circ$-Dalitz decay, with
$\eta$-Dalitz contributing about 15\,$\%$.  Errors arise from the
relative production cross section of the $\pi^\circ$ and from the
parametrisations of the input rapidity and transverse momentum
distributions.  Normalising the decay cocktail to the number
of charged particles is a very powerful means to keep the error in the ratio
$\left(n_{\pi^\circ}/n_{ch} \right)$ of eqn.~(4.26) small; it is
estimated to be at most 5$\%$. The rapidity distribution is uncritical
since it is taken directly from pion data; the errors should be less
than 3\,$\%$. The transverse momentum distribution, due to the
single-electron $p_t$ cut, is a little more critical. However, quite
different assumptions on the shape of the $p_t$ spectrum above the
cut, e.g., using $h^-$ from CERES rather than $\pi^-$ from NA49, lead
to differences in yield of only very few percent. This error is
therefore also estimated to be below 5$\%$. Assuming that the
systematic errors are uncorrelated, a total systematic error of 8$\%$
is obtained for the low-mass region.

In the high-mass region, the errors are dominated by those of the
relative production cross sections of the higher-mass mesons and of
the detailed properties of the electro-magnetic decays. Since most of
the particle yields relevant for the cocktail have not directly been
measured, or like the $\phi$, suffer from experimental
controversy~\cite{jouan-falco98,puehlhofer98}, the statistical model
predictions have been used instead, and it is therefore their
uncertainties which enter. Judging the average fit quality to measured
particle ratios~\cite{chem-eq99}, we estimate these errors to be
20$\%$. The uncertainties in the branching ratios and in the
transition form factors have been discussed in detail in
Ref.~\cite{neutral-meson-pBe}. They contribute $\approx$\,15$\%$ for
$m<$\,450\,MeV/$c^2$, $\approx$\,30$\%$ in the mass range of
450-750\,MeV/$c^2$, and 6$\%$ for $m>$\,750\,MeV/$c^2$.  Taking these
supposedly independent sources of uncertainty together, an overall
systematic error of 30$\%$ is estimated as the weighted average for
the high-mass region.

The integral number of electron pairs per charged particle in the
CERES acceptance expected from hadronic sources under standard cut
conditions $p^e_t \geq$~200~MeV/$c$ and $\Theta_{ee}\geq$~35~mrad is
\begin{eqnarray}
\left[\frac{\langle dN_{ee}/dm\rangle}{\langle N_{ch}\rangle}\right]_{Genesis}
& = & \nonumber \\
\left\{ \begin{array}{ll}
         (9.27\pm 0.74)~10^{-6} & ~~~m<~200~{\rm MeV}/c^2 \\
         (2.27\pm 0.7)~10^{-6} & ~~~m\geq~200~{\rm MeV}/c^2, 
         \end{array}
      \right.
\end{eqnarray}
where we have quoted the systematic errors of 8$\%$ and 30$\%$ for the
two mass regions, respectively.
\section{Results}
\label{sec:6}
\subsection{Samples of reconstructed pairs}
\label{subsec:24}

For an overview, the pair samples reconstructed from the two data sets
of 158\,GeV/n Pb-Au collisions taken by the CERES Collaboration in
1995 and 1996 are listed in Table~\ref{pair-samples}.  All analyses
apply standard cuts on track $p_t$ and and pair opening angle
$\Theta_{ee}$.

The table lists the pair signal (S) obtained by subtracting the
combinatorial background (B) from the measured numbers of unlike-sign
pairs, for the mass ranges below (Dalitz pairs) and above
200~MeV/$c^2$ (open pairs).  In this Table, pair yields are not
corrected for reconstruction efficiency (other than in the following
figures). In analysis no.\,5, event mixing was employed to obtain
combinatorial background of unlike-sign pairs. Quoted errors are
absolute statistical errors in the respective sample numbers amounting
to $\sigma_S=\sqrt{\,N_{+-}\,+\,B}$. The pair reconstruction
efficiencies are given in the last column.
\begin{table*}[tt!!]
\begin{center}  
     \begin{tabular}{ | l | c | l | l | r | l | l |}  
\hline
\raisebox{0pt}[13pt][7pt]{Work} &
\raisebox{0pt}[13pt][7pt]{Events} &
\raisebox{0pt}[13pt][7pt]{Dalitz} &
\raisebox{0pt}[13pt][7pt]{S/B} &
\raisebox{0pt}[13pt][7pt]{Open} &
\raisebox{0pt}[13pt][7pt]{S/B} &
\raisebox{0pt}[13pt][7pt]{Open/Dalitz} \\
{} &{} &{Pairs} &{ } &{Pairs} &{ } &{Pair Efficiency} \\
\raisebox{0pt}[13pt][7pt]{Year of run} &
\raisebox{0pt}[13pt][7pt]{${\rm \langle dN_{ch}/d\eta\rangle}$ } &
\raisebox{0pt}[13pt][7pt]{} &
\raisebox{0pt}[13pt][7pt]{ } &
\raisebox{0pt}[13pt][7pt]{} &
\raisebox{0pt}[13pt][7pt]{ } &
\raisebox{0pt}[13pt][7pt]{} \\
         \hline
         \hline
       1.~~~1995a &8.55$\cdot10^6$&  1038$\pm$55    &   1.05     &
        648$\pm$105  &       1/8     &      10.9/11.0\,$\%$  \\
       Refs.~49,54 &  {220}           &    {}           &   {}       &
                { }  &    {}         &          {}    \\
         \hline
       2.~~~1995b &8.55$\cdot10^6$&  1044$\pm$53    &   1.0     &
        468$\pm$104  &       1/11     &     ~8.0/10.9\,$\% $  \\
       Ref.~55  &  {220}           &    {}           &   {}       &
                { }  &    {}         &          {}    \\
         \hline
       3.~~~1996a &42.0$\cdot10^6$& 5631$\pm$129    &   1.03     &
       2018$\pm$237  &       1/13.4     &      ~8.8/11.9\,$\%$  \\
       Refs.~50,51,56 &  {250}           &    {}           &   {}       &
                { }  &    {}         &          {}     \\
         \hline
       4.~~~1996b &42.0$\cdot10^6$& 5013$\pm$128    &   0.88     &
       1722$\pm$236  &     1/15.7     &      ~9.5/10.8\,$\%$  \\
       Ref.~55  &  {250}           &    {}           &   {}       &
                { }  &    {}         &          {}     \\
         \hline
       5.~~~1996c &41$\cdot10^6$& 3537$\pm$103    &   1.04     &
       1305$\pm$194  &     1/13.9     &      ~4.3/~5.6\,$\%$    \\
       Ref.~57  &  {250}           &    {}           &   {}       &
                { }  &    {}         &          {}     \\
         \hline
         \end{tabular}
\caption{ \label{pair-samples} 
Analysis results of the 158~GeV/n Pb+Au data sets of 1995 and 1996
with total number of events analysed and mean charged density of
trigger.  Listed are the signals $S\pm \sigma_S$ of reconstructed
`Dalitz' and `Open' pairs of mass below and above 200~MeV/$c^2$,
respectively, after background subtraction.  Signal-to-background
ratios S/B are quoted. Pair reconstruction efficiencies $\langle
1/\epsilon\rangle^{-1}$ from open-pair and Dalitz-pair analyses are
given in the last column. Yields are not efficiency corrected.
} 
\end{center}
\end{table*}

We note a considerable spread among the results of different analysis
efforts. This originates from the different values of the rejection
cuts used along the various analysis chains. However, {\it normalised
pair yields}, i.e. numbers of pairs per event, per charged particle,
and corrected for pair efficiency, as will be shown below, are very
stable: the relative spread in the number of Dalitz pairs turns out to
be less than 15$\%$.

\subsection{Inclusive mass spectra}
\label{subsec:25}

\begin{figure}[t!]
  \resizebox{0.5\textwidth}{!}{%
  \includegraphics{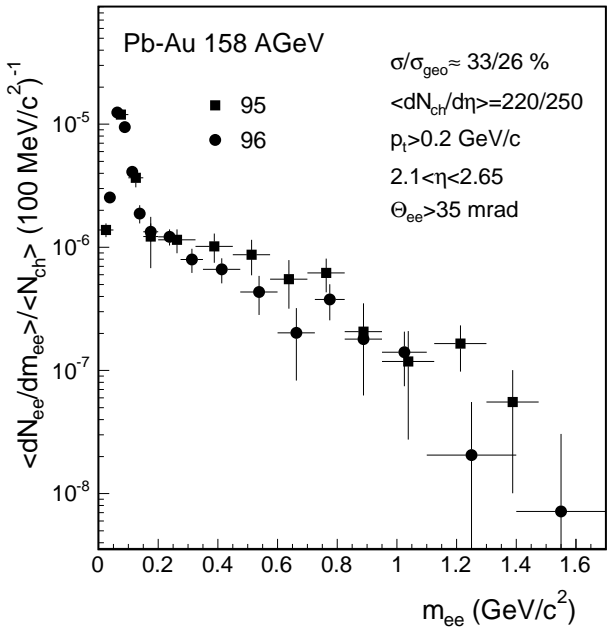}}
  \caption{\label{fig:6.1}Inclusive e$^+$e$^-$ mass spectra from '95
  and '96 data analyses. The pair yield is 
  efficiency corrected and normalised to the number of charged
  particles in the acceptance. Standard analysis cuts on track $p_t$
  and pair opening angle have been applied. Data correspond to the top
  33$\%$ and 26$\%$ of $\sigma_{inel}$ for the '95 and '96 data sets,
  respectively.  Vertical bars give statistically independent
  bin-to-bin errors.}
\end{figure}

Figure \ref{fig:6.1} shows the mass spectrum of
Ref.~\cite{plb422,voigt-phd} from 1995 together with that of
Ref.~\cite{lenkeit-qm99,lenkeit-paris,lenkeit-phd} from the 1996 data
set. The trigger centralities correspond to the most central 33$\%$ and
26$\%$ of the inelastic cross section for 1995 and 1996, respectively.
\noindent
The signal is obtained by subtracting the smoothed like-sign pair
background from the spectrum of unlike-sign pairs. The differential
pair yield $\langle dN_{ee}/dm_{ee}\rangle$ per event is normalised to
the mean number $\langle N_{ch}\rangle$ of charged particles in the
acceptance.  The brackets indicate particle yields per event measured
within the CERES acceptance. Reconstruction efficiency has been
corrected event-by-event by weighting pairs with the value of the
inverse pair efficiency at the particular centrality. Shown are the
bin-to-bin statistical errors of eqn.~(3.18). To include the
normalization errors discussed in sect.~\ref{subsec:18}, as it may be
relevant when individual data points are compared to other data, as in
this figure, or to theoretical-model predictions, the errors should be
multiplied by 1.4. Note, however, that these larger errors are no
longer statistically independent.

It is apparent that the '95 data points lie systematically higher than
the '96 data points. Apart from the small difference in mean trigger
centrality there is no relevant change in setup.  The '95
analysis~\cite{voigt-phd,ravinovich-qm97,plb422} followed the strategy
to optimise the quantity $S_{eff}\approx\varepsilon^2/B$. But while
the signal efficiency $\varepsilon$ was determined by Monte-Carlo, the
background $B$ was taken as the measured like-sign sample; this
strategy was discarded when the full MC simulation became available
for the '96 analysis. Although we could find absolutely no indication
that statistical fluctuations in $B$ might have steered the analysis
towards a `better' final sample, such possibility cannot be strictly
excluded.

The re-analysis of the '95 data~\cite{socol-phd} resulted in a mass
spectrum closer in absolute yield and shape to the '96 mass spectrum
shown in Fig.~\ref{fig:6.1}. In this analysis, any involuntary bias
was avoided by sampling the distributions in a random automatic
variation of all cut settings simultaneously, and then choosing the
centres of gravity.  However, seeing no direct evidence for a biased
tuning of cuts, we have chosen to keep the original '95 analysis to
expose our actual systematic uncertainties. We return to this issue
below.

\subsection{Centrality dependence}
\label{subsec:26}

\begin{figure}[t!]
   \resizebox{0.35\textwidth}{!}{%
    \includegraphics{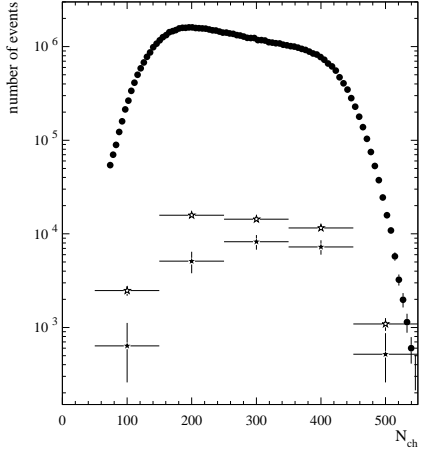}}       
    \caption{\label{fig:6.2} The $dN_{ch}/d\eta$ distribution of the trigger
    (full circles), and of events containing a Dalitz (open circles)
    or an open pair candidate (filled squares). Centres of gravity 
    are marked by
    the respective symbols in a box at the bottom.}
\end{figure}

The efficiency-corrected yield of Dalitz pairs is a solid reference
for linear $N_{ch}$ dependence. We have a first look at the centrality
dependence of the open pair yield by comparing in Fig.~\ref{fig:6.2}
the centres of gravity of three $N_{ch}$ distributions: that of
triggered events without any further condition, and two others taken
from events which contain a Dalitz or an open pair candidate,
respectively.  Because of the large open-pair background, the signal
is obtained by subtracting the $N_{ch}$ distribution of like-sign-pair
events from that of unlike-sign-pair events.  The $N_{ch}$-dependent
pair detection efficiency has been corrected for using the curve of
Fig.~\ref{fig:4.2}.

We see from Fig.~\ref{fig:6.2} that the centres of gravity of Dalitz
and open-pair samples are shifted progressively upward; the mean
multiplicities are 285 and 310, respectively, compared to
$\langle\,N_{ch}\rangle$=\,250 of the trigger distribution. The
averages of the trigger distribution {\it calculated} for events
depending {\it linearly} or {\it quadratically} on $N_{ch}$ are 287
and 315, respectively.  We conclude that the increase of open-pair
production is better described by a quadratic dependence on
$N_{ch}$, than by a linear dependence.  We will have a closer look at
the differential centrality dependence in sect.~\ref{sec:7} but note
that the current finding bears no reference to the hadronic cocktail.

\subsection{Invariant transverse-momentum spectra}
\label{subsec:27}

\begin{figure}[t!]
   \resizebox{0.48\textwidth}{!}{%
    \includegraphics{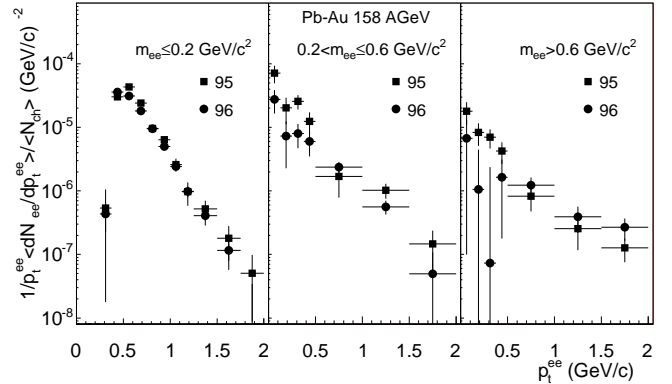}}    
\caption{ \label{fig:6.3}
Invariant pair transverse momentum spectra for three mass ranges and
both data sets ('95 squares, '96 circles). Statistical errors
only. Trigger centrality 33$\%$ and 26$\%$ for the '95 and '96 data
sets, respectively. }
\end{figure}

\noindent
Considerable physics potential resides in the spectra of
invariant transverse pair momentum, $p_t^{ee}$, as displayed in
Fig.~\ref{fig:6.3} for the two data samples and three mass bins.
The spectra are normalised and corrected for reconstruction
efficiency. 
\begin{figure}[b!]
  \resizebox{0.35\textwidth}{!}{%
    \includegraphics{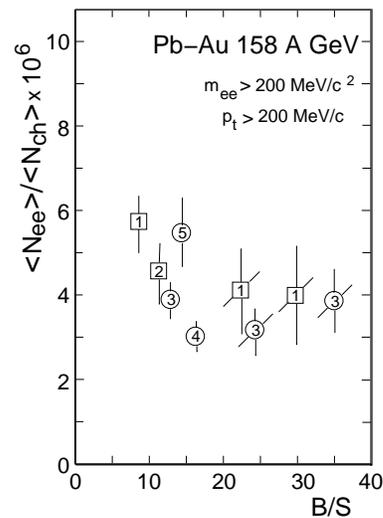}}    
    \caption{\label{fig:6.4} Results of all CERES electron-pair
    analyses for 158~GeV/n Pb+Au.  Plotted are normalised pair yields
    integrated from 200~MeV/$c^2$ until 1.1~GeV/$c^2$ {\it vs}
    background-to-signal ratio. Errors are statistical.  Slashed 
entries refer to
    relaxed rejection or quality cuts in otherwise identical
    analyses. Enclosed numbers refer to Table~\ref{pair-samples}.}
\end{figure}
In the Dalitz region, at low $p_t^{ee}$ the spectrum is void due to
the analysis cut; both samples agree very well.  Data points scatter
considerably for the two other mass bins. The full physics relevance
of these spectra emerges only in comparison to the expectations for
hadronic sources.

\subsection{Stability of results}
\label{subsec:28}

The optimisation of quality and rejection cuts was done with no
feedback from the signal itself not to exploit statistical
fluctuations likely in a sample that small. Still, at a
signal-to-background level of order 1/10 one is concerned
\begin{table*}[b!]
\begin{center} 
     \begin{tabular}{| l | l | l |}  
         \hline
       \ \  {Source of systematic error}         &\  {Comment}
     & {Relat. error}            \\
         {}                        & {}     &{on ${\rm N_{ee}/N_{ch}}$} \\
         \hline                 
                 \hline
{I.}         & {}              & {}      \\
{   MC efficiency correction}         & {}              & {}      \\
{   SiDC's and Pad Chamber }         & {3.0~$\%$ per hit on track}
     & { $\pm$ ~\,7.3~$\%$}            \\        
{   RICH's }   & {5.0~$\%$ per ring on track}      & { $\pm$~10.0~$\%$}    
   \\
                 \hline         
{Sum detector efficiencies } & {}          &\  {$\pm$~12.4~$\%$}  \\
                 \hline
                 \hline
{II.}         & {}              & {}      \\
{ Efficiency of analysis cuts }         & {}              & {}      \\
{ \ \ 11 matching cuts} & {1~$\%$ per cut on track } & { $\pm$\ ~4.7~$\%$} 
\\
{ \ \ 2 quality cuts} & {2~$\%$ per cut on track } & { $\pm$\ ~4.0~$\%$}  
\\
{ \ \ 7 rejection cuts}  & {1~$\%$ per cut on track } & { $\pm$\ ~3.7~$\%$}
\\
\hline          
{Sum analysis cuts} & {}          &\ {$\pm$~~7.2~$\%$} \\ 
         \hline
         \hline 
{III.}         & {}              & {}      \\
{ Triggered mean ${\rm N_{ch}}$}  & {comparison to other} & 
{ $\pm$~10.0~$\%$}    \\
{}          & {SPS experiments}              & {}              \\
{ No of events analysed}  & {}  & { $\pm$~~\,3.0~$\%$}            \\
         \hline 
         \hline
         {Total}                & {}    &\  {$\pm$~17.7~$\%$}      \\
         \hline
         \hline
         \end{tabular}
\end{center}
\caption{ \label{syst-uncorr} Estimates of systematic errors in normalised
 yields of Dalitz pairs. The uncertainties in
 efficiency for the two tracks of the pair are assumed to be {\it
 un}correlated and are added in quadrature like the contributions from
 different cuts or detectors. }
\end{table*}
how reliable and stable a signal can be that results from subtracting
two almost equal large numbers.  Even by careful inspection of raw
mass spectra by shape, or by magnitude, it would be quite hard to
distinguish above 200~MeV/$c^2$ the signal from background.

Because of cuts and kinematics, it so happens that the com\-binatorial
back\-ground spectrum peaks around $m\approx 2\,p^{cut}_t/c$=
400~MeV/$c^2$ and is surprisingly similar in shape to the low-mass
enhancement.

This raises the question, whether combinatorial background has been
insufficiently subtracted.  Aside from careful studies of analysis
cuts, rejection cuts, the determination of the reconstruction
efficiency as a function of centrality, we like to address this
question in a quantitative way.  A most likely cause leading to wrong
subtraction of background is an undetected asymmetry in reconstruction
efficiencies, i.e. for like-sign as compared to unlike-sign pairs. It
is shown in Appendix C that a 5$\%$ asymmetry is required to fake an
apparent enhancement factor of three over the hadronic sources,
assuming the S/B= 1/13 situation of the '96 analysis. However, this
asymmetry was measured to be less than 1$\%$, with confidence limit of
90$\%$~\cite{lenkeit-phd}, as reported in sect.~\ref{subsec:17}.

If the low-mass enhancement is faked by leakage of some amount of
combinatorial background into the spectrum of signal pairs, the
enhancement should increase with the amount
added. Figure \ref{fig:6.4}, summarising all CERES pair analyses for
158~GeV/n Pb-Au, does not show that. Rather, we see a pair signal
which is, in view of the errors, surprisingly stable despite very
large variations in background level.
 
\subsection{Systematic errors}
\label{subsec:29}

Estimates of systematic errors in the
mass-integrated, normalised yield of `Dalitz' 
pairs\footnote{Note that the use of calligraphic $\cal{N}$ for
normalised pair yields is restricted to this section.},
\begin{equation} 
{\cal N}(\rm `Dalitz')= 
\int_{0}^{0.2}dm~\langle dN_{ee}(m)/dm\rangle/\langle N_{ch}\rangle,
\end{equation} 
is given in Table~\ref{syst-uncorr}.  The definition of the `Dalitz'
sample includes the opening angle cut of 35~mrad and the $p_t$ cut at
200~MeV/$c$. The largest contributions of about 10$\%$ each arise from
detector efficiencies and the triggered centrality $\langle
N_{ch}\rangle$, while the uncertainties due to matching and rejection
cuts account for about 7$\%$. The resulting total of
18$\%$\footnote{If errors from both tracks would
be correlated, they would add to a total of 24$\%$ instead.} is larger than the rms deviation of Dalitz pair
yields from all our analyses which amounts to 12$\%$.

The normalised yields of {\it signal} or  {\it open} pairs 
\begin{equation}  
{\cal N}(\rm `Open')= 
\int_{0.2}^{1.1}dm~\langle dN_{ee}(m)/dm\rangle/\langle N_{ch}\rangle
\end{equation} 
plotted in Fig.~\ref{fig:6.4} display a relative spread of 24$\%$,
twice that of the Dalitz sample and now well above the
systematic Dalitz-sample error. We presume that the larger systematic
uncertainties in open pair yields are caused by the large combinatorial
background. We take the relative sample error of 24$\%$ in number of
normalised open pairs as a reliable measure of the magnitude of
systematic uncertainties, arguing that they were obtained in rather
independent efforts and by using diverse strategies.

\section{Final Results and Comparison to Hadronic Sources}
\label{sec:7}
\subsection{Inclusive mass spectra}
\label{subsec:30}

\begin{figure}[t!]
\sidecaption
   \resizebox{0.48\textwidth}{!}{%
   \includegraphics{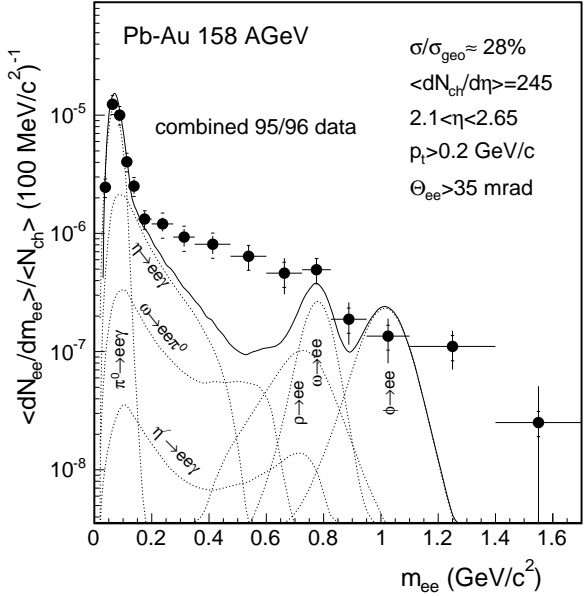}} 
\caption{\label{fig:7.1}
   Unified mass spectrum corresponding to average trigger centrality
   of 28$\%~\sigma/\sigma_{inel}$.  Yields are corrected for
   reconstruction efficiency and normalised to $\langle
   N_{ch}\rangle$. Statistical bin-to-bin errors are shown by vertical
   bars, systematic errors by brackets.  Horizontal bars indicate size
   of mass bins. Also shown is cocktail of hadron decays and its sum
   (solid line).  }
\end{figure}

The invariant mass spectrum of $e^+e^-$ pairs produced in 158~GeV/n
Pb-Au collisions as obtained from a weighted average of the individual
data sets, henceforth called `unified mass spectrum', is shown in
Fig.~\ref{fig:7.1}.  The squares of the inverse relative statistical
errors have been used as weights. Plotted is the differential yield
per event corrected for pair efficiency and normalised to the number
of charged particles in the acceptance, $\langle N_{ch}\rangle\approx
133$, which corresponds to the rapidity density $\langle
dN_{ch}/d\eta\rangle\approx 245$ and the most central 28$\%$ of the
reaction cross section.

The mass spectrum is compared to the expectation of hadron decays
modelled with the updated GENESIS code.  The resonance
structure of the light vector mesons is hardly visible in the measured
spectrum, and within statistical errors, the data points are
compatible with a smooth curve.  Note that the resonances were clearly
visible  in the p-Be and p-Au invariant mass spectra(Fig.~\ref{fig:1.1})
 despite a comparatively poor resolution of 9$\%$ at
$\rho/\omega$. Note also that the $\rho/\omega$ resonance seems to
become visible above the continuum in the mass spectrum selected for
$p_t^{ee}>$\,500\,MeV/$c^2$ of Fig.~\ref{fig:7.5}, to be
discussed below.

In the $\phi$ region, the expected mass resolution of 7$\%$ being
potentially sufficient, statistics hampers further conclusions.  We
nevertheless quote the ratio of the observed yield (in the mass region
0.95$< m <$\,1.1\,GeV/$c^2$) relative to the cocktail which is
dominated by the statistical-model result for the $\phi$ in this
region. We obtain a value of 0.68$\pm$\,0.28. With one standard
deviation below unity, the statistical accuracy is insufficient to settle
the pending controversy on the
$\phi$~\cite{jouan-falco98,puehlhofer98}.

\subsection{Enhancement factors}
\label{subsec:31}

All CERES results for Pb-Au collisions show pair yields in the
$\pi^\circ$-Dalitz region which are in good agreement with
predictions from known hadron decays. For masses above
200~MeV/$c^2$, however, the data overshoot the expectation from hadron
decays significantly. The largest enhancement over the
hadronic cocktail is in the mass range between 400 and 600~MeV/$c^2$
where it reaches a magnitude of six.

\begin{figure}[b!!]
   \resizebox{0.48\textwidth}{!}{%
   \includegraphics{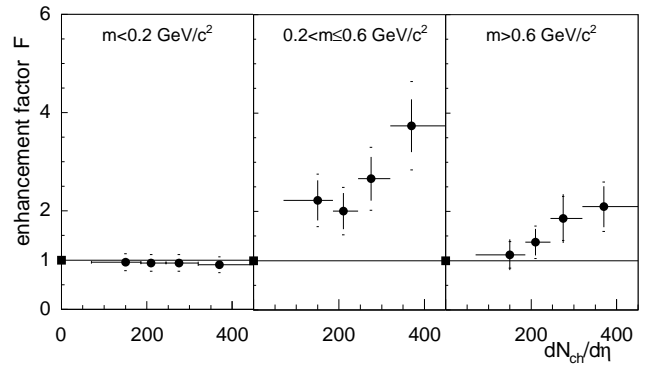}} 
\caption{\label{fig:7.2}
   Enhancement factor $\cal F$ {\it vs} $dN_{ch}/dy$ for three mass
   regions, unified data. The horizontal line indicates ${\cal F}$=~1
   of hadronic sources. The centre plot exhibits the largest effect,
   and the data points are consistent with a straight line that passes
   at $ dN_{ch}/d\eta=~0$ through `1' of the ordinate.}
\end{figure}

Integrating the normalised yields up to 200~MeV/$c^2$,
for the fraction dominated by $\pi^\circ$-Dalitz pairs (A),
and from 200\,MeV/$c^2$ upward for open pairs (B), we obtain
\begin{eqnarray}
\langle N_{ee}\rangle/\langle N_{ch}\rangle & = &\nonumber \\\left\{
\begin{array}{ll}
(8.52\pm 0.20~[stat.]\pm 1.54~[syst.])\,\times 10^{-6} & (A)\\
(5.25\pm 0.43~[stat.]\pm 1.26~[syst.])\,\times 10^{-6} & (B)
\end{array}
     \right . .
\end{eqnarray}
We have quoted the statistical and the systematic errors
of 18$\%$ and 24$\%$, respectively.

The ratios of the measured data to the integrated yields of the decay
cocktail (given by eqn.~(5.33)) for the two mass regions are
\begin{eqnarray}
{\cal F}= \frac{\langle N_{ee}\rangle/\langle N_{ch}\rangle}{
\langle N_{ee}\rangle/\langle N_{ch}\rangle|_{decays}} = &\nonumber \\
 \left\{
\begin{array}{ll}
0.92\pm0.02\,[stat]\pm0.17\,[syst]\pm0.07\,[decays]&\,(A)\\
2.31\pm0.19\,[stat]\pm0.55\,[syst]\pm0.69\,[decays]&\,(B). 
\end{array}
     \right . ,
\end{eqnarray}
where our estimate of the systematic error in the decay cocktail is
given separately.  For the low-mass continuum region 200~$\leq
m\leq$~600~MeV/$c^2$, the enhancement is even larger,
\begin{equation}
{\cal F}= 2.73\pm 0.25~[stat]\pm 0.65~[syst]\pm 0.82~[decays].
\end{equation}

\subsection{Centrality dependence}
\label{subsec:32}

The enhancement factors $\cal F$ are plotted {\it vs} charged-particle
density for three mass regions in Fig.~\ref{fig:7.2}.  The
enhancement for the mass region 200$\leq m\leq$~600~MeV/$c^2$ reaches
\begin{figure}[tt!!]
        \resizebox{0.48\textwidth}{!}{%
        \includegraphics{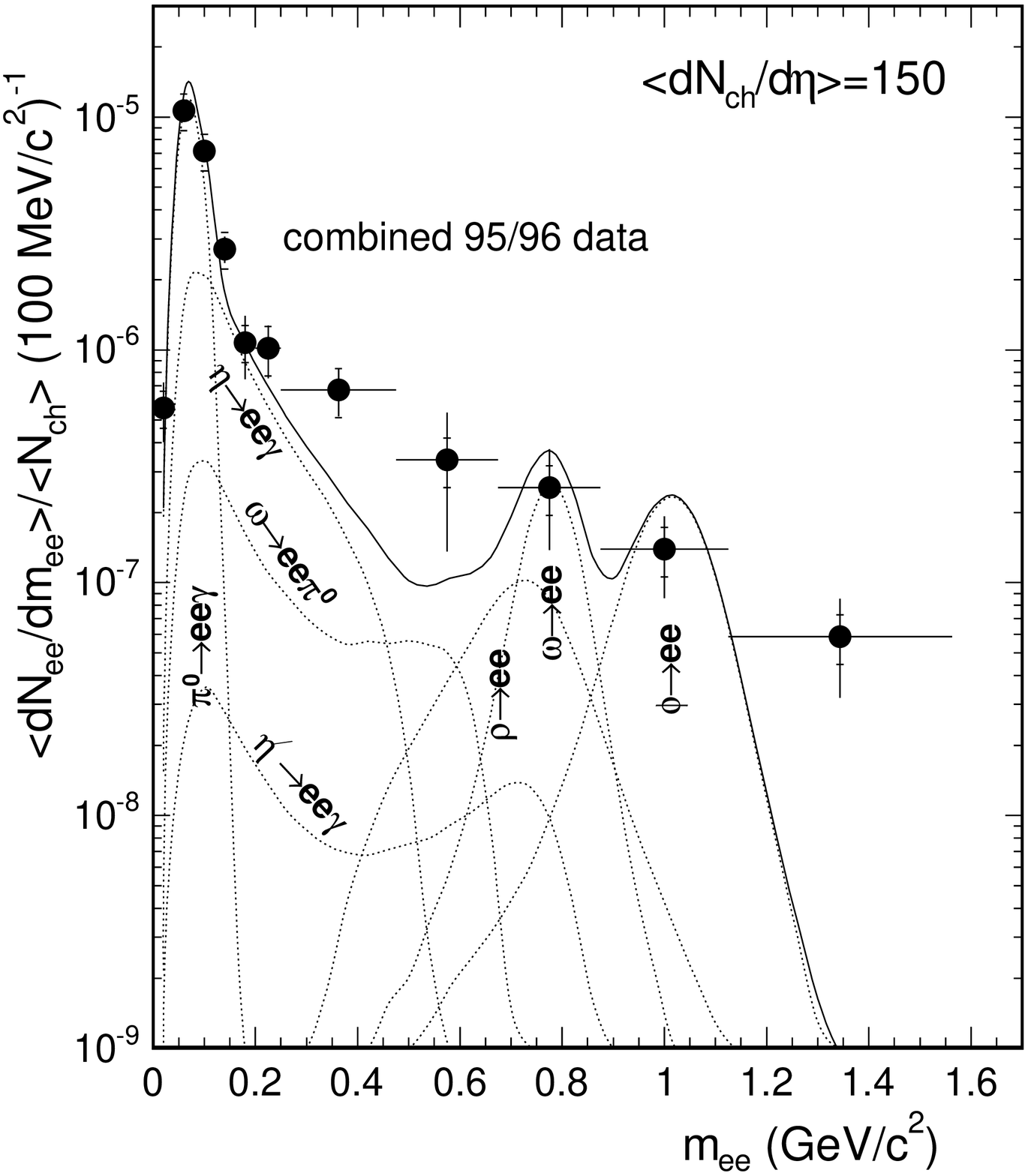}
        \includegraphics{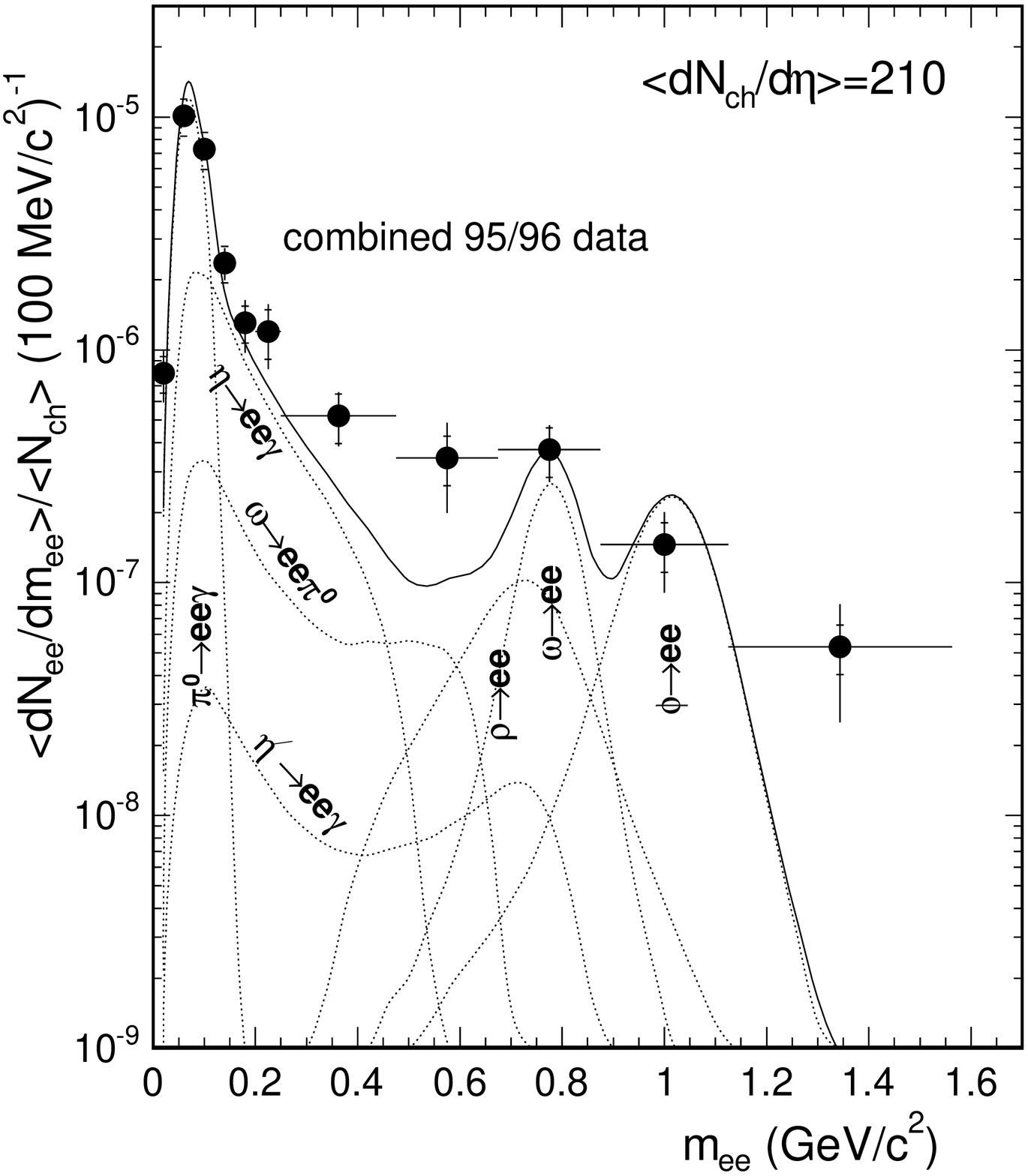}}
        \resizebox{0.48\textwidth}{!}{%
        \includegraphics{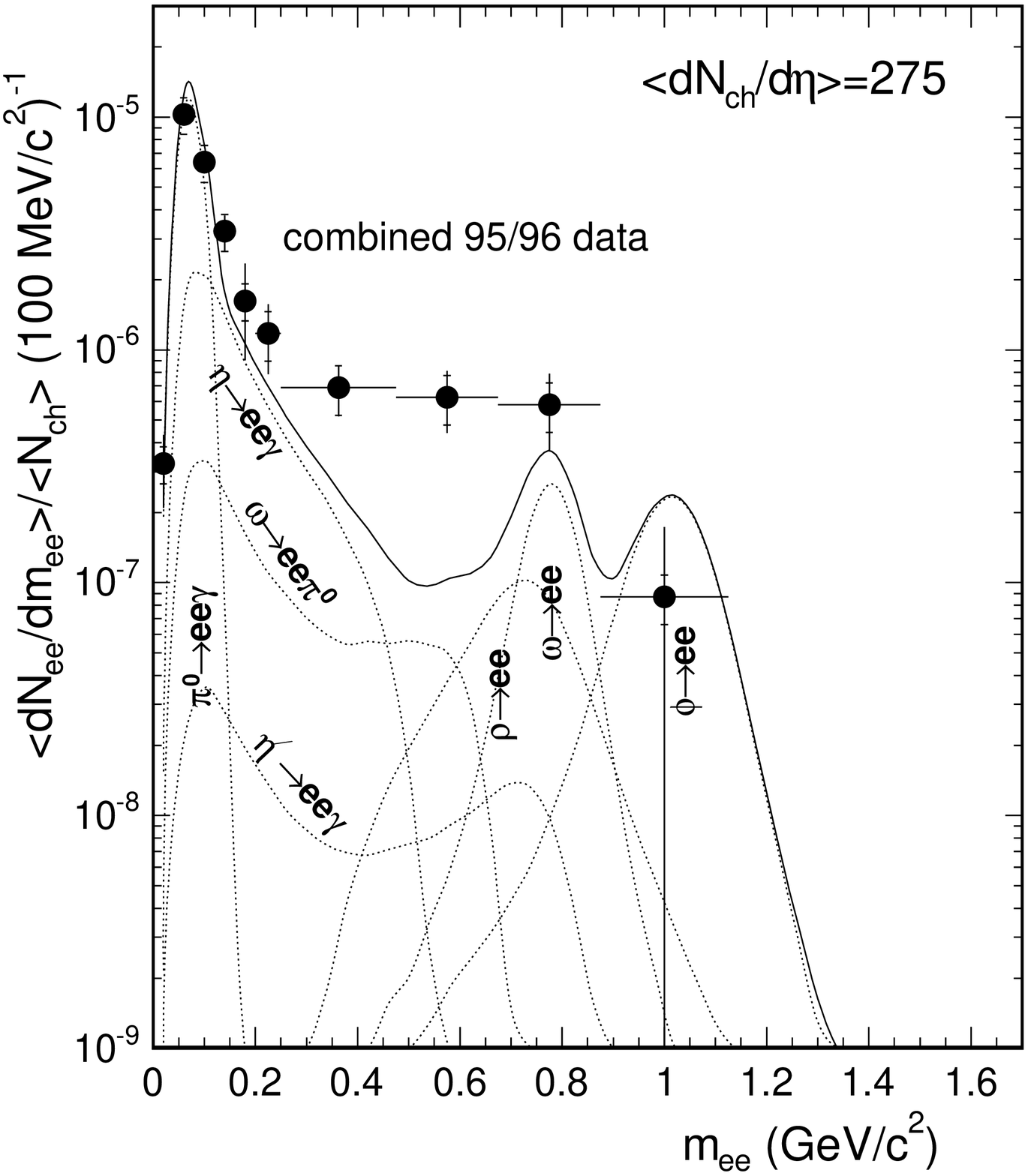}
        \includegraphics{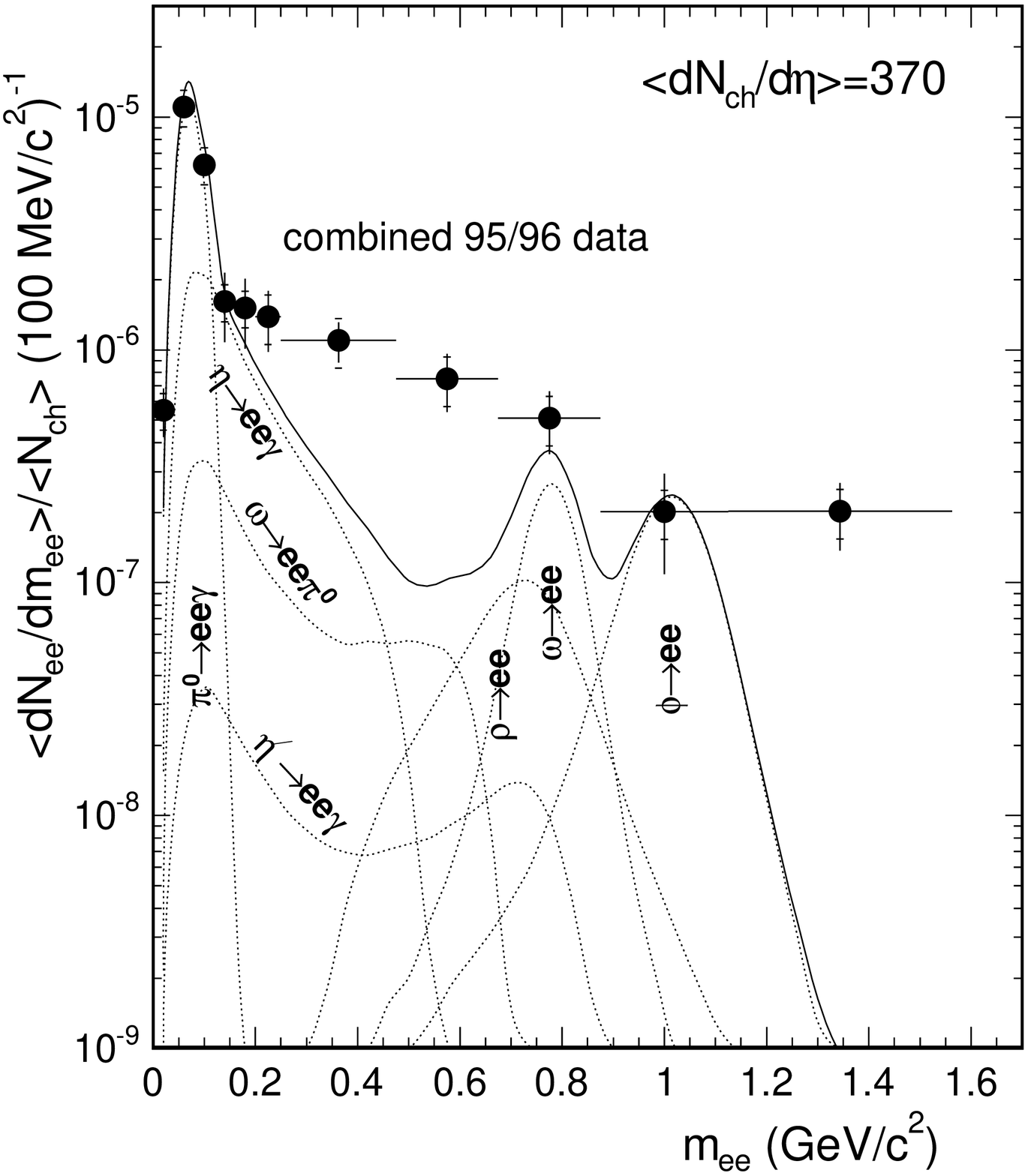}}
        \caption{\label{fig:7.3} Unified mass spectra for selected
        contiguous ranges in $N_{ch}$, $\langle
        N_{ch}\rangle$= 150, 210, 275, and 370 (from upper left to
        lower right). Unified data, standard analysis cuts.}
\end{figure}
about 4 at the most central collisions, and it is seen to rise about
linearly with charged-particle density; a straight-line fit to the
five data points gives a slope value deviating from zero by 4 standard
deviations.  This establishes that the integrated pair yield itself
has a stronger-than-linear dependence on charged particle density. (By
construction, the cocktail yield per $N_{ch}$ does not depend on
$N_{ch}$). This enhancement extends to the resonance region, although
with reduced significance and smaller values of the enhancement
factor. Unfortunately, statistical uncertainties, do not allow to
confirm or refute interesting details in the growth of the normalized
yield with $\langle N_{ch}\rangle$, like a threshold effect or a
saturation behaviour.

To see how the enhancement-typical spectral shape evolves with
increasing centrality, unified mass spectra for different ranges in
$N_{ch}$ are shown in Fig.~\ref{fig:7.3}. The series of spectra
clearly demonstrates that the excess over the cocktail for all four
spectra occurs between the two-pion threshold and the
$\rho/\omega$-resonance position, increasing with centrality.  The
spectra, being statistically independent samples, corroborate that the
largest enhancement is around 500~MeV/$c^2$, well below the
$\rho/\omega$ position.  With respect to the central issue of
in-medium modifications, one should be careful not to establish a
direct link to the enhancement factor as we defined it. The reason is
that pion annihilation {\em per se}, i.e.  with a vacuum $\rho$,
proceeding in the hot fireball is already a large, if not the
dominant, contribution to the enhancement factor.

\subsection{Pair transverse momentum}
\label{subsec:33}

\begin{figure}[t!]
   \resizebox{0.48\textwidth}{!}{%
    \includegraphics{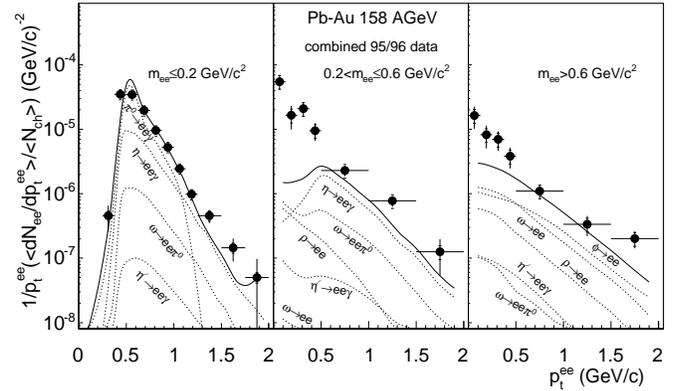}}   
    \caption{\label{fig:7.4}Unified invariant pair-p$_t$ spectra  
$m_{ee}\!\!\!\!\!\leq$\,0.2\,(left),~0.2$<\!\!\!\!\!
m_{ee}\!\!\!\leq$\,0.6\,(centre),
and  $m_{ee}>$~600\,MeV/$c$~(right). 
 28$\%~\sigma/\sigma_{inel}$.}  
\end{figure}

The Lorentz-invariant transverse momentum distributions of produced
electron pairs, observed first in the 1995 data
analysis~\cite{voigt-phd}, are shown in Fig.~\ref{fig:7.4} for the
combined 95/96 data for the $\pi^\circ$-Dalitz region and for open
pairs. At low pair $p_t$, denoted by $p_t^{ee}$, the yield is
suppressed due to the single-track $p_t$ cut at 200\,MeV/c.  As to be
expected, there is good agreement between data and hadronic cocktail
in the Dalitz region. However, the information from the pair $p_t$
distribution in the open-pair mass range is striking: the enhancement
grows towards small pair $p_t^{ee}$ despite the $p_t\geq$200~MeV/$c$
condition on single electron tracks. The surplus originates from
decays of virtual photons within the fireball that are created
favourably at rest. The enhancement above 500~MeV/$c$ pair transverse
momentum is considerably reduced. Whether this observation trivially
reflects the annihilation kinematics in a thermal pion gas, or
contains information about modified hadron properties, will be
discussed in sect.~8.

\begin{figure*}[tt!!]
\begin{center}
   \resizebox{0.70\textwidth}{!}{%
    \includegraphics{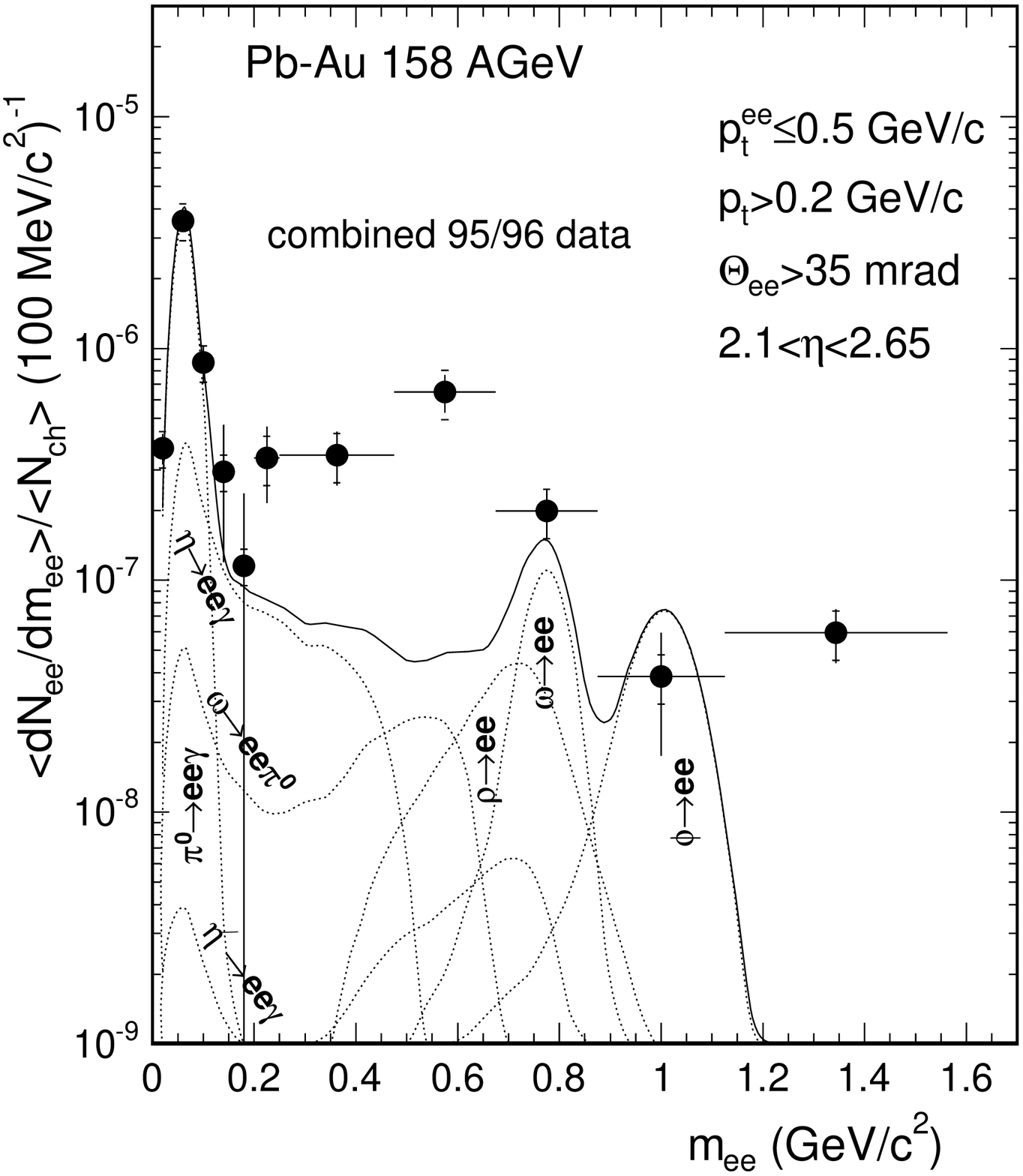}
    \includegraphics{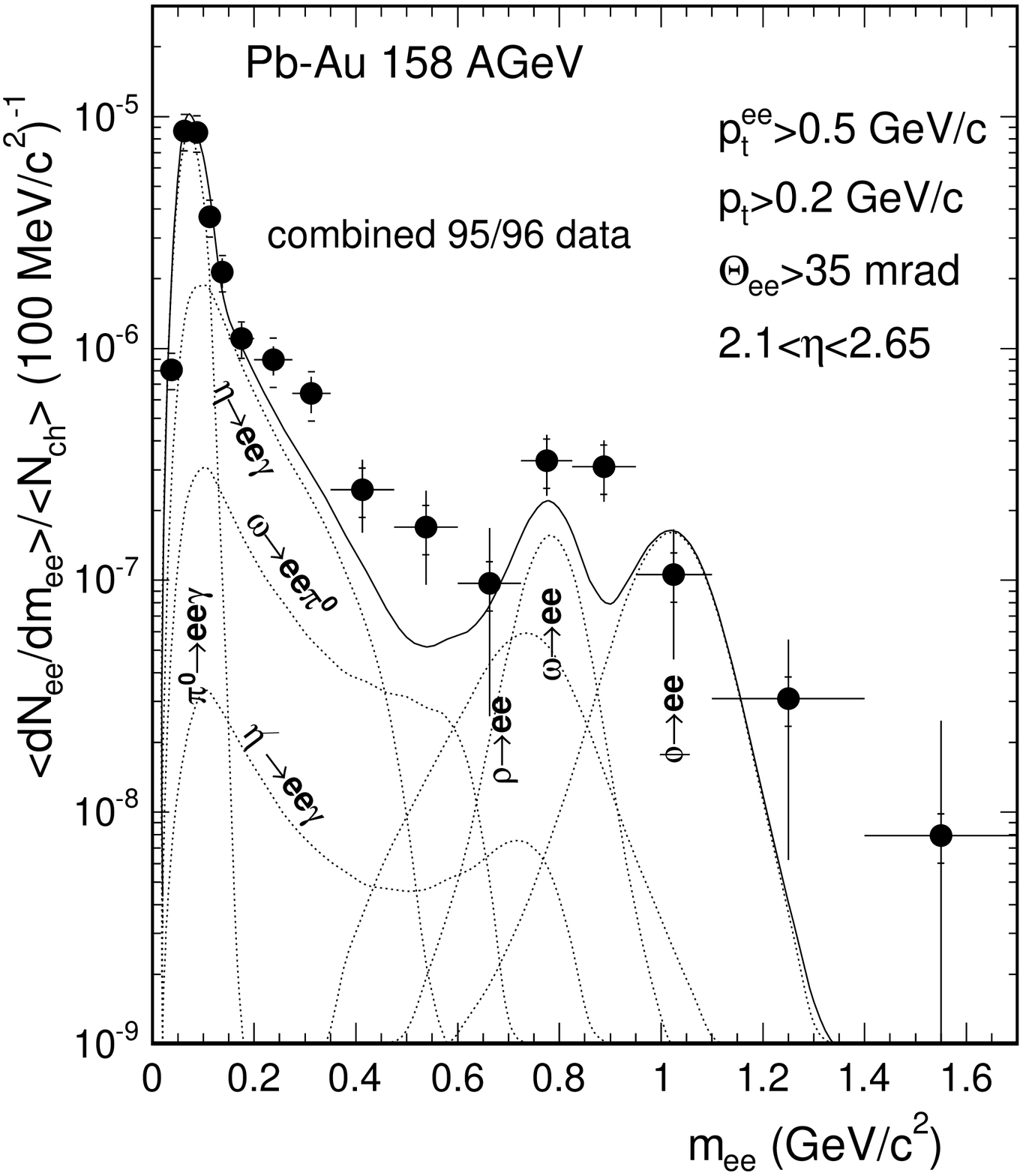}}   
\end{center} 
\caption{Unified mass spectra for two selected
   ranges of transverse pair momentum, $p_t^{ee}\leq$~500~MeV/$c$ (left),
and $p_t^{ee}>$~500~MeV/$c$ (right). Centrality 
 28$\%~\sigma/\sigma_{inel}$.
\label{fig:7.5}}
\end{figure*}

Conversely, the size of the average transverse pair momentum has a
remarkable influence on the shape of the inclusive mass spectra. We
observe in Fig.~\ref{fig:7.5} that the measured yield dramatically
overshoots the hadron decay contributions when low pair momenta are
selected. The enhancement in the mass range 500-700~MeV/$c^2$ reaches
locally an order of magnitude.
In contrast, for larger transverse pair momenta, the measured yield
comes pretty close in shape and magnitude to the cocktail
expectations; the resonance region becomes visible as it gains in
yield, even on absolute scale, while the previously amplified continuum 
at 400$\leq m\leq $600~MeV/$c^2$ is being deflated.

\section{Physics discussion}
\label{sec:8}

The two CERES runs of 1995 and 1996, combined to the largest
statistics sample of 158~GeV/n Pb-Au collisions taken so far,
corroborate the enhancement in low-mass electron-pair production over
that from hadron decays originally observed in 200\,GeV/n S-Au
collisions by CERES~\cite{prl1995} and in S-W collisions by
Helios-3~\cite{masera1995}.  Two additional CERES Pb-Au runs have been
performed since then with the new TPC and improved mass resolution,
one at the reduced energy of
40~GeV/n\,\cite{adamova-ee03,appelshauser02,wessels02} the results of
which are shortly addressed below, the other at
158~GeV/n\,\cite{marin04} with quantitatively consistent results.

In the region of the $\pi^\circ$-Dalitz decay, $m\leq$\,200~MeV/$c^2$,
the yield of the unified sample agrees with the cocktail within the
estimated systematic uncertainties.  In the mass range upward of
200~MeV/$c^2$, the normalised pair yield of the unified data analysis
significantly exceeds the yield from hadron decays per charged
particle, and the enhancement factor is given together with our
estimates of statistical and systematic errors in
sect.~\ref{subsec:31}.

The systematic errors of the decay cocktail were discussed in
sect.~4. To address the reliability of the generator in physics terms,
the situation has remarkably improved since the practice of scaling
yields up from p-p to Pb-Au collisions was abandoned in favour of
using particle ratios from statistical model systematics based on
Pb-beam data itself. Still, we shortly recall here some particle
ratios that have become subject to speculations of being grossly
enhanced (i.e.\@ by factors more than two) in nucleus-nucleus compared
to p-p reactions, even beyond the statistical-model systematics.

First, the $\eta/\pi^\circ$ ratio received interest from the fact that
the $\eta$-Dalitz decay is the most important single component of the
hadronic cocktail being suspect of enhanced production. It was
argued~\cite{drees-eta}, however, that an $\eta/\pi^\circ$ ratio
sufficiently large to explain the electron pair enhancement had not
been observed in the photon yield measured by WA80~\cite{wa80}. 

In the region below the $\rho/\omega$, the gap between cocktail and
data might be filled by raising the $\omega$-Dalitz decay contribution
as has been proposed by V.~Koch some time ago~\cite{koch99}. We hold
\begin{figure*}[tt!]
\begin{center}
   \resizebox{0.60\textwidth}{!}{%
    \includegraphics{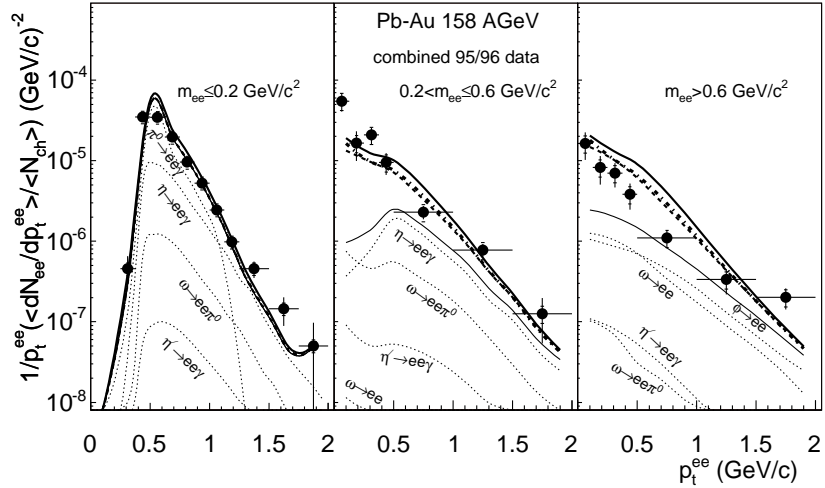}}   
\end{center}
    \caption{\label{fig:8.1} The $p_t^{ee}$ spectra of 
    Fig.~\ref{fig:7.4} compared to (i) free
    hadron decays without $\rho$ (thin solid line), (ii)
    model calculations with a vacuum $\rho$ spectral function
    (thick dashed line), 
   (iii) with dropping in-medium $\rho$ mass
    (thick dashed-dotted line), (iv) with a medium-modified $\rho$
    spectral function (thick solid line). For a cocktail
    including the $\rho$ decay see Fig.~\ref{fig:7.4}.
}
\end{figure*}
against that there is no indication otherwise for a strongly enhanced
$\omega$ production.  In addition, a boost in $\omega$-Dalitz
production is limited to an upper margin set by the data of
Fig.~\ref{fig:7.1} to the {\it direct} decay $\omega\rightarrow
e^+e^-$. Eventually, there is absolutely no reason to believe that
open-charm production in nucleus-nucleus collisions should be as
enormously enhanced as to explain the low-mass
enhancement~\cite{open-charm98}.

We return to state that the enhanced production of low-mass electron
pairs cannot be attributed to decays of produced hadrons. The excess
has to originate from processes which are active during the lifetime
of the fireball, i.e.\@ between the onset of hadronisation and
kinetic freeze-out, if of hadronic origin, and before hadronisation,
if created during the plasma phase. Present theoretical studies
allocate only a very minor fraction to the partonic part which reflects the
supposedly small 4-volumes of deconfined matter at SPS energies.

After having tried to give an overview of theoretical models of
dilepton production in the introduction, this discussion will be
guided by a few representative theoretical models: the spectral
function approach~\cite{rapp-wambach2000}, the dropping-mass
scenario~\cite{b-r03}, and pion annihilation with a vacuum $\rho$.  In
all calculations to be shown, the same fireball model has been used to
describe the space-time evolution~\cite{rapp-special}.

Which are the experimental signatures of pion annihilation, the
process most widely ascribed to take place in the dense hadronic
fireball~? As for any binary process, the annihilation rate is expected to
scale with the squared density of the particles annihilating in the
fireball (of course, the argument applies as well to $q\bar q$
annihilation). Unfortunately, the centrality dependence of the
dilepton yield is a topic barely addressed by full transport
calculations, except~\cite{cassing-brat1999}.

We take the observed stronger-than-linear scaling of the pair yield (in
the mass region of the strongest enhancement) with $N_{ch}$ as strong
evidence in support of the two-body annihilation reaction. This
deserves some words of justification. An increase in $N_{ch}$ signals
a larger pion density only to the extent that it is not compensated by
an associated increase in volume, maybe even in lifetime, of the
fireball. Therefore, pion annihilation does not necessarily go along
with quadratic scaling in $N_{ch}$. Conversely, however, an observed
stronger-than-linear scaling of the pair yield with $N_{ch}$ is
sufficient reason to infer a binary reaction at work, i.e. strongly
suggesting pion annihilation in hadronic matter, or $q\bar q$
annihilation in the plasma phase.

As other annihilation processes in a thermal medium, $\pi\pi$
annihilation takes place favourably at small relative momentum
of the constituents.
This behaviour was indeed encountered already in the {\it
pair transverse momentum spectra} of Fig~\ref{fig:7.4}. From
Fig~\ref{fig:8.1} shown here, we learn in addition that pion
annihilation with a vacuum $\rho$ is hard to distinguish, by yield
and shape of its $p_t^{ee}$ spectrum, from the medium-modified
spectral function and the dropping-mass approaches.

\begin{figure}[tt!]
\begin{center}
   \resizebox{0.48\textwidth}{!}{%
    \includegraphics{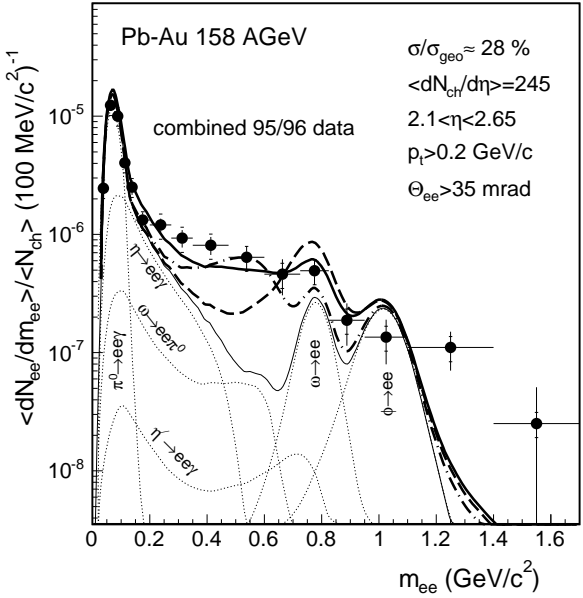}}  
\end{center} 
    \caption{\label{fig:8.2} Comparison of the inclusive mass
    spectrum of Fig.~\ref{fig:7.1} to (i) free
    hadron decays without $\rho$ decay (thin solid line), (ii)
    model calculations with a vacuum $\rho$ spectral function
    (thick dashed line), (iii) with dropping in-medium $\rho$ mass
    (thick dashed-dotted line), (iv) with a medium-modified $\rho$
    spectral function (thick solid line). A corresponding cocktail
    including the $\rho$ decay is shown in Fig.~\ref{fig:7.1}.
}
\end{figure}

The {\it invariant mass spectra} tell more about the physics processes
involved. In Fig.~\ref{fig:8.2}, the measured invariant mass spectrum
is compared to the model calculations.  It is evident at first sight
that pion annihilation with a vacuum $\rho$ (thick dashed line) does
not describe the shape of the spectrum; rather the calculations
overshoot the data at the nominal $\rho$ position by about a factor of
2 and under-predict the data in the continuum region by about a factor
of 3 -- yet, the integral yield comes out about right.

In-medium modifications produce a dramatic change:
both the calculations with an in-medium modified $\rho$ spectral
function (thick solid line), and with a dropping in-medium $\rho$
mass (thick dashed-dotted line) describe very well the marked increase
of the continuum yield around 500~MeV/$c^2$ as well as the
depletion at the vacuum $\rho/\omega$ position; the differences
among the competing approaches again are rather subtle. We remark that
calculations adopting the chiral reduction formalism approach
give very similar results, except that the depletion at
the vacuum $\rho/\omega$ position is absent~\cite{steele-zahed99}.

Before discussing medium modifications in some detail, let us examine
the contributions of $\rho$ and $\omega$ to the cocktail (for
$m>$400~MeV/$c^2$, say), and more speculative, to in-medium pair
production.  While the $\omega$ with its direct decay and part of its
Dalitz decay clearly dominates over the $\rho$ in the cocktail (see
Fig.~\ref{fig:7.1}), it is the $\rho$ which provides essentially all
of the enhancement.  One might wonder by which mechanism such drastic
change should be accomplished.

Some estimates based on vacuum properties are collected in the
appendix. The yield from mesons formed initially by {\it
hadronisation} is topped by the $\omega$ by its larger
electro-magnetic branching ratio.  In contrast, by
$e^+e^-$ decays from the {\it in-medium} meson
population, the $\rho$ wins by a large margin due to
its much larger $\pi\pi$ width. Altogether, we find that the
$\rho$/$\omega$ ratio of pair yields from secondary, in-medium
generated mesons, is by orders of magnitude larger than for initially
produced mesons.

 This digression illustrates from another point of view what we know
 already: the  exceptional potential of the $\rho$ propagator to
 dominate electron-pair production quite unrelated to its share in
 the cocktail -- and that there is no way to describe the
 surplus of electron pairs other than by direct or thermal radiation
 out of the fireball, i.e. from mesons which are regeneratively
 produced in $\pi\pi$ annihilation. The chances are feeble to
 observe such radiation from the $\omega$, primarily because of its
 extremely weak coupling to the hadronic medium.  Medium
 modifications, however, might be observed from primary $\omega$ and
 $\phi$ mesons by the fraction that decays within the lifetime of the
 fireball.

We return to compare theory to our data. Figure \ref{fig:8.1} shows
that the enhancement at low pair $p_t^{ee}$ over the cocktail which is
seen in the continuum region and to a lesser extent also in the
resonance region is a feature present in all model calculations, with
minor differences only between a dropping, broadening, or vacuum $\rho$
propagator. We meet here the governing features of pion annihilation,
rather than of medium modifications proper.

It is the mass spectrum which uncovers the characteristic of medium
modifications as distinct from pion annihilation with vacuum $\rho$
propagator as seen from the comparison of the data with the three
modell calculations in Fig.~\ref{fig:8.2}. The differences between the
two models incorporating medium modifications, however, are rather
subtle.

\begin{figure*}[tt!]
\begin{center}
   \resizebox{0.35\textwidth}{!}{%
    \includegraphics{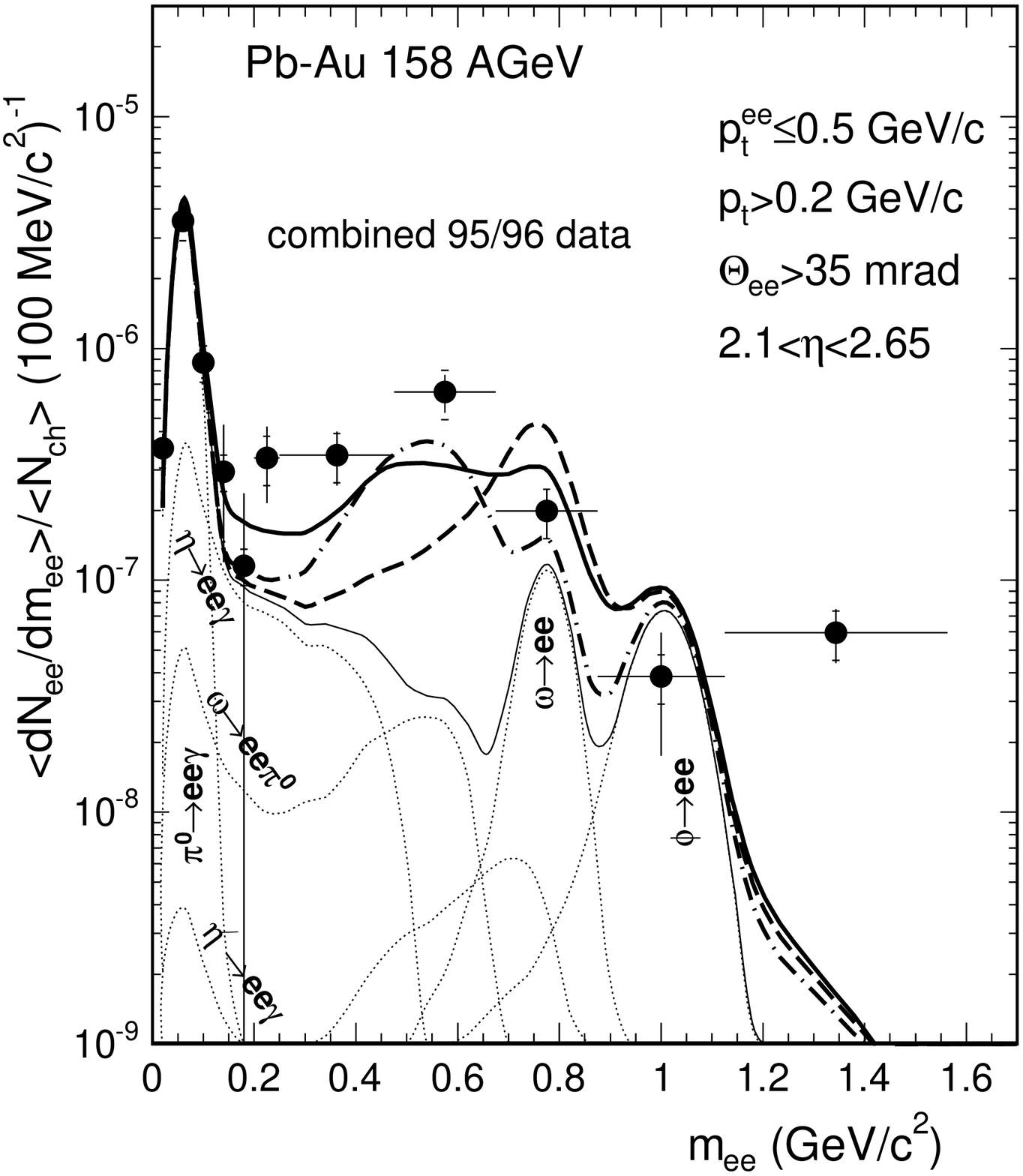}}   
   \resizebox{0.35\textwidth}{!}{%
    \includegraphics{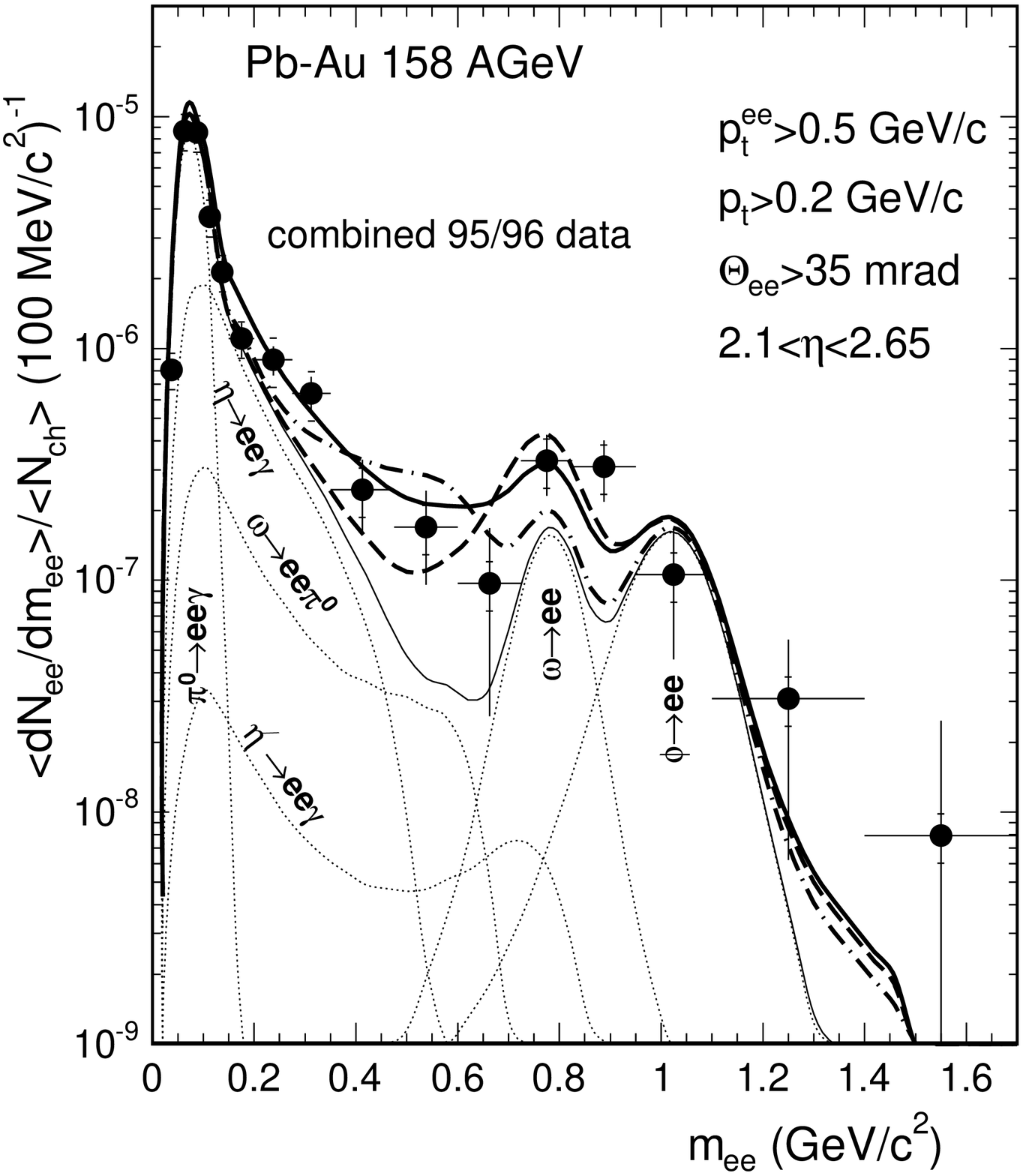}}  
\end{center} 
\caption{Comparison of the $p_t^{ee}$-selected mass spectra 
 of Fig.~\ref{fig:7.5}, $p_t^{ee}\leq$~500~MeV/$c$ (left),
 $p_t^{ee}>$~500~MeV/$c$ (right) to (i) free hadron decays without
$\rho$ (thin solid line), (ii) spectral function  with vacuum $\rho$
 (thick dashed line), (iii) in-medium dropping
 $\rho$ mass (thick dashed-dotted line), (iv) in-medium $\rho$ spectral 
function (thick solid line). 
\label{fig:8.3}}
\end{figure*}

Let us inspect the model calculations of $p_t^{ee}$-selected mass
spectra compared in Fig.~\ref{fig:8.3} to the data. The drastic
impact on the shape of the mass spectrum the selection of low
$p_t^{ee}$ has, is also present in the two model calculations with
modified $\rho$. In these models, the effect maybe somewhat weaker,
yet locally the enhancement over the cocktail reaches 10 (see
Fig.~\ref{fig:7.5} for the complete cocktail).  The vacuum-$\rho$
calculation is only weakly affected by the $p_t^{ee}$-selection.  For
larger $p_t^{ee}$, data and model calculations come much closer to the
decay cocktail.

The processes causing the in-medium changes of the $\rho$ spectral
function, or the dropping mass of the $\rho$, clearly also favour low
pair $p_t^{ee}$. Such behaviour would arise most naturally in the
dropping-mass scenario from the Boltzmann (or Bose) factor producing
the largest gain for small in-medium masses at vanishing 3-momentum
(see App.~A). For the spectral function approach, the observation had
its impact to install the $s$-wave N(1520) $\rho$-nucleon
resonance~\cite{peters98,rapp-wambach2000} as the moving agent in
place of the $p$-wave N(1720) resonance which had pioneered the
importance of $\rho\,N$ resonances for medium
modifications~\cite{friman-pirner97}; this change also met
requirements by photo absorption data to soften the form
factor~\cite{urban98}.

The effective downward shift of strength to lower masses in
the melting-$\rho$ treatment~\cite{rapp-chanfray-w1997} is largely due
to strong meson-baryon coupling, and most approaches agree to its
importance for generating in-medium
effects~\cite{rapp-chanfray-w1996,klingl-weise1996,steele-zahed99}; a
finite nucleon chemical potential is required also for some
meson-meson mixing effects to take place in approaching chiral
symmetry restoration~\cite{chan-del-erik98,theo01}. Only very small
effects of baryon density have been reported for UrQMD transport
calculations~\cite{bleicher00}.\footnote{This conclusion rests on the
(false) premise that the data have been satisfactorily described
without medium modifications using boosted $\omega$ and $\eta$ Dalitz
decays.}  We have remarked that the CERES run at reduced SPS energy of
40\,GeV/n observed an even larger enhancement as in 158\,GeV/n
collisions~\cite{adamova-ee03} reaffirming the
conclusion~\cite{rapp-nassau} that the increase in baryon density
seems to have won over the reduced pion density, or lower
temperature. As it is the total baryon density that matters - vector
mesons interact symmetrically with baryons and
anti-baryons~\cite{rapp-prc01}- the situation at RHIC energies will
not be greatly different from top SPS energy (despite vanishing net
baryon density).

In concluding this review of selected theory descriptions of the
CERES Pb-Au dilepton data, we like to add that the spectral function
approach had also other successes. Within the same framework, the
intermediate-mass enhancement observed by NA50~\cite{abreu00} has been
successfully described as thermal radiation with a 30$\%$ share
of the quark gluon plasma~\cite{rapp-shuryak00}; there
was no need to invoke open-charm enhancement. This finding may be seen
as the first glimpse of light in the long search for thermal $\bar qq$
radiation from the quark-gluon plasma\cite{shu80,mclerran85}.

Still within the same framework, the $p_t$ spectra of
photons in 158\,GeV/n Pb-Pb collisions measured by\\
WA98~\cite{aggarwal02} have been reproduced; up to transverse momenta
of about 1.5~GeV/$c$, thermal emission from the expanding hadronic
fireball has been found to dominate with photons mainly of baryonic
origin~\cite{turbide04}, fully consistent with the closely related
calculations that describe the low-mass dilepton excess observed by
CERES.

\section{Conclusion and Outlook}
\label{sec:9}

The analysis of the large unified data sample has substantiated the
earlier finding of a strong source of continuum electron pairs which
contributes to the invariant mass spectrum beyond the decays of
produced mesons, most strongly around 500~MeV/$c^2$.  There is ample
evidence that we observe dilepton radiation from the inte\-rior of the
hadronic fireball in which pion annihilation mediated by the $\rho$
propagator plays a major role. The excess yield rises significantly
steeper than linearly with charged-particle density, consistent with
the binary annihilation process. Another piece of circumstantial
evidence for $\pi\pi$ annihilation is delivered by the invariant pair
transverse-momentum spectra for the continuum pairs of masses between 
200 and 600\,MeV/$c^2$: the dramatic enhancement over the
cocktail occurs at very low $p_t^{ee}$.

Full calculations with vacuum $\rho$ describe the measured yield about
correctly, but fail to account for the characteristic shape of
the spectrum of excess dileptons.  The $\rho$ propagator is manifestly
modified in the medium which is well described by two theories
incorporating medium modifications of the $\varrho$, which are,
however, very different in concept: while the Brown-Rho scaling
hypothesis explicitly refers to restoration of chiral symmetry, the
many-body spectral-function approach of Rapp and Wambach, although
tracing some of the induced mixings, does not have chiral symmetry
restoration as a ruling concept. That it may very well be implicitly
included is inferred from the fact that the hadronic dilepton rates
extrapolated up to $T_c$ (`bottom up') come out very similar to the
(`top down') extrapolated perturbative QGP rates~\cite{rw98}.  Both
theories also give a good description of the mass spectra for selected
ranges of pair transverse momentum $p_t^{ee}$, where an additional
preference of the mechanism generating the medium modifications has
shown up.
 
Three unsolved issues remain. To the present data accuracy, it is not
possible to decide between one or the other of the two competing
theories so that the role of chiral symmetry restoration for in-medium
modifications remains unclear. 

Patience seems also advised in localising the source of
medium-modified electron-pair production within the phase diagram. On
one hand, preformation of vector and axial-vector correlator strengths
in the non-perturbative plasma might influence dilepton production
across the\\ phase boundary~\cite{lee98,jaikumar02}. On the late end of
the time scale, the impact of a growing pion chemical potential for
dilepton production might not be entirely settled yet.

An issue widely overlooked is whether the hadronic fireball is
`boiling' long enough to radiate sufficient amounts of dileptons.  The
time spent by the system between chemical and kinetic freeze-out came
under scrutiny on the basis of recent pion interferometric
data~\cite{adamova03-hbt} which indicated only a 30$\%$ change in
volume. If, in addition, chemical freeze-out should occur essentially
at the phase boundary between hadronic and quark matter, as suggested
by the asymptotic statistical-model result of
$T_{chem}\approx$~170\,MeV at higher energies~\cite{pbm04}, the purely
hadronic origin of the low-mass dilepton enhancement might have to be
negotiated again.

We look out to further progress that can only be expected from
radically better data, with greatly improved statistics and less
combinatorial background, but also with improved mass resolution.
This will not be easy. But a first step in this direction is the CERES
2000 run with the new Time Projection Chamber; preliminary data have
been presented very recently~\cite{marin04}.

\acknowledgement{{\bf Acknowledgement}\\

We acknowledge the good performance of the CERN PS and SPS
accelerators and the excellent support for the central data recording
from the IT division. We are grateful to D.A.~Pinelli at BNL and
O.~Runolfsson at CERN for their delicate work in the assembly of
motherboards for SIDC's.  We acknowledge the support by Deutsches
Bundesministerium f\"ur Bildung, Wissenschaft, Forschung und
Technologie (BMBF), the U.S. Department of Energy, the Minerva
Foundation, the Israeli Science Foundation, and the German Israeli
Foundation for Scientific Research and Development.
}\\

\appendix
\section{Appendix $\rho\rightarrow e^+e^-$ decay rate}
\label{sec:appendixA}

The thermal emission rate of dielectrons from
two-body decay of rho mesons has been worked out from 
Ref.~\cite{song} by 
B.~Friman and J.~Knoll~\cite{frimann-knoll00}
as follows:

\begin{eqnarray}
        \frac{dR}{dM d^{\,3}\vec{q}} =  
\frac{\alpha^2m^4_{\rho}}{3(2\pi)^{4}}\frac{(1- 4m_{\pi}^2/M^2)^{3/2}}
{(M^2-m^2_{\rho})^2+M^2\Gamma_{tot}^{0^2}}\nonumber & \times &\\
~\left\{e^{-\sqrt{{\scriptstyle M^2}+ \vec{q}\,^2}/T} 
~\frac{M}{\sqrt{M^2+ \vec{q}\,^2}}\right\}.
\end{eqnarray}

Here,  $\mbox{\boldmath q}$ is the
\mbox{3-momentum} of the rho meson. The new expression
differs from the one used in the previous GENESIS code by the
Boltzmann-type phase space factor in the curled parentheses being
explicitly momentum dependent. To obtain the final invariant mass
distribution, the differential rate $dR/dM d^{3}\mbox{\boldmath q}$
has to be integrated over 3-momentum $\mbox{\boldmath q}$. The final
result used in the 2003 version of New GENESIS~\cite{genesis-new} for
comparison to the data reads

\begin{equation}
\frac{dR}{dM}=
\frac{\alpha^2 m_{\rho}^4}{3(2\pi)^4} \frac{(1- 4m_{\pi}^2/M^2)^{3/2}}
{(M^2-m_{\rho}^2)^2 + M^2\Gamma_{tot}^{0^2}} (2\pi MT)^{3/2}  e^{-M/T}.
\end{equation}

\section{Estimate of \Vr/\Vo ~ratios}
\label{sec:appendixB}

We estimate the $\rho/\omega$ ratio of electron-pair yields from
initially produced mesons and from those produced in a hadronic
fireball by using simplified order-of magnitude estimates
expressed solely by the (vacuum) particle properties~\cite{pdg2004}.

{\it Primary} mesons are produced during hadronisation. They decay
into electron pairs with a fraction given by the electro-magnetic
(e.m.) branching ratio $B_{ee}=
\Gamma_{ee}/\Gamma$. As in p-p collisions~\cite{neutral-meson-pBe},
we assume equal primary populations of $\rho$ and $\omega$. To the
mass range $m\geq$\,400~MeV/$c^2$, the $\omega$ contributes by its
direct decay and by about 15$\%$ of its Dalitz decay. In this mass
range, the ratio of the number of electron pairs emitted by the
initial population of $\rho$'s and $\omega$'s is
\begin{equation}
\left(\frac{Y_\rho}{Y_\omega}\right)_{\scriptstyle prim}
= \frac{B_{ee}(\rho)}{B_{ee}(\omega)}\approx 0.3.
\end{equation}
The $\rho$ looses in this comparison  of e.m. branching
ratios due its so much larger total
width of $\Gamma(\rho)$= 150~MeV, compared to $\Gamma(\omega)$=
8.4~MeV.

We are mostly interested in mesons that decay into electron pairs
within the lifetime of the fireball since the others cannot
probe medium modifications. Besides primary produced $\rho$
mesons, an additional source of in-medium decays are from

{\it Secondary} mesons. These are continuously produced in the
fireball by two-pion annihilation, and most of the time disintegrate
back into two pions as expressed by eqn.~(1.1).  We like to present
here a simple estimate of the relative in medium-pair production
yields from $\rho$ and $\omega$ which neglects all finer details of
the reaction dynamics by assuming that the ratio of rates is given as
the ratio of the 2-pion annihilation cross sections, which are
estimated from detailed balance, times the ratio of the e.m. decay
widths. The two vector mesons differ in an essential manner in
strength of coupling to the hadronic medium: while the $\pi\pi$
channel exhausts the large width of the \Vr,
$B_{\pi\pi}(\rho)\approx$\,1, the $\omega$ (with larger branching into
$3\pi$) is extremely weakly coupled to the $\pi\pi$ channel,
$B_{\pi\pi}(\omega)\approx$\,2\,$\%$.  With $Y_{med}\propto
\Gamma_{\pi\pi}\,\Gamma_{ee}$,
\begin{equation}
\left(\frac{Y_\rho}{Y_\omega}\right)_{\scriptstyle med}
\approx 4\times 10^{\textstyle 3}.
\end{equation}
The estimate demonstrates that by in-medium produced electron pairs,
the $\omega$ meson is completely outnumbered by the $\rho$, unless 
its in-medium properties should drastically be changed. 

\section{Instrumental asymmetries as source of enhancement?}
\label{sec:appendixC}

Let us assume there is an asymmetry in reconstruction efficiencies for
like-sign pairs (L) as compared to unlike-sign pairs (U) which would
affect the signal in a most direct way. Asymmetries of this kind might
be expected since rings tend to end up in a more or less dense
environment in RICH-2 depending on whether they are deflected away
from each other, or not.

Writing the pair efficiencies as $ \varepsilon_L\,=~\varepsilon_\circ,
~\varepsilon_U~=~\varepsilon_\circ\,(1\,+\,\delta)$,
the signal is given by
\begin{equation}
 S\approx(~U(1-\delta)-L)/\varepsilon~= S_\circ- \delta U/\varepsilon.
\end{equation}
To be definite, let us assume a signal ${\ S_\circ}$ three times the
hadronic background, i.e. an enhancement factor of three. The
enhancement will be gone if the asymmetry term reduces the signal to
one third of ${\ S_\circ}$. The corresponding asymmetry parameter
is
\begin{equation}
\delta=  2(U- L)/3U\approx \frac{2}{3}(U- L)/L.
\end{equation}
For a signal-to-background ratio  $(U-L)/L=~S/B=$\,1/13, we find that an
asymmetry of $\delta\approx$~5.1$\%$ in efficiencies for like-sign
and unlike-sign pairs is sufficient to fake an enhancement factor of 3.

\end{document}